\documentclass{ar-1col}

\usepackage{url} 
\usepackage{amsmath}
\usepackage{mathrsfs}
\usepackage[comma]{natbib}

\newcommand{\apj}{ApJ}
\newcommand{\apjl}{ApJL}
\newcommand{\apjs}{ApJS}
\newcommand{\aap}{A$\&$A}
\newcommand{\aapr}{A$\&$A}
\newcommand{\jcap}{J. Cosmol. Astropart. Phys.}

\newcommand{\araa}{ARA$\&$A}

\newcommand{\mnras}{MNRAS}
\newcommand{\pasj}{PASJ}
\newcommand{\prd}{Phys. Rev. D}
\newcommand{\prl}{Phys. Rev. Lett.}
\newcommand{\aj}{AJ}
\newcommand{\nat}{Nature}
\newcommand{\nar}{New Astron. Rev.}
\newcommand{\physrep}{Physics Reports}

\newcommand{\sovast}{Soviet Astronomy}

\newcommand{\pasa}{PASA}
\newcommand{\na}{New Astronomy}

\newcommand{\msun}{{\rm M}_\odot}
\newcommand{\rsun}{{\rm R}_\odot}
\newcommand{\zsun}{{\rm Z}_\odot}

\newcommand{\cc}{{\rm cm}^{-3}}
\newcommand{\mdot}{\dot m}
\newcommand{\msunyr}{{\rm M}_\odot~{\rm yr}^{-1}}
\newcommand{\kpc}{{\rm kpc}}
\newcommand{\mpc}{{\rm Mpc}}
\newcommand{\gpc}{{\rm Gpc}}

\newcommand{\K}{{\rm K}}
\newcommand{\kms}{{\rm km~s}^{-1}}
\newcommand{\pc}{{\rm pc}}

\newcommand{\beq}{\begin{equation}}
\newcommand{\eeq}{\end{equation}}

\newcommand\lsim{\mathrel{\rlap{\lower4pt\hbox{\hskip0pt$\sim$}}
        \raise1pt\hbox{$<$}}}
\newcommand\gsim{\mathrel{\rlap{\lower4pt\hbox{\hskip0pt$\sim$}}
        \raise1pt\hbox{$>$}}}

\jname{Annu. Rev. Astron. Astrophys.}
\jvol{58}
\jyear{2019}
\doi{10.1146/(please add article doi)}

\begin{document}

\markboth{Inayoshi et al.}{The First Massive Black Holes}

\hoffset=-1.0in
\voffset=-1.0in
%

\title{The Assembly of the First Massive Black Holes}

\author{Kohei Inayoshi,$^1$ Eli Visbal,$^2$ and Zolt\'an Haiman$^3$
\affil{$^1$Kavli Institute for Astronomy and Astrophysics, Peking University, Beijing 100871, China; email: inayoshi@pku.edu.cn}
\affil{$^2$Department of Physics \& Astronomy, University of Toledo, Toledo, Ohio 43606; email: Elijah.Visbal@utoledo.edu}
\affil{$^3$Department of Astronomy, Columbia University, New York, NY 10027; email: zoltan@astro.columbia.edu}}

\begin{abstract}
  The existence of ${\sim}10^9~\msun$ supermassive black holes (SMBHs)
  within the first billion year of the universe 
  has stimulated numerous ideas
  for the prompt formation and rapid growth of BHs in the early
  universe.  Here we review ways in which the seeds of massive BHs may
  have first assembled, how they may have subsequently grown as
  massive as ${\sim}10^9~\msun$, and how multi-messenger observations could 
  distinguish between different SMBH assembly scenarios.  We conclude
  the following:
 \begin{itemize}
    \setlength{\leftskip}{-5.5mm}
 \item The ultra-rare ${\sim}10^9~\msun$ SMBHs represent only the tip
   of the iceberg. \\
   Early BHs likely fill a continuum from
   stellar-mass ($\sim 10~\msun$) to the \\
   super-massive ($\sim10^9$)
   regime, reflecting a range of initial masses and \\
   growth histories.
 \item Stellar-mass BHs were likely left behind by the first
   generation of stars\\ 
   at redshifts as high as $\sim 30$, but their
   initial growth was typically\\
   stunted due to the shallow potential
   wells of their host galaxies.
 \item Conditions in some larger, metal-poor galaxies soon became
   conducive\\
   to the rapid formation and growth of massive `seed'
   holes, via gas \\
   accretion and by mergers in dense stellar clusters.
 \item BH masses depend on the environment (such as the number and\\
   properties of nearby radiation sources and the local baryonic
   streaming\\ velocity),
   and on the metal enrichment and assembly history
   of the host\\ galaxy.
 \item Distinguishing between assembly mechanisms will be difficult,
   but \\
   a combination of observations by LISA (probing massive BH
   growth\\ 
   via mergers) and by deep multi-wavelength
   electromagnetic observations\\
   (probing growth via gas accretion) is particularly promising.
 \end{itemize}
\end{abstract}

\maketitle

\tableofcontents

\section{INTRODUCTION}
\label{sec:intro}

\subsection{Summary of Observations}

Observations of high-redshift quasars ($z \gsim 6$) indicate that
supermassive black holes (SMBHs) with masses greater than
$\sim10^9~\msun$ formed within the first billion years after the Big
Bang. These objects are very rare (having number density of $\sim
1~{\rm Gpc}^{-3}$) and have so far been found in optical/infrared
(IR) surveys that cover very large portions of the sky.  The Sloan
Digital Sky Survey (SDSS) was the first to discover high-redshift
quasars \citep{Fan+2001,Fan+2003} and was followed by a number of
additional efforts such as the UKIRT Infrared Deep Sky Survey
\citep[UKIDSS;][]{Lawrence+2007}, the Canada-France High-redshift
Quasar Survey \citep[CFHQS;][]{Willott+2007}, and the Panoramic Survey
Telescope \& Rapid Response System 1
\citep[Pan-STARRS1;][]{Morganson+2012}.  These surveys have yielded
well over $100$ quasars with redshifts $z>6$, many of which have
inferred BH masses $M_\bullet >10^9~\msun$.  The most massive of
these, SDSS J010013.02+280225.8 \citep{Wu+2015}, has an estimated mass
of $1.2\times10^{10}~\msun$ at $z=6.3$.  The most distant, ULAS
J1342+0928 \citep{Banados+2018}, has a mass of $7.8\times 10^8~\msun$
at $z=7.54$.  These large surveys have enabled accurate
characterization of the bright end of the quasar luminosity function
(LF) \citep[e.g.][]{Jiang+2016}.  Recently, a large sample of
additional, lower-luminosity quasars have been uncovered in the Subaru
High-z Exploration of Low-Luminosity Quasars (SHELLQS), bringing the
total number of $z>6$ quasars to nearly 200, and extending the
constraints on the LF to fainter quasars \citep{Matsuoka+2018c}.
These observations are summarized in
\textbf{Figure~\ref{fig:mbh-vs-z}}, and will be discussed in detail in
\S~\ref{sec:obs} below.

In addition to optical/IR surveys, quasars have been observed across a
variety of wavelengths, from X-rays (with Chandra, XMM-Newton and
Swift-XRT) to radio (with the Very Large Array, Giant Metrewave Radio Telescope, 
and Murchison Widefield Array). For example, the FIRST survey \citep{Becker+1995}
found a number of quasars.  
Overall, observations across all wavelengths indicate that there is little
evolution in the physical properties of the brightest quasars or their host galaxies
over cosmic time, as inferred either from detailed optical/IR, X-ray~\citep{Nanni+2017},
or radio~\citep{Banados+2015} analyses, implying that the hosts of these objects 
formed early as well (see~\S~\ref{sec:obs}).

\subsection{Timescale Issues}

The presence of $>10^9~\msun$ SMBHs before the Universe was a billion
years old represents an intriguing puzzle.  \emph{How did the first
  SMBHs grow so large so fast?}  This question had been raised already
at the discovery of quasars at $4<z<5$~\citep{Turner1991}, but pushing
the redshift limits to $z>7$, and the correspondingly shorter cosmic
time available made it significantly more
intriguing~\citep{HaimanLoeb2001}.  A naive explanation is that these
early SMBHs were seeded by BH remnants of the first Population III
(Pop III) stars. Pop~III stars are expected to form in $\sim
10^{5-6}~\msun$ dark matter (DM) ``minihalos'' through primordial gas
undergoing molecular hydrogen (H$_2$) cooling.  The metal-free
primordial gas is significantly warmer (a few $100~\K$) than
star-forming molecular clouds in the interstellar medium (ISM) of
low-$z$ galaxies ($\sim 10~\K$).  The general expectation is that
inefficient cooling of the primordial gas leads to inefficient
fragmentation, making Pop~III stars unusually massive.  The initial
mass function of Pop~III stars remains uncertain, but simulations
suggest that it is indeed top-heavy, with a mass range of $10\lsim
M_\star/\msun \lsim 10^3$ \citep{Hirano+2014}.

If BH growth is dominated by Eddington-limited accretion, a seed will
grow exponentially with an $e$-folding time of $t_{\rm Edd} \approx
50~{\rm Myr} $, assuming a radiative efficiency of $\epsilon \approx
10\%$.  A comparison between the observed quasar activity across all
redshifts and the local population of remnant SMBHs \citep{Soltan1982}
implies that most low-$z$ SMBHs assembled the bulk of their mass at
$z=2-3$ at this efficiency \citep{Haehnelt+1998,YuTremaine2002,Shankar+2004}.  
This efficiency is also similar to the value ($\sim 0.06$) expected for
non-rotating BHs, based on their innermost stable circular orbit
(ISCO; \citealt{Rees1984}).  Assuming that the seeds of high-$z$ SMBHs
have a similar radiative efficiency, and that their accretion obeys
the corresponding Eddington limit, a $100~\msun$ Pop~III seed BH would
need to accrete for $\approx 0.8~{\rm Gyr}$ to reach $10^9~\msun$.
This is comparable to the age of the universe at $z\approx 6$, and
requires a duty cycle of near-Eddington accretion $f_{\rm duty}
\approx 1$ over eight orders of magnitude growth in mass.  Several
effects make such a high duty cycle for a Pop~III seed, sustained over
orders of magnitude growth in mass, unlikely, including feedback from
accretion onto the BH itself, as well as displacement of the gas
reservoir by UV radiation and supernovae (SN) explosions of the
Pop~III stars in the shallow gravitational potential wells of
minihalos
\citep[][see~\S\ref{sec:lightseed}]{JohnsonBromm2007,Whalen+2008,
  Milosavljevic+2009b,Alvarez+2009}.

A number of different scenarios have been put forward to ease these
timescale constraints and help explain the existence of $M_\bullet
\sim 10^9~\msun$ at $z=6-7$.  Generally, the two options are to
increase either the seed BH mass or the growth rate.  Before
enumerating these, it is worth making a few points.  First, even in
models with massive BH seeds, a high duty cycle is required, if
accretion is Eddington limited \citep{TanakaHaiman2009}.  While
feedback effects in minihalos make this unlikely, such efficient
accretion may be easier to maintain for larger seeds residing inside
more massive halos~(\citealt{DiMatteo+2008}; see \S~\ref{sec:BHIMF}).
Second, it is worth emphasizing that even for the most massive and
highest-redshift SMBHs, the time-averaged accretion rate needs to be
only modestly ($\sim$ twice) above the Eddington-limited rate
for most BH seeding models.
Moderately super-critical rates, at a few times the Eddington-limited
value, could be maintained with duty cycles of $\sim 20-30~\%$ in some
accretion disk models \citep[e.g.,][]{Sadowski2009,Madau+2014}.  Finally, we
emphasize that only a tiny minority of early BHs, born in highly
biased regions of the universe, grow to $\sim 10^9\msun$ by $z\gsim
6$.  The vast majority of massive BHs, born in more typical regions,
will remain far below this mass by this redshift.

\begin{figure}[t]
\includegraphics[width=3.5in]{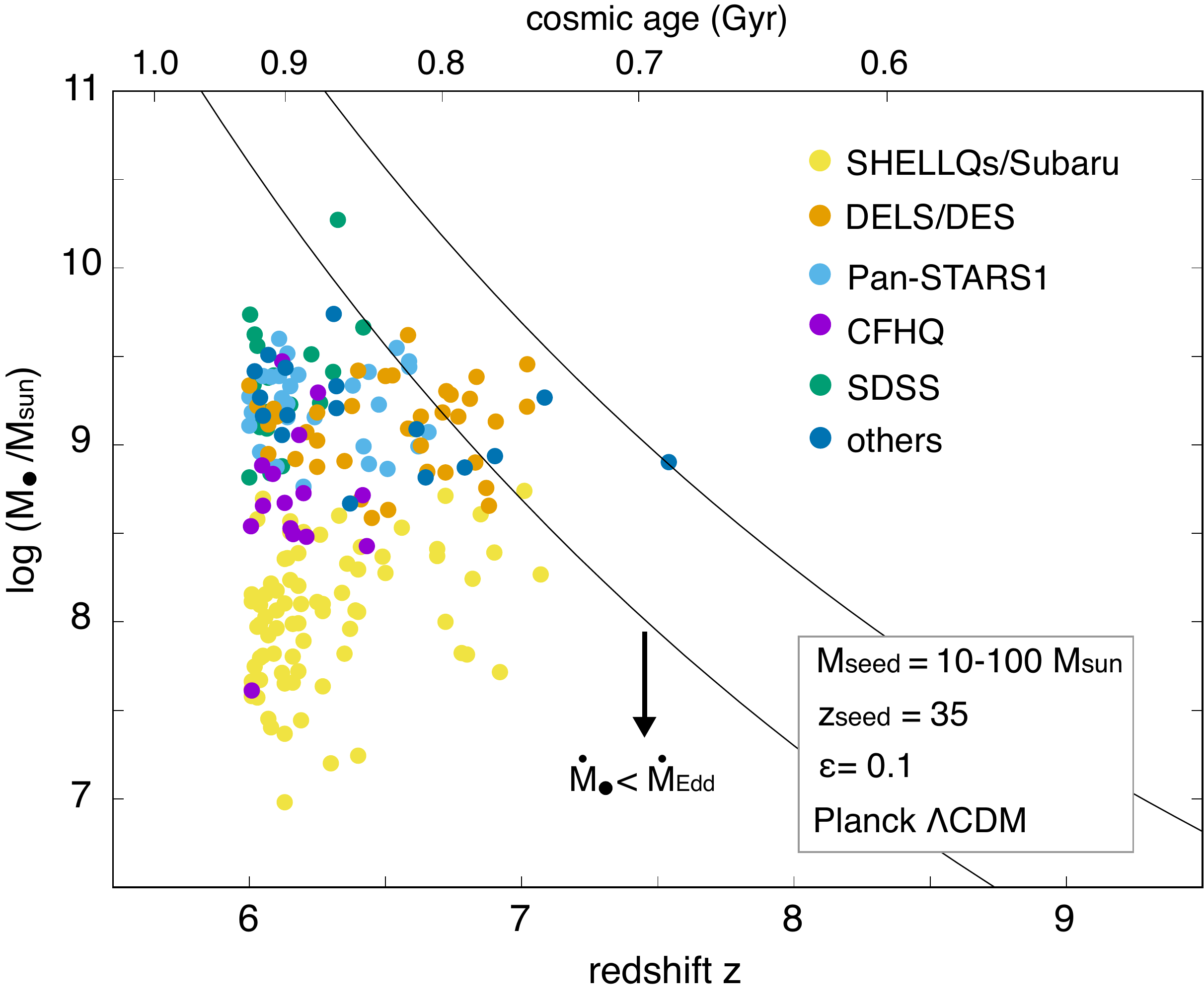}
\vspace{-1\baselineskip}
\caption{Masses and redshifts of quasars known at $z\geq 6$ to date.
  Of the 203 quasars shown, 79 sources were adopted from the
  compilation by \citet{Banados+2016} complete as of March 2016.
  Subsequently discovered quasars were added from
  Pan-STARRS (8 quasars; \citealt{Mazzucchelli+2017,Tang+2017,Koptelova+2017,Banados+2018}),
  VIKING (3 quasars; \citealt{Decarli+2018,Venemans+2019}),
  DES/SDSS combined with data from surveys in the IR by UKIRT, VISTA, and  {\it WISE} (37 quasars;
  \citealt{Wang+2016b,Reed+2017,Wang+2017,Reed+2019,Wang+2018b,
    Fan+2019,Yang+2019}),
  from the SHELLQs and other Subaru surveys (72 quasars;
  \citealt{Kashikawa+2015,Matsuoka+2018b,Matsuoka+2018a,Matsuoka+2019,Matsuoka+2019b})
  and from  VST-ATLAS combined with {\it WISE} (4 quasars;
  \citealt{Carnall+2015,Chehade+2018}).
  Masses were estimated from the rest-frame UV luminosity
  ($M_{1450}$), and assuming a constant bolometric correction and
  Eddington ratio ($f_{\rm Edd}=1$), except for the strongly lensed
  $z=6.51$ quasar~~\citep{Fan+2019} for which we adopted the published
  virial mass, including a magnification factor of 51.3.  Many of the
  least luminous quasars, discovered predominantly in the SHELLQs
  survey (shown in yellow), have Eddington ratios below unity; the
  masses for these least luminous sources are underestimated by the
  assumption of $f_{\rm Edd}=1$ (see text).  The pair of black curves
  show the mass of a BH, for reference, which grows continuously at
  the Eddington rate, with a radiative efficiency of $\epsilon=0.1$,
  starting from a stellar-mass seed BH of $M_\bullet = 10~\msun$
  (lower curve) or $100~\msun$ (upper curve) at $z=35$, in a flat
  concordance cosmology with $\Omega_\Lambda=0.69$, $\Omega_m=0.31$,
  $h=0.68$ \citep{Planck2018}.}
\label{fig:mbh-vs-z}
\end{figure}

\subsection{Accelerated Growth by ``Mergers and Acquisitions''}

We briefly enumerate several pathways to accelerate the assembly
of massive BHs in high-$z$ galaxies.  These pathways are illustrated in
\textbf{Figure~\ref{fig:rees}}, and will be discussed in detail in
subsequent sections.

One possibility is that a small fraction of ``lucky'' early Pop~III
seeds are able to sustain Eddington accretion over most of
the history of the Universe.  While this is unlikely for stellar-mass
seeds formed in the shallow potential wells of minihalos due to
negative feedback processes, it may be possible in rare massive halos
with $\gsim 10^8~\msun$ which form at redshifts as high as $z\gsim
30$~\citep{Tanaka14}.  Alternatively, BH assembly could be accelerated
if a significant fraction of the growth was due to mergers of compact
objects (mostly other BHs).  However, a possible issue with this
scenario is that BH mergers can lead to gravitational-wave induced
kicks which remove BHs from their reservoirs of dense gas
\citep{Haiman2004}.
 
Another, popular class of scenarios relies on the formation of
$\approx 10^5~\msun$ ``massive seed'' BHs, giving them a head start
toward the SMBH regime. Though there are a number of variations, these
models generally invoke the rapid collapse of chemically pristine
primordial gas in so-called ``atomic cooling halos'' (ACHs) with
virial temperature $T_{\rm vir} \sim 10^4~{\rm K}$. This is thought to
form a $10^{5-6}~\msun$ supermassive star (SMS), which then promptly
collapses to a BH with a similar mass.\footnote{These are often
  referred to as ``direct collapse black hole'' (DCBH) models,
  although arguably this is a misnomer, given the inevitable
  intermediate stage of a supermassive star. In this review, we
  therefore do not employ the otherwise very popular ``DCBH''
  terminology.} The key ingredient is the large accretion rate of the
protostar, which requires the collapsing gas to remain warm ($\gsim
5000~\K$).  This in turn requires avoiding efficient metal- or
H$_2$-cooling and fragmentation.  Several mechanisms/environments have
been put forward to lead to this thermodynamical state, including
exposing the ACH to intense H$_2$ photodissociating UV radiation in
the Lyman-Werner (LW) bands
\citep{Omukai2001,OhHaiman2002,BrommLoeb2003}; suppressing H$_2$
cooling \citep{Fernandez+2014} and/or heating the gas
\citep{Yoshida+2003,Wise+2019} in ACHs with a violently rapid merger
history; delaying H$_2$ cooling due to unusually high baryon-matter
streaming velocities~\citep{TanakaLi2014, Hirano+2017}; high-velocity
collisions of two halos near the atomic-cooling threshold
\citep{Inayoshi+2015}, or some combination of these effects.

An idea related to the above scenario is that some Pop~III remnant BHs
find themselves at the center of an ACH without prior star formation,
because of the above peculiar mechanisms/environments. Some of these
BHs would then accrete at super/hyper-Eddington rates and grow to
$\sim 10^{5-6}~\msun$~\citep{Pacucci+2015,Inayoshi+2016, Ryu+2016}.

A different possibility is that a $M_\bullet\approx 10^{3-4}~\msun$  intermediate-mass BH (IMBH) 
forms promptly through stellar
mergers in the core of an ultra-dense stellar cluster in a metal-poor
protogalaxy~\citep{Omukai+2008,DevecchiVolonteri2009}.  Direct
collisions can occur on a timescale shorter than the
lifetime of massive stars \citep{Katz+2015,YajimaKhochfar2016,Sakurai+2017}, 
especially if the cluster is still embedded in dense gas, where some of
the protostars are accreting at high rates and have bloated envelopes,
significantly increasing their geometric
cross-section~\citep{Reinoso+2018,Tagawa+2019}.

Finally, in the absence of a definitive conclusion for early massive
BH formation via the astrophysical scenarios above, it is worth
keeping in mind other, more exotic possibilities. These include
primordial BHs formed soon after the Big Bang, supermassive stars
sustained by dark-matter annihilation, efficient energy dissipation of
magnetic fields in ACHs, and fueling BHs by self-interacting DM
(see~\S\ref{sec:exotica}).

\subsection{Below the Tip of the Iceberg}
\label{sec:iceberg}

The current surveys of distant quasars can only detect unusually
bright and massive BHs that accrete near the Eddington limit.  This
makes the $\sim 10^9~\msun$ SMBHs ultra-rare objects, unrepresentative
of the underlying massive BH population.  Their hosts are also very
massive, highly evolved galaxies (see~\S~\ref{sec:obs} below), which
themselves must have formed in highly biased regions of the universe.
Of course, explaining the existence of such extreme objects is crucial
to improve our understanding of BH and galaxy formation in the early
universe.  However, it is arguably even more important to understand
the much larger population of massive BHs, currently hidden from our
view because of their lower masses, and/or lower accretion rates.
Many early BHs can also remain undiscovered because obscuration by
large amounts of gas and dust makes them too dim.  As discussed in
\S~\ref{sec:obsdiag}, the nature of this population of massive BHs
needs to be investigated in order to better constrain theoretical
assembly models with ongoing and future observational programs.  In
this review, we will focus on a theoretical framework of massive BH
formation and growth processes for a wide range of initial BH masses,
$10 \lsim M_\bullet /\msun \lsim 10^6$, addressing the formation of
the relatively typical BH population as well as of the extreme BHs.

\vspace{\baselineskip}
This review is organized as follows.
In \S\ref{sec:obs}, we summarize high-$z$ quasar observations,
including their current status and the most recent discoveries.
In \S\ref{sec:growacc}, we discuss the timescales for BH growth,
taking into account the physics of accretion flows over a wide range
of spatial scales, and summarizing the results of radiation
hydrodynamical simulations conducted in the past decade.  
We then specialize to applications to the high-$z$ universe.
In \S\ref{sec:angmom}, we discuss the physics of transferring angular
momentum, which is one of the biggest obstacles to maintain rapid
inflows from galactic scales down to the nuclear BH.  We then again
specialize to the high-$z$ universe.
In \S\ref{sec:BHIMF}, we summarize possible formation pathways of
massive seed BHs in high-$z$ protogalaxies, and discuss their
subsequent growth, as well as the evolution of the overall population
of massive BHs in the early universe.
In \S\ref{sec:exotica}, we briefly mention several more exotic ways
of producing massive BHs in the early universe.
Finally, we summarize future observational diagnostics of the
formation and growth processes of the early massive BH population in
\S\ref{sec:obsdiag}, and offer our main conclusions in
\S\ref{sec:conc}.

\vspace{\baselineskip}
Many previous reviews have addressed various aspects of the above
topics, including \citet{HaimanQuataert2004}, \citet{Volonteri2010},
\citet{Volonteri2012}, \citet{VolonteriBellovary2012},
\citet{Haiman2013}, \citet{JohnsonHaardt2016},
\citet{LatifFerrara2016}, \citet{Gallerani+2017} and \cite{Woods+2018}, as well as the book
edited by \citet{LatifSchleicher2018}.  Here we aim to provide a
comprehensive but concise up-to-date review, focusing especially on
the physics of BH formation and growth processes.

\section{OBSERVATIONS}
\label{sec:obs}

The first quasars were identified as quasi-stellar radio sources in
radio surveys in the 1950s.  Based on its optical spectrum, the radio
source 3C 273 was interpreted as the bright nucleus of a galaxy at
redshift $z=0.158$~\citep{Schmidt1963}. The large energy output and
short time-scale variability soon led to the consensus that quasars
are powered by massive black holes via an accretion disk
(see \citealt{Rees1984} for an early review).

Over several decades, increasingly large surveys (mainly in the
optical, but also in X-ray and radio bands) mapped out the luminosity
function (LF) of quasars.  These have revealed a clear evolution over
cosmic time, with quasar activity rising from early times, peaking
around $z\approx 2$, and falling again toward $z=0$.

This behavior is broadly consistent with a cosmological picture, in
which massive BH seeds grow primarily during brief episodes of
accretion. These episodes are expected to be often triggered by major
mergers of their parent halos
(e.g. \citealt{KauffmannHaehnelt2000,Volonteri+2003,Somerville+2008,Hopkins+2008}),
for which there is some observational support (see
\citealt{Goulding+2018} and references therein).  Furthermore,
assuming a radiative efficiency of $\epsilon$, i.e.  that the
accretion rate $\dot{M}$ produces a luminosity $L=\epsilon \dot{M}
c^2$ and a growth of the BH's rest mass at a rate of
$(1-\epsilon)\dot{M}$, the above picture directly links quasar
activity to the local population of remnant BHs
\citep{LyndenBell1969,Soltan1982}.  The total quasar light output,
measured by integrating the quasar LF over luminosity and redshift, is
consistent with the local nuclear BH mass density of $\approx 4\times
10^5~\msun~\mpc^{-3}$, 
measured using correlations between BH mass and global
galaxy properties \citep{KormendyHo2013}, and an average radiative
efficiency of $\epsilon\approx 10\%$ (with the latter depending on
luminosity; \citealt{YuTremaine2002,Shankar+2004}).
What this broad picture is missing is where the BHs with masses
of $\approx 10^{5-6}\msun$, corresponding to the low-mass end of the
SMBH mass function, come from.  Over the past two decades, beginning
with discoveries of distant quasars at $z\gsim 6$ in the Sloan Digital
Sky Survey (SDSS), it has become clear that such seeds must have
appeared very early on.

\begin{table}[t]
\tabcolsep7.5pt
\caption{List of surveys that have discovered high-$z$ quasars at redshift $z\geq 6$.}
\label{tab:high-z-surveys}
\begin{center}
\begin{tabular}{@{}l|c|c|c|c@{}}
\hline
Name & Bands & Area (deg$^2$) & \#QSOs & Refs.\\
\hline
SHELLQs (+ other Subaru)          & opt $g,r,i,z,y$     & 1,400     & 81   &   1,2,3,4,5\\
DELS (Dark Energy Survey)         & opt $g,r,z$ + IR    & 14,000    &  37   &  6  \\
Pan-STARRS1                       & opt $g,r,i,z,y$     & 31,000    & 29   &   7              \\
SDSS (Sloan Digital Sky Survey)   & opt $u,g,r,i,z$     & 15,000    & 20   &    8,9        \\
CFHQS (Canada-France-Hawaii T.)   & opt $g,r,i,z$       & 500       & 15   &    10,11  \\
UKIDSS (UKIRT)                    & IR  $z,Y,J,H,K$     & 7,000     & 7    &    12\\
VIKING (VISTA)                    & IR  $z,Y,J,H,K$     & 1,500     & 7    &   13,14 \\
VST-ATLAS (VLT)+{\it WISE}        & opt $u,g,r,i,z$ +IR & 4,700     & 6    &   15,16 \\
FIRST+NDWFS+FLAMEX                & 21cm + opt + IR & 4         & 1    &   17   \\
\hline
\end{tabular}
\end{center}
\begin{tabnote}
  Some quasars were discovered independently in more than one survey.
  References: 1=\citet{Matsuoka+2016}, 2=\citet{Matsuoka+2018a},
  3=\citet{Matsuoka+2018b}, 
  4=\citet{Matsuoka+2019b},
  5=\citet{Kashikawa+2015},
  6=\citet{Dey+2019},
  7=\citet{Chambers+2016}, 
  8=\citet{Jiang+2016}
  9=\citet{Wang+2016b},
  10=\citet{Willott+2007}, 
  11=\citet{Willott+2010},
  12=\citet{Lawrence+2007}, 
  13=\citet{Edge+2013}
  14=\citep{Venemans+2019},
  15=\citet{Carnall+2015},
  16=\citet{Chehade+2018}
  17=\citet{McGreer+2006}.
\end{tabnote}
\end{table}

\subsection{High-Redshift ($z\gsim 6$) Quasar Surveys}

Searching for high-$z$ quasars presents some technical challenges.
Because bright quasars, detectable at large redshifts, are rare, large
fractions of the sky need to be surveyed.  The primary means of
identifying quasars is based on their multi-color broad-band
photometry, which allows efficient separation from the stellar locus
in color space, particularly via the prominent Lyman-$\alpha$
break~\citep[e.g.][]{Warren+1987}.  At high redshifts, this requires
photometry at the reddest optical bands.  For example, the Lyman break
falls at the center of the common optical $u,~g,~r,~i,~z$ band
filters at $z=1.9,~2.9,~4.1,~5.3$ and $6.5$.  Finally, the large
amount of data requires efficient automated data-processing.

\begin{table}[t]
\tabcolsep5.5pt
\caption{List of $z\geq 7$ quasars.}
\label{tab:z7qsos}
\begin{center}
\begin{tabular}{@{}l|c|c|c|c|c@{}}
\hline
Name & Surveys & redshift & $M_\bullet/\msun$$^{\rm a}$ & $f_{\rm Edd}$ & Refs.\\
\hline
ULAS J1342+0928 & {\it WISE}/DELS/ & 7.541 & $7.8^{+3.3}_{-1.9}\times10^8$ & $1.5^{+0.5}_{-0.4}$   &  1   \\
 & UKIDSS & [CII]  & & & \\ \hline
HSC  J1243+0100  & SHELLQs           & 7.07    & $3.3^{+2.0}_{-2.0}\times10^8$ & $0.34^{+0.2}_{-0.2}$ &  2 \\
 & &  MgII & &   \\ \hline
ULAS J1120+0641 & UKIDSS              & 7.085  & $2.0^{+1.5}_{-0.7}\times10^9$  & $1.2^{+0.6}_{-0.5}$  &   3  \\
 & SDSS  & SiIII/CIII]/MgII  &   & \\  \hline
DELS J0038-1527 & DELS/{\it WISE}/  & 7.021 & $1.33^{+0.25}_{-0.25}\times10^9$ & $1.25^{+0.19}_{-0.19}$  &  4   \\
& Pan-STARRS1 &  MgII/OIII  &  & &   \\ \hline
DES  J0252-0503 & DES/VHS/ULAS/  & 7.021  & $\sim 1.6\times10^9$    & --      & 5    \\
&  {\it WISE}/VIKING & Ly$\alpha$/NV &  & &   \\ \hline
HSC J2356+0017 & SHELLQs & 7.01  & $\sim 5.5\times10^8$    & --      & 6    \\
&  & Ly$\alpha$ &  & &   \\
\hline
\end{tabular}
\end{center}
\begin{tabnote}
  $^{\rm a}$All masses are published estimates based on the MgII line
  width and a virial mass estimator~\citep{VestergaardOsmer2009},
  except for DES J0252-0503 and HSC J2356+0017, for which we use the proxies from the
  rest-frame UV luminosity ($M_{1450}$), assuming a constant
  bolometric correction and Eddington ratio. 
  Redshifts are based on the metal lines and/or Ly$\alpha$ listed in the
  third column. References: 1=\citet{Banados+2018},
  2=\citet{Matsuoka+2019}, 3=\citet{Mortlock+2011},
  4=\citet{Wang+2018a}, 5=\citet{Yang+2019}, 6=\citet{Matsuoka+2019b}
\end{tabnote}
\end{table}

These criteria were first met by the SDSS, resulting in the first
handful of quasars at $z\gsim 6$, beginning with \citet{Fan+2000}.
Large optical and infrared surveys have continued to dominate high-$z$
quasar searches in the past two decades (see
\textbf{Table~\ref{tab:high-z-surveys}}).  The SDSS~\citep{Jiang+2016}
and the Canada-France High-Redshift Quasar Survey (CFHQS;
\citealt{Willott+2007,Willott+2010}) have together found several dozen
quasars out to $z\lsim 6.5$, limited by their reddest bands. The
addition of a $y$ filter extends this redshift range to $z\sim 7.2$,
and has resulted in discoveries of many of the highest-$z$ quasars by
the Pan-STARRS1
survey~\citep{Chambers+2016,Banados+2016,Tang+2017,Mazzucchelli+2017},
the Subaru High-z Exploration of Low-Luminosity Quasars
project~(SHELLQs;
\citealt{Matsuoka+2016,Matsuoka+2018a,Matsuoka+2018b,Matsuoka+2019}),
and by the Dark Energy Survey~\citep{Reed+2015,Reed+2017,Reed+2019},
particularly the Dark Energy Camera Legacy Survey (DELS;
\citealt{Wang+2018a,Wang+2018b,Yang+2019}), combined with near- and
mid-IR data from several other surveys.  In this combination, the
quasar is a drop-out in the optical bands, but detected in the IR.
Many of the highest-$z$ sources have indeed been recently discovered
by such combinations from multiple surveys, which included the United
Kingdom Infrared Telescope (UKIRT) Infra Red Deep Sky Surveys (UKIDSS;
\citealt{Lawrence+2007,Mortlock+2011}), the UKIRT Hemisphere Survey
\citep[UHS;][]{Wang+2018a}, the Dark Energy Spectroscopic Instrument
Legacy Imaging Surveys (DELS; \citealt{Dey+2019}),
VIKING~\citep{Edge+2013,Venemans+2013},
VST-ATLAS~\citep{Carnall+2015,Chehade+2018}, and the VISTA Hemisphere
Survey (VHS;~\citealt{McMahon+2013,Venemans+2015,Pons+2019}).  The
only $z\gsim 6$ quasar whose discovery involved other wavelengths is a
radio-loud quasar at $z=6.1$, found by matching optical with radio
data in the FIRST survey (\citealt{McGreer+2006}; although this
quasar, too, was independently discovered in optical-only;
\citealt{Stern+2007}). \textbf{Table~\ref{tab:high-z-surveys}}
summarizes results from these surveys and
\textbf{Figure~\ref{fig:mbh-vs-z}} shows the redshifts and inferred
masses\footnote{Virial or other mass estimates have only been
  published for a fraction of the quasars, but absolute rest-frame
  1450\AA\ magnitudes (${\rm M_{1450}}$) are available for all
  quasars.  We therefore obtained masses by assuming a constant
  (product of the) bolometric correction and the Eddington ratio:
  $\log_{10}(M_\bullet)= (-{\rm M_{1450}} - 3.46)/2.5$, which yields,
  on average, the published virial mass estimates.}  of the 203
currently known quasars at $z\geq 6$.  The full list is provided in
the \textbf{Supplemental Material} (follow the \textbf{Supplemental
  Material link} from the Annual Reviews home page at
http://www.annualreviews.org.)
  
At the time of this writing, only six quasars are known at $z>7$
(listed in \textbf{Table~\ref{tab:z7qsos}}), which will surely soon
change with large forthcoming IR surveys, such as Euclid and WFIRST.

Among these discoveries, the SHELLQs survey stands out as being deeper
than other large-solid angle optical surveys, and therefore able to
find less-luminous quasars.  This has allowed a determination of the
$z\sim 6$ luminosity function over an unprecedentedly broad range of
luminosities~\citep{Matsuoka+2018c}, and has led to the important
finding that these somewhat fainter objects have Eddington ratios that
are typically lower than unity. \citet{Onoue+2019} measured virial
masses from the MgII line in deep optical spectra for six of the least
luminous SHELLQs quasars at $6.1 \lsim z \lsim 6.9$. They found
$f_{\rm Edd}\approx 1.1$ for one source, but $0.16\leq f_{\rm Edd}\leq
0.43$ for the other five. This appears significantly lower than the
Eddington ratios $f_{\rm Edd}\sim 1$ typically found in the past for
more luminous quasars at these redshifts, and also somewhat lower than
measured previously for 10 faint CFHQs quasars~\citep{Willott+2010c}.
\citet{Shen+2019} recently presented virial masses for a sample of 50
$z>5.7$ quasars with a range of luminosities, and found a median value
of $f_{\rm Edd}\sim 0.3$ and \citet{Mazzucchelli+2017} found an
average $f_{\rm Edd}\sim 0.4$ at $z>6.5$.  These recent results
together suggest that the global Eddington ratio distribution at
$z\sim 6$ is broader than previously measured, similar to that of
low-$z$ quasars, and with many of the most massive $\sim 10^9~\msun$
SMBHs at $z>6$ either accreting or shining at sub-Eddington rates.

\begin{figure}[t]
  \includegraphics[width=3.4in,trim={1.4cm 6.5cm 3.0cm 7.0cm},clip]{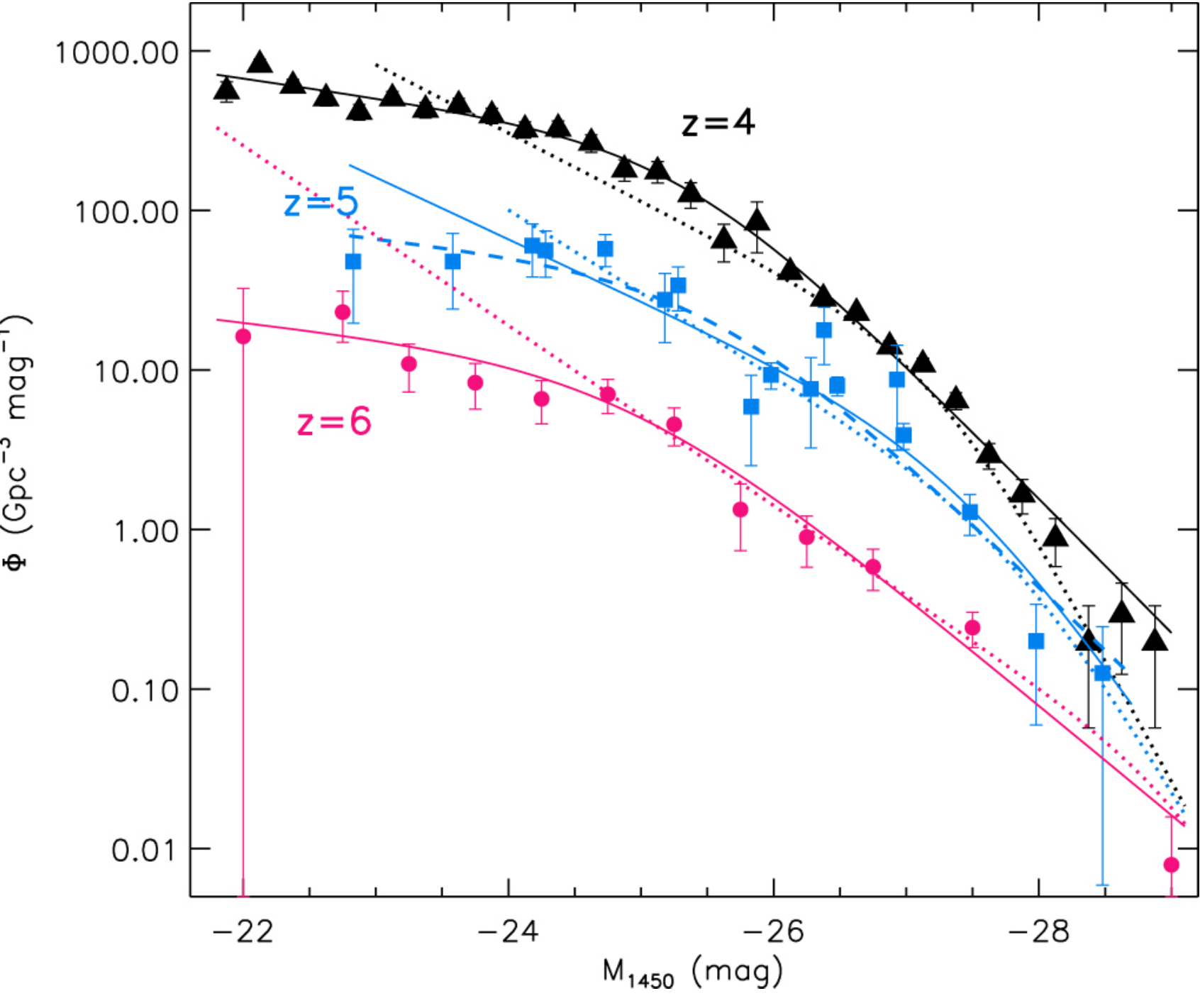}
  \vspace{-1\baselineskip}
\caption{High-$z$ quasar luminosity functions from SHELLQs
  ($z=6$, red circles; \citealt{Matsuoka+2018c}, CFHTLS
  ($z=5$, blue squares; \citealt{McGreer+2018}) and a combination
  of data from SDSS and Subaru's Strategic Program Wide survey ($z=4$,
  black triangles; \citealt{Akiyama+2018}).  The dotted lines show
  broken-power law fitting functions. Figure adopted from
  \citet{Matsuoka+2018c}.}
\label{fig:qso-lf}
\end{figure}

The most striking feature of the high-$z$ quasars is their large BH
masses.  The masses are estimated based on virial
relations~\citep{VestergaardOsmer2009}, typically using the width
$\Delta v$ of the CIV or MgII line.  Inferring the BH mass from a
virial relation of the form $\Delta v^2\sim GM/r$ requires knowledge
of the size $r$ of the broad line region (BLR), which, in
turn, is calibrated on low-$z$ quasars. Such calibrations are
performed with quasars less luminous than their high-$z$
counterparts~\citep[e.g.][]{Peterson2006}.  Thus, the high-$z$ mass
estimates rely on extrapolating these relations in both redshift and
luminosity.  This is somewhat justified by the fact that apart from
much stronger absorption from the intergalactic medium (IGM), the
spectra of the $z\sim 6$ quasars are indistinguishable from low-$z$
quasars~(\citealt{Fan2006}; but see details below).  This similarity,
particularly in the line-to-continuum ratios, also makes it
implausible that high-$z$ quasars are preferentially beamed towards us
(resulting in overestimates of their luminosities).  In fact, the lack
of obvious spectral differences, which include both the shape of the
continuum, and the strength of metal lines, more generally implies
that the birth environments of quasars are established very early
on~\citep{Shen+2019}.

One obvious question is whether luminosities (and thus masses) may
have been overestimated significantly due to gravitational lensing.
Strong lensing along a random line of sight to $z\sim 6$ is apriori
very unlikely (probability of $\sim 10^{-3}$), but if the intrinsic
(unlensed) $z\sim 6$ quasar LF is steep and/or extends to faint
magnitudes, then magnification bias can boost the probability of
strong lensing to even order
unity~\citep{Comerford+2002,WyitheLoeb2002}.  In the majority of
cases, strong lensing would produce two detectable and resolvable
images at the sensitivity and spatial resolution of the {\it Hubble
  Space Telescope}~\citep{Keeton+2005}, yet a search among $\sim 200$
quasars have not revealed any multiple
images~\citep{Richards+2006,McGreer+2014}.  This would naively rule
out the possibility that most $z\sim 6$ quasars are strongly lensed.
However, a strongly lensed quasar has recently been discovered at
$z=6.51$, with three images and an inferred total magnification of a
factor of $\approx 50$~\citep{Fan+2019}. This source was lensed by an
unusually faint foreground galaxy, whose starlight did not
significantly contaminate the quasar's spectrum.  Intriguingly, this
suggests that for a typical, brighter lens, the background lensed
quasar would not be identified as a quasar by traditional
color-selection criteria. A significant population of strongly lensed,
high-redshift quasars could therefore still be missing from the
existing surveys~\citep{Fan+2019,PacucciLoeb2019}.

\subsection{Properties of the $z\sim 6$ Quasar Population}

Luminosity functions have been measured in several surveys
~\citep{Willott+2010,Jiang+2016}, with the most complete
determination, extending to the lowest luminosities by the SHELLQs
project on the Subaru telescope~\citep{Matsuoka+2018c}.  As shown in
\textbf{Figure~\ref{fig:qso-lf}}, the LF at $z=6$ is roughly
consistent with an extrapolation from lower redshifts, with its shape remaining 
self-similar (well-fit by a broken power-law), but the normalization dropping 
steeply with redshift, with quasars at
$z\approx 6$ about 100 times less abundant than at $z\approx 4$.  
One important finding from the SHELLQs project is that the quasar LF
flattens significantly toward lower luminosities so that the total (faint+bright) quasar 
population could not provide enough photons to keep the IGM ionized (see also 
\citealt{Jiang+2008}), even assuming a clumping factor of $\approx 1$ \citep{Madau+1999}. 
This indicates that quasars are not a major contributor to cosmic reionization.

However, it is also important to consider how complete the current
quasar surveys may be, as we could be missing many lensed quasars (see
above) as well as a population of heavily obscured quasars.  The
latter can be an especially large effect, since the optical selection
of the current $z>6$ quasars is highly biased against obscured
quasars. The first and only known highly-obscured quasar candidate at
this redshift, detected in X-rays~\citep{Vito+2019}, is optically
classified as a Type 1 AGN.  A comparison of the extrapolation of the
$z=3-6$ X-ray AGN LF (which does not select against obscured quasars)
with the optically-selected $z>6$ LF \citep{Matsuoka+2018c} suggests
that as much as 80-90\% of all $z>6$ quasars may be obscured and
missed by optical/IR surveys \citep{Vito+2018}.

Overall, the high-redshift quasars individually look very similar to
their low-$z$ counterparts.  In particular, the hosts of the $z\gsim
6$ quasars contain significant amounts of metals and dust, including
the host of the most-distant quasar at
$z=7.54$~\citep{Venemans+2017b, Novak2019}.  Copious amounts of metals are
revealed by observations of molecular lines (e.g. CO, CII), in the ISM
of the hosts on kpc scales
\citep{Bertoldi+2003,Walter+2003,Walter+2009,Wang+2013,
  Willott+2015,Venemans+2017,Venemans+2019}.  The amount of cool
molecular gas is $\sim 10^{9-10}~\msun$ \citep{CarilliWalter2013}.
The FIR continuum in these observations likewise reveal a large amount
($\sim 10^{7-8}~\msun$) of warm, thermally emitting dust.

The highest-angular resolution observations by instruments such as
ALMA and IRAM have spatially resolved the hosts of many luminous
high-redshift quasars, and found a diverse range, which include
compact dispersion-dominated systems, rotationally supported galaxies,
as well as isolated galaxies, major mergers, and close companions in
some cases~\citep{Decarli+2017,Neeleman+2019}.

On larger scales, one would naively expect that the luminous quasars
at any redshift should reside in the most massive halos, which are in
the most overdense environments (although cosmological simulations
suggest that this is not strictly true; \citealt{Fanidakis+2013}).
Several surveys have looked for a corresponding excess overdensity of
galaxies around high-$z$ quasars on Mpc scales. However, the evidence
is inconclusive: the environments of some of the quasars show galaxy
overdensities, and some do
not~~\citep{Kim+2009,Utsumi+2010,McGreer+2014,Mazzucchelli+2017,Balmaverde+2017,Ota+2018}.

The overall strength and kinematics of the molecular lines and of the
continuum dust emission are together consistent with these hosts being
analogs of low-redshift starburst galaxies, with the dust being heated
by star formation on kpc scales, at star-formation rates of up to a
few $1000~\msunyr$.  
Likewise, the nuclei of these hosts appear highly
enriched on $\ll$ pc scales, as evidenced by broad metal emission
lines, such as CIV in their rest-frame UV spectra, which are similar
to those of their low-$z$
counterparts~\citep{DeRosa+2014,Mazzucchelli+2017,Reed+2019,Shen+2019}.
The FeII/MgII line ratio, a proxy for the chemical abundance of the
BLR gas in bright quasar hosts, also shows no
redshift-evolution~\citep{deRosa+2011}.  The CIV lines do show unusually large
blueshifts, indicating that winds driven out of the nuclear disks may
be especially strong in the luminous, high-$z$ quasars.  
In addition to the systematically larger blueshifts of broad emission lines at 
higher redshift \citep{Meyer2019}, there have been tentative claims for a 
systematically larger fraction of weak-line quasars at $z>6$ 
\citep{Banados+2016,Shen+2019,Meyer2019}.
{\it In total, the evidence above points to an early, rapid assembly of the
  massive host galaxies of the highest-redshift quasar BHs.}

Finally, an interesting tentative difference in the hosts of the
highest-$z$ quasars is the ratio of their BH and galaxy masses.  The
resolved kinematics of the ISM, especially from the strongest [CII]
emission line, tend to yield dynamical masses for the fainter,
lower-luminosity quasars that obey local scaling relations, albeit
with a larger scatter~\citep{Willott+2017,Izumi+2018}.  However, the
hosts of more luminous high-$z$ quasars appear to have dynamical
masses an order of magnitude below the corresponding low-$z$
relations~\citep{Wang+2016,Decarli+2018,ShimasakuIzumi2019}.  If
confirmed, this suggests that the most massive SMBHs at $z\gsim 6$ got
a ``headstart'' over the growth of their host galaxies, which is
perhaps in slight tension with the high metal and dust-enrichment of
these hosts. The clear caveats are that gas tracers can underestimate
dynamical masses and that the brightest QSOs can suffer from a
selection bias that picks out preferentially massive BHs (see
\citealt{VolonteriStark2011,Lupi+2019} and references therein for
discussions of such biases).  We also note that the SMBHs at the high
end of the locally measured $M_\bullet-\sigma_\star$ relation also
tend to have higher masses, but there is a similar tentative
``upturn'' in the Faber-Jackson relation between host luminosity
$L_{\rm gal}$ and $\sigma_\star$, so that the most massive BHs are not
outliers in the $M-L_{\rm gal}$ relation~\citep{Lauer+2007}.  It is
possible that high-$z$ quasars fit the same trend. However, we
emphasize that the local relations relate BH masses to properties of
the bulge component, about which we have no information at high
redshift. More generally, a key missing piece of evidence is the
direct observation of starlight from the high-$z$ quasar host galaxies
at rest-frame UV to near-IR wavelengths~\citep{Fan+2019b}. The one
exception is the UV starlight detected in a $z=6.2$
system~\citep{Decarli+2019}, interpreted to be a merger.

\section{ACCRETION AND RADIATIVE FEEDBACK}
\label{sec:growacc}

\begin{figure}[t]
\includegraphics[clip,width=5.0in]{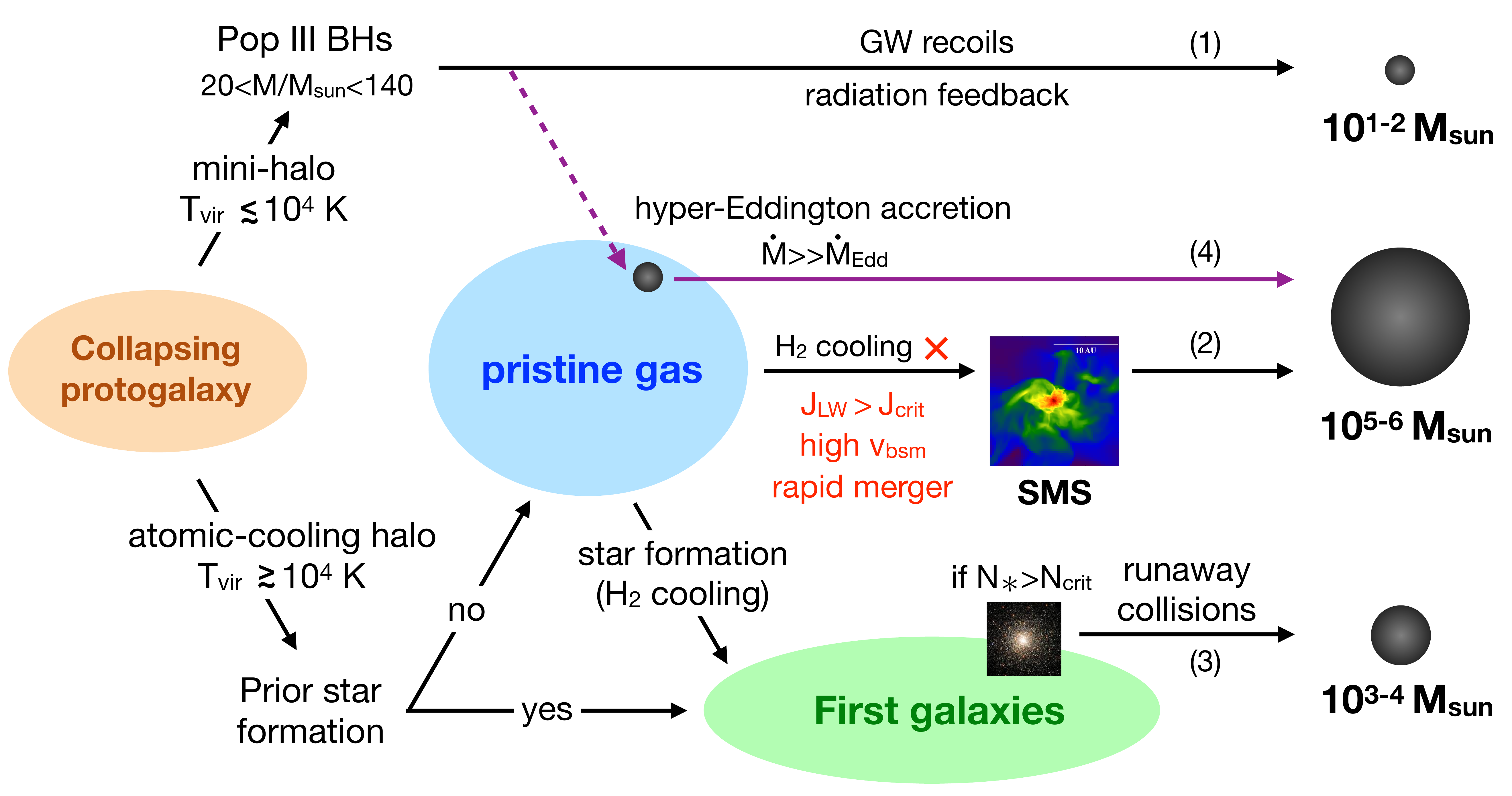}
\caption{Formation pathways of
  seed BHs in early protogalaxies:
  (1) Pop~III remnant BHs with a mass of $M_\bullet \approx
  10^{1-2}~\msun$,
  (2) massive seed BHs with $M_\bullet \approx
  10^{5-6}~\msun$ in ACHs under peculiar conditions such as strong 
  LW radiation ($J_{\rm LW}>J_{\rm crit}$),
  high baryon-DM streaming velocity, and rapid mergers of DM
  halos, and
  (3) relatively massive seeds with $M_\bullet \approx
  10^{3-4}~\msun$ via runaway collisions in ultra-dense stellar
  clusters.
  (4) Hyper-Eddington accretion onto stellar-mass BHs 
  ($\dot{M}_\bullet \gg \dot{M}_{\rm Edd}$) would effectively result in a
  massive seed at the center of a dense pristine gas cloud.}
\label{fig:rees}
\end{figure}

In this section, we review the theoretical framework of BH accretion
and discuss rates at which pre-existing BHs can grow by accretion, in
particular in the face of radiative feedback.  This is motivated by
the natural availability of stellar-mass BHs in the early universe,
left behind by the first generation of stars.  As we argue, it is
possible for BHs to grow at highly super-Eddington rates, which
represents one of the pathways for rapid BH assembly in the early
universe.  These pathways are illustrated in
\textbf{Figure~\ref{fig:rees}}, along with other possibilities that
will be discussed in the sections below.  However, we emphasize that
no self-consistent calculation to date has included all of the
necessary multi-scale physics {\em and} followed the BH growth over
several orders of magnitude in mass.  We first focus on the basic
underlying physics (\S~\ref{subsec:MdotEdd}) and then discuss
applications to the high-$z$ universe (\S~\ref{sec:lightseed} and
\S~\ref{subsec:BHaccretion-apps}).

\begin{figure}[t]
\includegraphics[clip,width=4.7in]{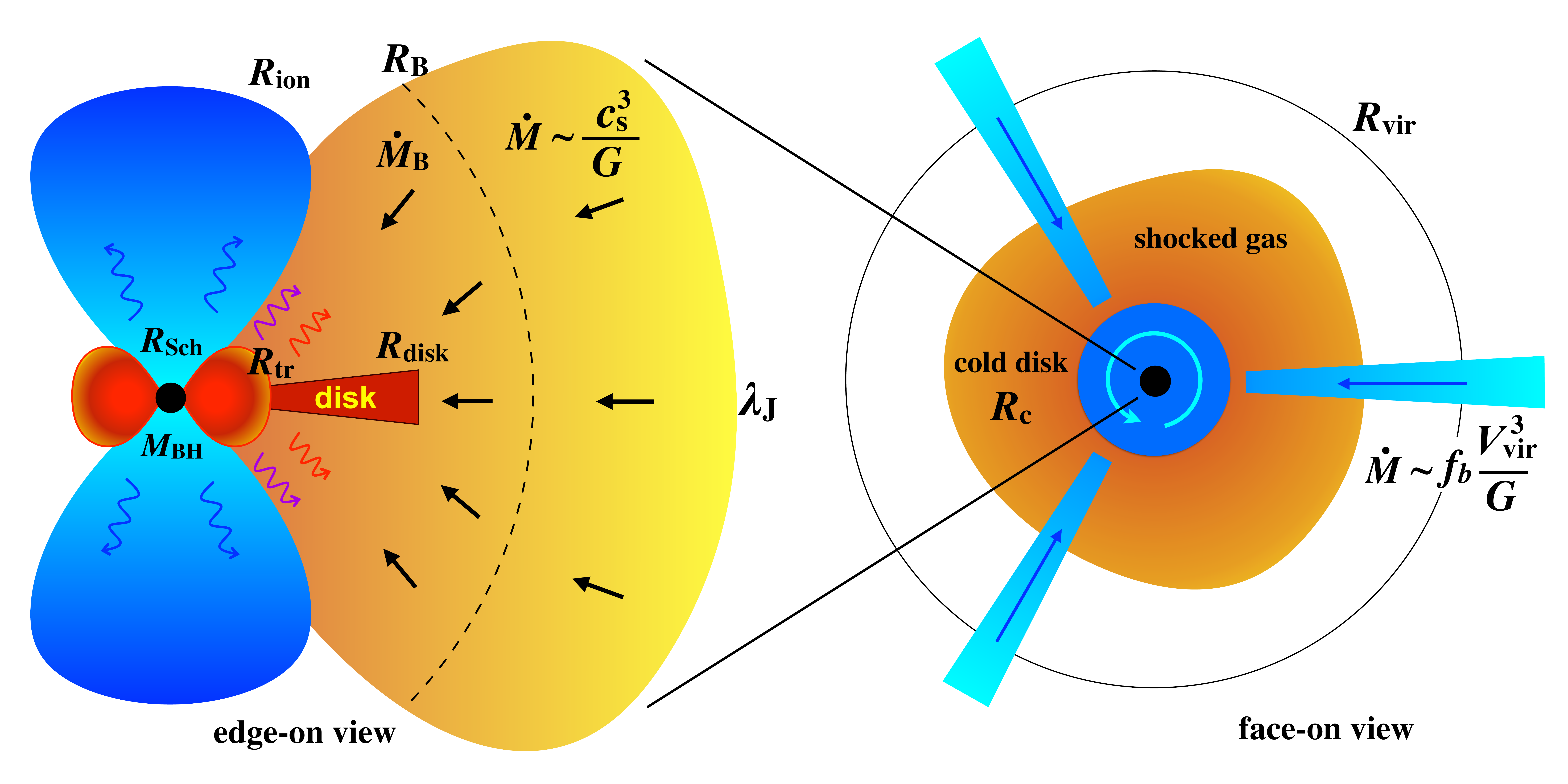}
\vspace{-1\baselineskip}
\caption{Schematic illustrations of the accretion flow on to a massive
  BH with a mass of $M_\bullet$ at a rate significantly exceeding
  $\dot{M}_{\rm Edd}$ (edge-on view; left) and of the early
  protogalaxy that hosts the accreting BH (face-on view; right).  The
  characteristic radii and mass accretion rates of this system, shown
  in the illustrations, are summarized in the accompanying \textbf{Table~\ref{tab:phys}}, 
  along with their definitions and fiducial values.
 }
\label{fig:phys_scale}
\end{figure}

For convenience, \textbf{Figure~\ref{fig:phys_scale}} illustrates the
structure of accretion flows on to a BH embedded in a protogalaxy,
with the characteristic physical scales and mechanisms relevant to the
discussions below (with their definitions and fiducial values listed
in \textbf{Table~\ref{tab:phys}}).

\subsection{Growing BHs by Accretion: Is There an Eddington Limit?}
\label{subsec:MdotEdd}

Assuming that high-$z$ SMBHs grow mostly via rapid gas accretion and
radiate $\sim 10\%$ of the rest mass energy of accreting matter as
low-$z$ quasars do on average \citep{Soltan1982,
  YuTremaine2002,Ueda+2003}, the outward radiation pressure force on
the infalling gas, through electron scattering, matches the inward
gravitational force at the critical accretion rate of $\dot{M}_{\rm
  Edd} \equiv 10~L_{\rm Edd}/c^2$, where $L_{\rm Edd}=4\pi cGM_\bullet
/ \kappa_{\rm es}$ is the Eddington luminosity.\footnote{ This
  definition includes a fiducial factor of 10, which assumes a
  radiative efficiency of 10\%. We will employ this definition
  throughout this review, but we caution the reader that an equally
  common definition in the literature is $\dot{M}_{\rm Edd}\equiv
  L_{\rm Edd}/c^2$, i.e. excluding this factor.}  If accretion is
limited to this rate, the time-scale for growth to the BH mass
$M_\bullet$ becomes as long as
\begin{equation}
t_{\rm grow}\approx \frac{0.45~\epsilon}{(1-\epsilon) f_{\rm duty}}
~\ln \left(\frac{M_\bullet}{M_{\rm seed}}\right)~{\rm Gyr} \approx 0.81~{\rm Gyr}.
\end{equation}
where $f_{\rm duty}$ is the duty cycle of accretion, $M_{\rm seed}$ is
the initial seed mass, and the last step adopts the fiducial values
$\epsilon=0.1$, $f_{\rm duty}=1$, $M_{\rm seed}=100~\msun$, and
$M_\bullet=10^9~\msun$.  This estimate shows that the growth timescale
is comparable to the age of the universe at $z\sim 6$, even when
continuous and rapid gas supply is assumed
\citep{HaimanLoeb2001,MadauRees2001,Volonteri+2003,Li+2007}.  This, in
turn, raises basic questions, such as: {\em what is the radiative
efficiency of early BHs? Can accretion occur at rates exceeding
the fiducial Eddington-limited value? Can the required fuel
supply be maintained over several orders of
magnitude growth in mass?}

\begin{table}[t]
\tabcolsep5.5pt
\vspace{\baselineskip}
\caption{List of relevant physical scales and related quantities discussed in this review.}
\label{tab:phys}
\vspace{-0.2\baselineskip}
\begin{center}
\begin{tabular}{@{}lccc@{}}
\hline
Quantity & Symbol & Approx.$^{a}$\\
\hline
\vspace{2mm}
Jeans mass	& $M_{\rm J} \equiv \rho \lambda_{\rm J}^3$ & $2\times 10^4~n_{\rm H, 4}^{-1/2} T_3^{3/2}$\\
\vspace{2mm}
Eddington accretion rate 	& $\dot{M}_{\rm Edd}\equiv \frac{L_{\rm Edd}}{0.1c^2}$     &
$2.3\times 10^{-5}~ M_{\bullet, 3}$ \\
\vspace{2mm}
Bondi accretion rate 	& $\dot{M}_{\rm B}\equiv \pi e^{3/2} \rho \frac{G^2M_\bullet ^2}{c_{\rm s}^3}$     &
$4.5\times 10^{-3}~ n_{\rm H, 4} T_3^{-3/2} M_{\bullet, 3}^2$ \\
\vspace{2mm}
Accretion rate in an unstable cloud	& $\dot{M}\sim \frac{c_{\rm s}^3}{G}$   &
$4\times 10^{-3}~ T_3^{3/2}$ \\
\vspace{2mm}
Mass inflow rate from galactic scales	& $\dot{M}\sim f_{\rm b}\frac{V_{\rm vir}^3}{G}$ & 
$6\times 10^{-2}~T_{\rm v,4}^{3/2}$\\
\vspace{2mm}
Schwarzschild radius		& $R_{\rm Sch}\equiv \frac{2GM_\bullet}{c^2}$	& $2\times 10^{-3}~M_{\bullet, 3}~[{\rm AU}]$ & \\
\vspace{2mm}
Photon trapping radius		& $R_{\rm tr}\equiv \frac{\kappa_{\rm es}\dot{M}_{\bullet}}{4\pi c}$		& 
$0.01~M_{\bullet, 3} \left(\frac{\dot{m}}{100}\right)~[\rm AU]$	& \\
\vspace{2mm}
Bondi radius				& $R_{\rm B}\equiv \frac{GM_\bullet}{c_{\rm s}^2}$		& $0.6~T_3^{-1}M_{\bullet, 3}~[\pc]$\\
\vspace{2mm}
Jeans length				& $\lambda_{\rm J}\equiv \sqrt{\frac{\pi k_{\rm B}T}{G\mu m \rho}} $ & $4~n_{4}^{-1/2} T_3^{1/2}~[\pc]$\\
\vspace{2mm}
Centrifugal radius (halo scale)	& $R_{\rm c}\equiv \lambda R_{\rm vir}$     	&
$26~\lambda_{0.05}T_{\rm v,4}^{1/2} \left(\frac{1+z}{16}\right)^{-3/2}~[\pc]$ \\
\vspace{1mm}
Halo virial radius			& $R_{\rm vir}$     	& $520~T_{\rm v,4}^{1/2} \left(\frac{1+z}{16}\right)^{-3/2}~[\pc]$\\
\hline
\hline
\end{tabular}
\end{center}
\begin{tabnote}
$^a$ The units for mass and accretion rate are $\msun$ and
  $\msunyr$.  The BH mass is $M_\bullet =10^3M_{\bullet, 3}~\msun$, gas density
  $n_{\rm H}=10^4n_{\rm H, 4}~\cc $, gas temperature $T=10^3T_3~\K$, DM halo
  virial temperature $T_{\rm vir}=10^4T_{\rm v,4}~\K$, DM halo spin
  parameter $\lambda=0.05\lambda_{0.05}$, and $\dot{m}\equiv
  \dot{M}_\bullet /\dot{M}_{\rm Edd}$ is the dimensionless BH
  accretion rate normalized by the Eddington rate (at 10\% radiative
  efficiency, as defined in the second row).
   \end{tabnote}
\end{table}

\subsubsection{Photon trapping on small scales near the BH}

Gas accreting onto a BH releases a large amount of energy at the
vicinity of the BH event horizon, $R_{\rm Sch}\equiv 2GM_\bullet / c^2
\approx 300~{\rm km} ~(M_\bullet/100~\msun)$.  The intense radiation
would naively limit the BH growth below the critical Eddington value.
However, does the Eddington limit really matter for the accretion
rate?  In general, active galactic nuclei (AGN) appear to obey the
Eddington limit on the luminosity, based on BH masses from
reverberation measurements~\citep[e.g.][]{Peterson+2004}.  However,
some X-ray binaries, such as SS 443 in our Galaxy
(e.g. \citealt{Okuda+2002}) and some ultra-luminous X-ray sources,
which are suspected to contain stellar-mass BHs
\citep{King+2001,Winter+2006}, are believed to accrete at
super-Eddington rates~\citep{Poutanen+2007,Kawashima+2012}.  
Several ULXs have been observed to pulsate on a timescale of $\sim$1s,
implying a stellar-mass source and thus favoring super-Eddington 
accretion~\citep[e.g.][and references therein]{King+2017}. In the
supermassive BH regime, narrow-line Seyfert-1 galaxies are presumed to
be super-Eddington accretors
\citep{Mineshige+2000,WangNetzer2003,CollinKawaguchi2004,Du+2014}.

The possibility of super-Eddington accretion has been explored
theoretically by many authors.  A basic reason why this may be
feasible goes as far back as \citet{Begelman1979}.
In a spherically symmetric accretion flow at a rate of $\gg
\dot{M}_{\rm Edd}$, in which the radiation pressure force is supposed to halt the
inflow, the emergent radiation flux is reduced by photon trapping in
the optically-thick accreting matter.  This trapping effect operates
when the radial gas inflow speed is faster than the outward photon
diffusion speed, i.e., $|v_r| > c/\tau$, where 
$\tau ~(=\rho \kappa_{\rm es}r)$ is the optical depth to 
electron scattering. This condition is satisfied within the so-called 
``trapping radius'' defined by
\begin{equation}
R_{\rm tr} \equiv \frac{\kappa_{\rm es}}{4\pi c}\dot{M}_\bullet 
=5\dot{m}R_{\rm Sch},
\end{equation}
where $\dot{m}\equiv \dot{M}_\bullet / \dot{M}_{\rm Edd}$.  The
trapping effect becomes physically relevant when this radius is
outside $R_{\rm Sch}$, i.e. for $\dot{m}\gsim 0.2$.  Since most of the
radiation produced inside $R_{\rm tr}$ is advected with the flow due
to electron scattering, the diffusive luminosity seen at larger radii
is limited to $L \lsim GM_\bullet \dot{M}_\bullet/R_{\rm tr}=L_{\rm
  Edd}$, {\em independent of the mass inflow rate}.  Therefore, the BH
growth rate is unlimited, and can exceed the Eddington value by an
arbitrary factor, as long as a correspondingly large amount of
inflowing gas is maintained from larger scales down to the vicinity of
the BH (see also \cite{Begelman1978} who constructed a global
spherical accretion solution for ionized gas at $\dot{M}_\bullet \gg
\dot{M}_{\rm Edd}$).

\begin{figure}[t]
\includegraphics[clip,width=3.6in]{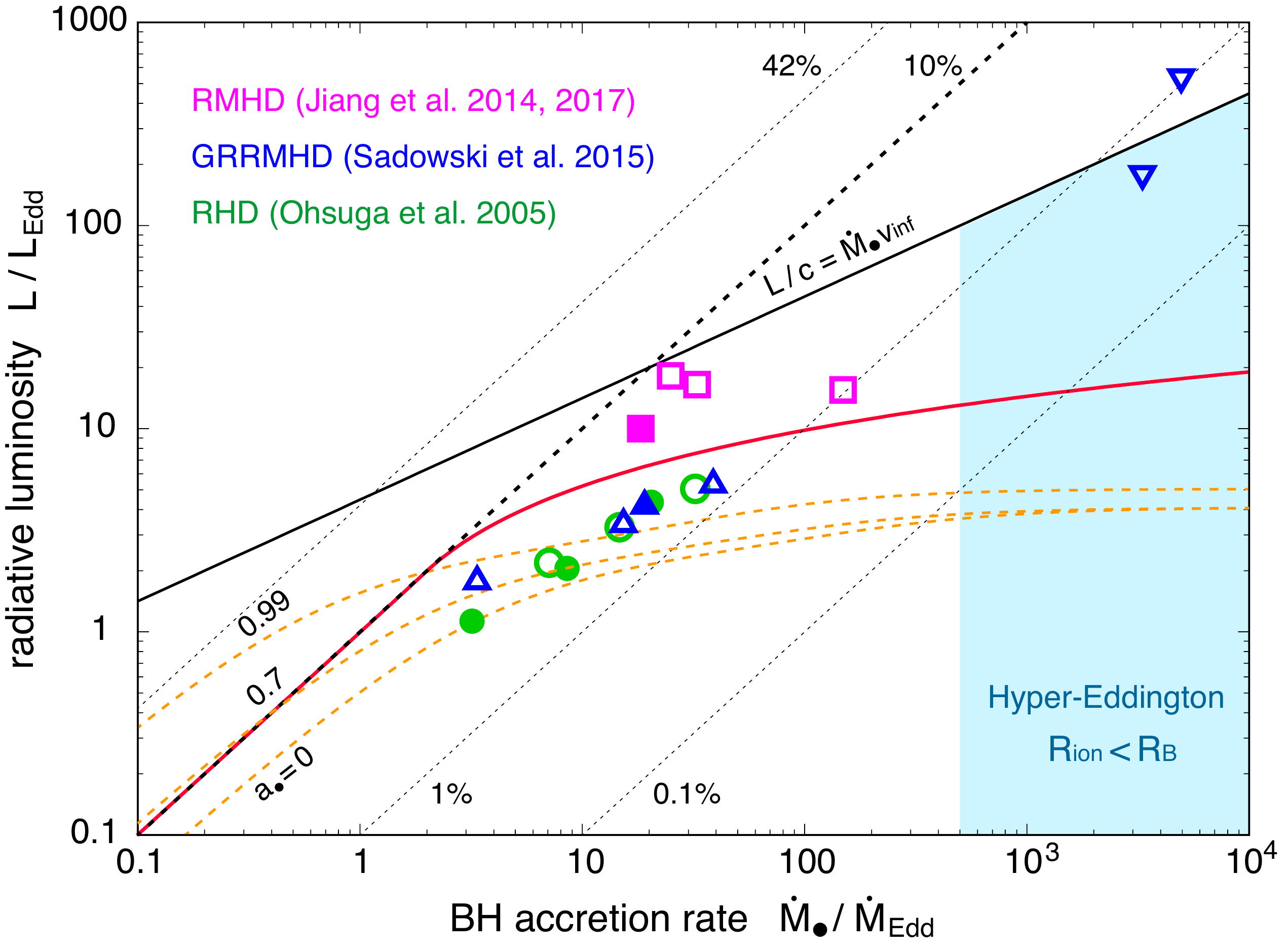}
\vspace{-0.5\baselineskip}
\caption{Summary of theoretical results for radiative luminosity
  vs. BH accretion rate.  In the analytical, ``slim-disk'' model
  assuming a pseudo-Newtonian potential (\citealt{Watarai2000}; red)
  or taking into account the GR effect around a BH with a spin of
  $a_\bullet$ (\citealt{Sadowski+2015}; orange), the radiative
  luminosity gradually increases as $L \propto \ln(\dot{m}$) at high
  rates.  Simulation results are shown by green circles (RHD;
  \citealt{Ohsuga+2005}) for metallicities $Z=0$ (filled) and
  $Z=Z_{\odot}$ (open), blue triangles (GRRMHD;
  \citealt{Sadowski+2015}) for $a_\bullet=0$ (open) and $0.9$
  (filled), and magenta squares (RMHD;
  \citealt{Jiang+2014,Jiang+2017}) for stellar-mass BH (filled) and
  SMBH/AGN (open). For a highly magnetized accretion disk around a
  rapidly spinning BH $a_\bullet=0.9$ (inverted triangle), the disk
  transits into a magnetically arrested disk (MAD) state, producing
  higher radiative luminosity but with a lower efficiency.
  Metallicity and BH spin both impact the structure of the flow, via
  opacity and radiative efficiency near the BH, respectively.  These
  simulations find self-consistent super-Eddington accretion on small
  scales with lower values of the radiative efficiency below $10\%$
  (thick dashed), but are numerically limited to model only short
  durations and small scales.  In the shaded region, hyper-Eddington
  accretion from the BH sphere of influence $R_{\rm B}$ would be
  realized and sustained, because the ionized region is smaller than
  the Bondi radius ($R_{\rm ion}\lsim R_{\rm B}$), and radiative
  feedback is therefore unable to suppress the inflow.  The efficient
  growth phase can stably exist unless the outward momentum $L/c$
  dominates the inward ram pressure of the rapidly accreting gas
  (black solid).}
\label{fig:slim_disk}
\end{figure}

The above consideration holds, however, only in spherical symmetry,
and ignores the question of the stability of the flow.  Subsequent
analytical work~\citep[e.g.][]{QuataertGruzinov2000,BlandfordBegelman2004}, as
well as early multi-dimensional hydrodynamical
simulations~\citep[e.g.][]{Stone+1999,IgumenshchevAbramowicz1999},
including those with magnetic fields \citep[e.g.][]{Igumenshchev+2003}
suggested that when $\dot{M}_\bullet \gg \dot{M}_{\rm Edd}$, these
so-called radiatively inefficient accretion flows (RIAFs) become
unstable to outflows, and only a small fraction of the mass reaches
the event horizon.

The photon-trapping effect has more recently been incorporated into
accretion disk models including direct radiation hydrodynamical (RHD)
simulations.  In \textbf{Figure~\ref{fig:slim_disk}}, we summarize
theoretical predictions of the radiative luminosity, as a function of
the dimensionless accretion rate captured by the BH, based on both
analytical work and RHD results.  In the slim-disk analytical model
(red curve), the radiative luminosity is proportional to
$\dot{M}_{\bullet}$ in the sub-Eddington regime and gradually
increases as $L/L_{\rm Edd}\propto \ln (\dot{m})$ in the
super-Eddington regime \citep{Abramowicz+1988,Watarai2000}.  Rotating
BHs produce radiation more efficiently at lower accretion rates
($\epsilon \approx 0.42$ for the dimensionless spin parameter
$a_\bullet = 0.99$), but the luminosity is still saturated at
$L/L_{\rm Edd}\sim 3$ at higher rates of $\dot{m}\gsim 10$
(\citealt{Sadowski2009}; orange curves, for three different spin
parameters).  In the past decade, RHD simulations including magnetic
fields (RMHD) and general relativistic effects (GRRMHD) have revealed
the properties of rapidly accreting gas within a few 100 $R_{\rm Sch}$
of the BH \citep{Ohsuga+2005,Jiang+2014,McKinney+2014,Fragile+2014,
  McKinney+2015,Sadowski+2015,TakahashiOhsuga2015}.  The radiative
efficiency modestly decreases with the accretion rate down to
$\epsilon \approx 1-5\%$ at $3 \lsim \dot{m} \lsim 150$.

The numerical results are overall qualitatively consistent with the
analytical model, but have some discrepancies.  In fact, the
efficiency of photon trapping is significantly reduced due to
non-inflowing gas motion caused by radiation pressure and magnetic
buoyancy \citep{Jiang+2014}, which are not taken into account in the
analytical models.  Importantly, the radiative efficiency obtained in
simulations with approximate numerical algorithms for radiative
transfer that impose local closure relations between the radiation
pressure tensor and radiation energy density, such as the flux-limited
diffusion (FLD; green) or the so-called M1 closure (blue)\footnote{The
  two approximated treatments cannot capture the angular distribution
  of photons near the photosphere accurately. Since the radiation flux
  in FLD points toward {\it any} gradient of radiation energy density,
  unphysical radiation flux will be produced and thus such a structure
  would likely be smeared out. In M1 closure method, the collimation
  level of the radiation-driven outflow might be affected, since
  photons in the outflow will be merged into a single beam near the
  photosphere.}  is systematically lower than those with a more
accurate numerical algorithm to solve the time-dependent radiative
transfer equations directly (magenta).  In the latter case, rapid gas
accretion is still allowed in the equatorial region, but a large
amount of radiation emerges with $1\lsim L/L_{\rm Edd}\lsim 20$ toward
the polar conical regions.

Of particular interest to us are simulations at the highest accretion
rates.  It would be a remarkable coincidence if the mass supply rate
from large scales precisely tracked $\sim \dot M_{\rm Edd}$ as a BH
grows by orders of magnitude in mass.  More likely, the mass supply
rate is initially much larger, and then gradually decreases when
measured in Eddington units.  \cite{Jiang+2017} have explored a case
with $\dot{m}\approx 150$, where even the polar funnel regions become
optically thick.  Since the disk has inflows instead of launching
strong outflows, radiation is effectively trapped and advected towards
the BH, with a small fraction of the radiation able to diffuse
outward, matching the expectations of the analytical models.
\cite{Sadowski+2015} reported a radiative efficiency as low as
$\epsilon \approx 0.01$ at their very high accretion rates of
$\dot{m}\approx 5\times 10^3$.  In this latter simulation, the BH was
assumed to be rotating with $a_\bullet =0.9$ and the initial poloidal
magnetic field in the disk was assumed to be in a so-called
magnetically arrested disk (MAD) state. In this state, the accreting
gas drags the poloidal magnetic field to the center such that the
accumulated, strong field disrupts the inflow structure and is likely
to produce outflows and/or jets.  Once the disk turns into a MAD
state, the luminosity becomes as high as $\sim 100~L_{\rm Edd}$, but
the radiative efficiency is still as low as $<1\%$ even for a rapidly
rotating BH (see also \citealt{Narayan+2003,Tchekhovskoy+2011} and
\citealt{McKinney+2015}).

It is worth noting that these simulations have explored the properties
of accretion flows on small scales, assuming a compact torus in
hydrostatic equilibrium as the initial state, or adopting mass-input
boundary conditions.  Even though the BH feeding rate is high, as
shown in \textbf{Figure~\ref{fig:slim_disk}}, a steady accretion disk
forms only within $R_{\rm disk}\sim 20-100~R_{\rm Sch}$, which
corresponds to the (half) radius of the location of the density peak
of the initial torus.  Outside the steady disk, a significant fraction
of the gas is ejected in a wind, and in fact the mass-loss rate
dominates significantly over the BH feeding rate.  Imposing mass-input
from the outer boundary, the mass inflow rate decreases toward the
center owing to strong outflows ($\dot{M}_{\rm in}\propto r^s$, where
$0\lsim s \lsim 1$; see \citealt{BlandfordBegelman2004}), and thus
only a fraction $(R_{\rm disk}/R_{\rm tr})^s$ of the inflowing gas
reaches a steady accretion disk and is ultimately contributing to the
growth of the BH.  The existence of outflows launched from a rapidly
accreting BH seems ubiquitous and could potentially even reverse the
inflow.  Therefore, it is crucial to address how these super-Eddington
accretion simulations are connected with the outer boundary conditions
on larger scales, and to assess whether, and by how much, the
radiative and mechanical outputs might suppress the gas inflow at the
BH's horizon.

In summary, high accretion rates exceeding the Eddington value are 
possible but produce intense radiation flux toward the polar directions 
with $L\approx O(1-10)~L_{\rm Edd}$.  These results, however, are 
valid only {\it as long as a sufficient amount of gas at rates of 
$\dot{M}\gg \dot{M}_{\rm Edd}$ is supplied from larger scales without 
being impeded by the strong radiation feedback}.

\subsubsection{Inflow from large scales}
\label{sec:largescale}

Gas inflows from larger scales ($r \gg R_{\rm Sch}$) can be triggered
by several physical processes.  First, baryons accrete into a DM halo
along well-defined cold filamentary streams connected with the
large-scale cosmic web.  The mass accretion rate averaged over 
cosmological timescale is approximately given by
\begin{equation}
\dot{M}\approx f_{\rm b}\frac{V_{\rm circ}^3}{G} \approx 
0.3~\msunyr \left(\frac{V_{\rm circ}}{20~\kms}\right)^3,
\end{equation}
where $f_{\rm b}\approx 0.16$ is the global mean baryon fraction and
$V_{\rm circ}$ is the circular velocity of the halo.  Such large-scale
inflows are expected to be triggered by major mergers of two galaxies
\citep{Springel+2005,HopkinsQuataert2010, Mayer+2010} or in massive DM halos in
which the gas cooling timescale is significantly shorter than the
dynamical timescale
\citep{BirnboimDekel2003,Keres2005,DekelBirnboim2006,DiMatteo+2012}.
Strong perturbations in both gas and stars in a merging galaxy lead to
non-axisymmetric spiral structures, which transport angular momentum
and induce mass accretion down to smaller scales (see detailed discussion in
\S\ref{sec:angmom} below).

Second, the rapidly accreted pristine gas settles into a compact
circum-nuclear disk, which becomes gravitationally unstable and thus
leads to fragmentation and clump formation
\citep{OhHaiman2002,LodatoNatarajan2006,Dekel+2009}.  Since primordial
gas is as warm as $T \sim 10^3~\K$ due to the absence of metal cooling
\citep{Palla+1983}, massive self-gravitating clumps form with a Jeans
mass of $M_{\rm J} \approx 2\times 10^4~\msun ~n_{\rm H, 4}^{-1/2}
T_3^{3/2}$ and collapse at rates of
\begin{equation}
\dot{M} 
\approx \frac{M_{\rm J}}{t_{\rm ff}}
\approx \frac{c_{\rm s}^3}{G}
\approx 4\times 10^{-3}~\msunyr T_3^{3/2},
\label{eq:mdot_sg}
\end{equation}
where $n_{{\rm H}, x} \equiv n_{\rm H}/(10^x~\cc)$ and $T_y \equiv T/(10^y~\K)$.
Note that the accretion rate depends only on the temperature and not
on the density \citep{Larson1969,Penston1969}.  The physical size of
the collapsing clump is given by the Jeans length, 
$\lambda_{\rm J}\approx 4~\pc ~n_{\rm H, 4}^{-1/2} T_3^{1/2}$, 
which is substantially smaller than the halo scale but still far away from the central BH
itself.  If a seed BH is embedded in such an unstable cloud, the mass
accretion onto the BH is much higher than the Eddington rate, namely
$\dot{M}/\dot{M}_{\rm Edd} \approx 2\times 10^3 ~T_3^{3/2}
(M_\bullet/100~\msun)^{-1}$.

Third, on smaller scales, the dynamics of accreting gas is finally
influenced by gravity of the central BH.  The characteristic scale is
the so-called Bondi radius, defined by $R_{\rm B} \equiv GM_\bullet /
c_{\rm s}^2\approx 0.06~\pc~T_3^{-1}(M_\bullet/100~\msun)$, where
$c_{\rm s}$ is the sound speed of the gas.  Gas is captured by the BH
and begins to accrete from the Bondi radius.
If the specific angular momentum of the gas is sufficiently small, a
centrifugally supported disk forms only inside the Bondi radius, and
gas flows inward at supersonic velocities. The characteristic
accretion rate at this radius is the Bondi rate,
\begin{align}
\dot{M}&\approx \dot{M}_{\rm B} \equiv \pi e^{3/2} \rho \frac{G^2M_\bullet ^2}{c_{\rm s}^3},\nonumber\\
&\approx 4.5\times 10^{-3}~\msunyr ~ n_{\rm H, 6} ~T_3^{-3/2} \left(\frac{M_\bullet}{100~\msun}\right)^2,
\end{align}
where $\rho$ is the mass density at $r=R_{\rm B}$, the gas is assumed
to be isothermal \citep{Bondi1952}, and $\dot{M}/\dot{M}_{\rm
  Edd}\approx 2\times 10^3~n_{\rm H,
  6}~T_3^{-3/2}(M_\bullet/100~\msun)$.  The Bondi rate should in
general be considered an upper limit on the accretion rate, because it
assumes free-fall of gas from the Bondi radius.  Negative effects
associated with BH feedback, gas rotation and MHD-winds reduce the
inflow rate
\citep{ProgaBegelman2003,Sijacki+2007,Li+2007,Tchekhovskoy+2011,Sadowski+2015},
and even in the absence of these effects, when the gas is very cold,
it is susceptible to gravitational perturbations which determine the
inflow rate~\citep{HopkinsQuataert2010}.  Importantly, the Bondi
radius is generally much larger than the trapping radius, namely
$R_{\rm B}/R_{\rm tr} \approx 7\times 10^3
~(\dot{m}/10^3)^{-1}T_3^{-1}$ (see \textbf{Figure~\ref{fig:phys_scale}}).

\subsubsection{Photoionization and heating}
\label{sec:photoheat}

\begin{figure}[t]
\includegraphics[clip,width=2.9in]{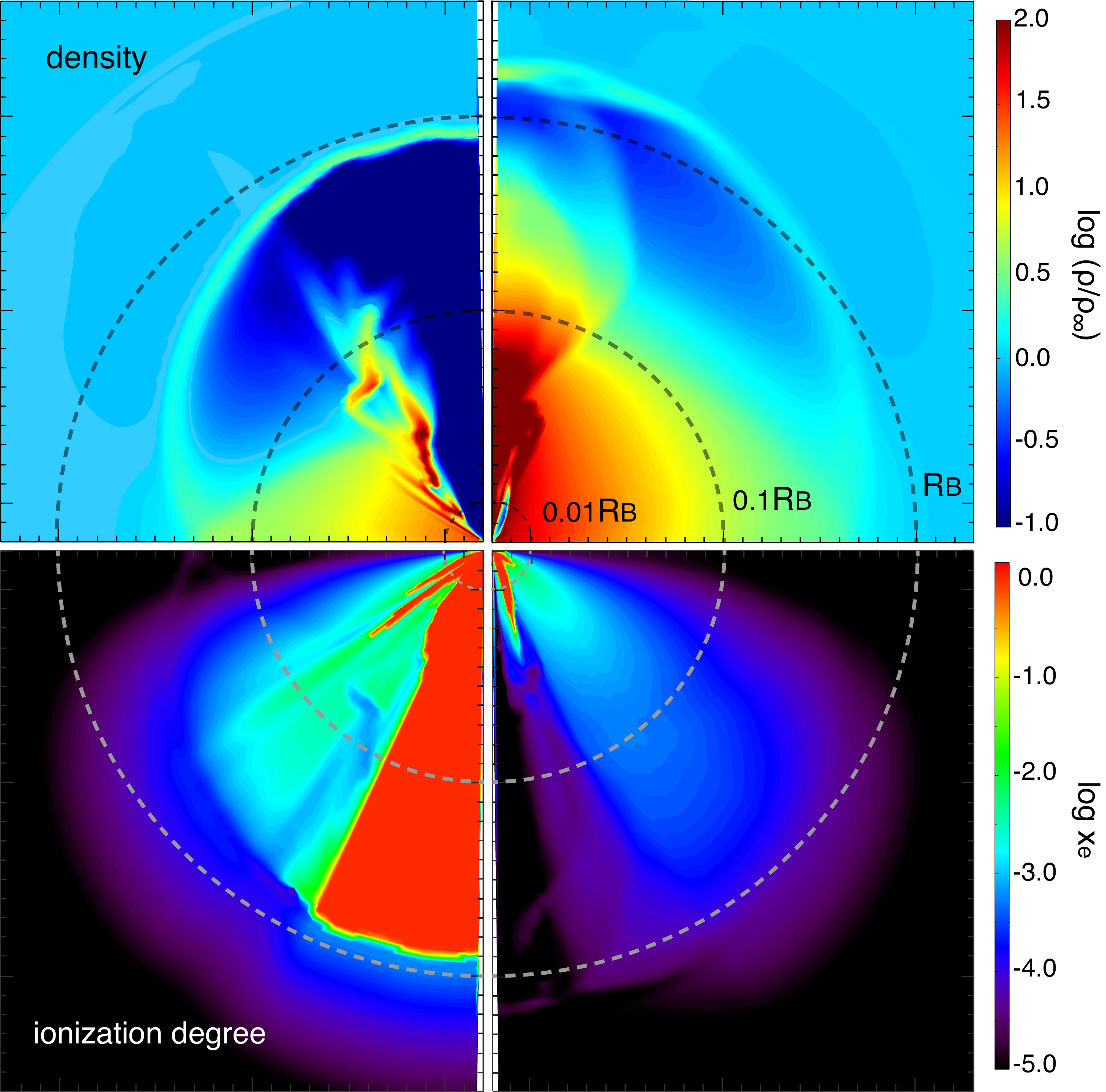}
\caption{Two-dimensional distributions of gas density (top panels) and
  ionization fraction (bottom panels) in accretion flows onto a BH in
  the $\log r$ -- $\theta$ plane.  The concentric dashed circles
  indicate constant fractions of the Bondi radius: $r/R_{\rm B}=0.01$,
  $0.1$ and $1.0$.  The left panels show the accretion flow during a
  transition to the hyper-Eddington regime.  Even though the
  ionization front reaches the Bondi radius in the polar regions,
  rapid gas accretion begins to occur through the equatorial plane,
  with intense ram pressure.  The right panels show the accretion flow
  after the transition, when the HII region collapses and the entire
  flow becomes neutral.  The accretion rate dramatically increases
  from time-dependent oscillations at $\langle \dot{M} \rangle \lsim
  \dot{M}_{\rm Edd}$ to steady hyper-Eddington accretion at $\dot{M}
  \sim \dot{M}_{\rm B} (\sim 500~\dot{M}_{\rm Edd}$).  The figure is
  based on the simulation data from \cite{Takeo+2018}.  }
\label{fig:hyper_Edd}
\end{figure}

Radiative feedback associated BH accretion can play a crucial role on
the inflow rate from the Bondi radius, where gas is only marginally
bound to the BH.  Even if a BH is embedded in a self-gravitating
cloud, the ratio of the thermal energy to the gravitational energy is
as high as $\gsim 5/\pi^2\approx 0.5$ \citep{Larson1969,Truelove+1997}.
Therefore, photoionization and heating by the central BH can unbind
the gas, and suppress gas inflow from large scales.  Unfortunately, no
multi-dimensional simulation can self-consistently resolve all the
relevant scales from the event horizon to the Bondi radius.  To
roughly quantify this effect, let us instead approximate the size of
the ionization bubble by the Str\"omgren radius in a uniform medium,
\begin{equation}
R_{\rm ion}\approx 
\left(\frac{3Q_{\rm ion}}{4\pi \alpha_{\rm B}n_{\rm H}^2}\right)^{1/3} 
\propto M_{\bullet,2}^{1/3}n_{\rm H,7}^{-2/3} f_{\rm Edd}^{1/3},
\end{equation}
where $Q_{\rm ion}$ is the ionizing photon flux, $\alpha_{\rm B}$ is
the case-B recombination rate, and $f_{\rm Edd}\equiv L/L_{\rm Edd}
\sim O(1)$, providing $R_{\rm ion}/R_{\rm B}\approx f_{\rm
  Edd}^{1/3}~n_{\rm H, 7}^{-2/3}(M_\bullet/100~\msun)^{-2/3}$.  Thus, for a
$100~\msun$ stellar-remnant BH, embedded in a gas cloud with $n_{\rm
  H}<10^7~\cc$, the ionization front expands outside the Bondi radius
($R_{\rm ion}\gsim R_{\rm B}$). Accretion becomes intermittent and the
time-averaged rate is orders of magnitude below the original Bondi
rate without feedback, remaining in the $\langle\mdot\rangle \lsim 0.5$ regime
(\citealt{Milosavljevic+2009,Milosavljevic+2009b,ParkRicotti2011,ParkRicotti2012};
see also \citealt{CiottiOstriker2001}).  Therefore, super-Eddington
accretion with a large photon trapping radius can not be realized in
this regime.

However, when the ambient gas is sufficiently dense that the Bondi
rate exceeds $\sim 500~\dot{M}_{\rm Edd}$,
one-dimensional RHD simulations find that the inflowing gas structure
approaches a steady state without time-dependent oscillations,
yielding hyper-Eddington accretion \citep{Inayoshi+2016}.
Two-dimensional RHD simulations have confirmed this conclusion
\citep{Takeo+2018} and yielded a more detailed understanding of the
accretion morphology \citep{Sugimura+2017}.
\textbf{Figure~\ref{fig:hyper_Edd}} shows the
two-dimensional distribution of gas density (top panels) and
ionization fractions (bottom panels) of accretion flows in this
high-density regime.  In the early stage (left panels), ionizing
radiation does not reach the Bondi radius due to effective
recombination ($R_{\rm ion} \lsim R_{\rm B}$; the outer-most dashed
circle).  Then, neutral gas accumulated at $R_{\rm ion}< r < R_{\rm
  B}$ forms a dense shell and collapses onto the center without being
prevented by radiative feedback, but rather lead to collapse of the
ionized region (right panels).  As a result, steady, isothermal
accretion at $\dot{M}\approx \dot{M}_{\rm B}~\gsim 500~\dot{M}_{\rm
  Edd}$ is achieved.

This hyper-Eddington accretion solution is stable against radiative
and mechanical feedback because photon trapping reduces the emergent
luminosity, unless the outward momentum $L/c$ dominates the inward ram
pressure of neutral gas (black solid line in
\textbf{Figure~\ref{fig:slim_disk}}).  \citet{Sakurai+2016} have found
that the inflow rate is not suppressed unless the luminosity emerging
at the photosphere is a factor of $10-100$ above $L_{\rm
  Edd}$\footnote{ Radiation heating plays an important role on
  suppressing inflow gas from the Bondi scale before the transition
  where $\langle L \rangle \lsim L_{\rm Edd}$.  Therefore, the
  radiation force onto inflow gas exerted through electron scattering
  and even bound-free absorption by neutral hydrogen is subdominant
  until the transition to a hyper-Eddington accretion phase.}.
\textbf{Figure~\ref{fig:hyper_Edd}} also shows that the anisotropic
radiation field has a large impact on the gas distribution near the
polar regions, but much less in regions near the equatorial plane,
i.e., gas flows inward through a disk (see also
\citealt{Sugimura+2017, Sugimura+2018}).  The size of the ionized
bubble depends on the spectrum of the radiation emerging from the
photosphere, with harder spectra easing the criterion for
hyper-Eddington accretion \citep{Takeo+2019}.  In addition, dust in
the accreting gas softens the spectral shape due to UV attenuation,
making the ionized regions smaller \citep{Yajima+2017}.  Thus, rapid
accretion is triggered even for lower BH masses or ambient density
unless $Z>10^{-2}~\zsun$ where super-Eddington accretion is prevented
by the radiation pressure of diffuse infrared light on dust grains
\citep{Toyouchi+2019}.

\subsubsection{Mechanical feedback}
\label{sec:mechanical feedback}

In addition to radiative feedback, BHs accreting at super-Eddington
rates can exert negative feedback via winds and jets.  In fact, most
numerical simulations focusing on the dynamics of a BH accretion disks
on small scales ($\lsim 100~R_{\rm Sch}$) find outflows/jets driven by
a strong radiation flux and/or a strongly arrested magnetic field.
The mechanical feedback associated with BH feeding could play an
important role, similarly to the low-$z$ AGN population, which is
believed to affect large-scale environments, such as star formation on
galactic scales \citep[e.g.,][]{Fabian2012,HeckmanBest2014}.  Although
mechanical feedback has not received much attention in the context of
BH growth in high-$z$ protogalaxies, this effect could limit their
growth significantly.

Recently, \cite{Regan+2019} have investigated the effect of jets
launched from an accreting seed BH on gas inflows in an ACH,
performing cosmological simulations that resolve the BH gravitational
sphere of influence.  They found that the momentum injection by jets
evacuates a region of approximately $\approx 0.1$ pc surrounding the
BH seed, but cannot break out of the halo.  Since the impact of the
feedback is limited to the vicinity of the BH, the heated and kicked
gas will cool and fall back to the center, leading to burst-like
accretion episodes as seen in RHD simulations that take into account
radiative heating (see \S\ref{sec:photoheat}).  As a result, the
time-averaged accretion rate over one dynamical timescale ($\approx
100$ kyr) at $r\approx R_{\rm B}$ becomes as low as $\approx
0.2-0.8~\dot{M}_{\rm Edd}$ (note that their definition of
$\dot{M}_{\rm Edd}$ is 1.6 times higher than ours).  This simulation
result suggests that a modestly high level of BH accretion at $\approx
\dot{M}_{\rm Edd}$ is possible unless the jets impact the gas near the
Bondi scale $\approx 1$ pc.  In a longer-duration simulation ($t \gg
100$ kyr), a substantial fraction of the gas could remain within the
central pc, and the accumulated mass within the BH influence radius
could fall back to the central BH at an even higher rate because ram
pressure of inflows would dominate the momentum output of the jets.
Although there are still several caveats on the prescriptions for jet
mechanical feedback (e.g., mass loading factor and jet energy
efficiency), future studies should address these issues and improve
our understanding of the early stage of BH growth.

\subsection{Stellar-mass Black Hole Remnants of the First Stars}
\label{sec:lightseed}

Having established that super- or hyper-Eddington accretion can occur
in principle, we next examine how this may be realized and lead to
rapid growth of BHs in the early universe.  A natural and attractive
candidate for the initial seed BHs for such rapid accretion is
remnants formed in the gravitational collapse of massive Pop III stars, 
which are the first-generation stars in the universe
(\citealt{Carr+1984,OmukaiNishi1998,Abel+2002,Bromm+2002,
  Yoshida+2006,Yoshida+2008,BrommYoshida2011}; see also the review by
\citealt{Greif2015} and references therein).

In the framework of hierarchical structure formation in the
$\Lambda$CDM model, the first collapsed baryonic objects are expected
to form at $z\gsim 20$ in DM minihalos with masses of
$10^5-10^6~\msun$~\citep{Haiman+1996a,Tegmark+1997}.  The virial
temperature of these halos is $T_{\rm vir}=$ few$\times 10^2~\K$,
which is sufficiently high to excite line emission from H$_2$, formed
via gas-phase reactions in the pristine metal-free gas (see more
discussion in \S~\ref{subsec:warmgas}).  This emission allows gas to
cool and condense in the halo, but since the collapsing gas remains
relatively warm ($\gsim 100~\K$), it has a large Jeans mass ($M_{\rm
  J}\propto T^{3/2}\sim 10^3~\msun$), and protostellar cores have a
large accretion rate ($\sim 10^{-3}~\msunyr$; see
equation~\ref{eq:mdot_sg}).  As a result, it has long been thought
that the first stars were unusually massive - gaining masses of a few
100 $\msun$ in their Kelvin-Helmholtz contraction time of $\sim10^5$
years.  On the other hand, a growing Pop~III protostar has a high
effective surface temperature and emits copious UV radiation. The
corresponding ionization and heating of the surrounding gas
self-regulates their growth, and limits their final masses to $\sim
100~\msun$, with the precise value depending on the ambient temperate
and accretion
rate~\citep{McKeeTan2008,Hosokawa+2011,Stacy+2012,Susa+2014}.  Recent
cosmological simulations of Pop~III star formation in minihalos have
suggested that the initial mass function indeed tends to be overall
top-heavy, with a nearly flat mass distribution in the range $10\lsim
M_\star/\msun \lsim 300$ \citep{Hirano+2014,Hirano+2015,Stacy+2016}, and with the
upper end limited by feedback from the protostar's own UV radiation.

Massive Pop~III stars are expected to be short lived ($\sim 10^6$ yr)
and to promptly endow their parent minihalos at $z\gsim 20$ with
$M_{\rm seed}\approx 10-100~\msun$ seed BHs.  Note that very massive
nonrotating Pop~III stars with $140\lsim M_\star/\msun \lsim 260$ may
not leave any remnant because of energetic pair-instability supernova
explosions \citep[PISNe;][]{HegerWoosley2002}.  Stellar rotation would
affect the final fate of massive Pop~III stars due to rotation-induced
mixing and extension of the He core.  However, the range of zero-age
main sequence mass where PISNe are predicted to occur is shifted lower
only by 10\% \citep{Takahashi+2018}.

While the early appearance of such seed BHs is good news, their
subsequent rapid growth in minihalos unlikely.  Their Pop~III stellar
progenitors (as well as any other Pop~III star(s) in the same halo)
irradiate and blow the ionized gas out of the minihalo, because the
gravitational potential well is not sufficiently deep: the sound speed
of ionized gas, $\sim 10~{\rm km~s^{-1}}$, exceeds the escape
velocity, $\sim 1~{\rm km~s^{-1}}$, from
minihalos~\citep{Kitayama+2004,Whalen+2004,JohnsonBromm2007}.  Some
remnants are left after energetic supernova explosions, which likewise
quickly evacuate the gas from the minihalos
\citep{Kitayama+2005,Whalen+2008,Ritter+2012}.  Therefore, the remnant
BHs are likely to typically find themselves in exceptionally
low-density environments, and cannot accrete efficiently.  Even if the
BHs were to avoid such starvation, and began to accrete, the UV/X-ray
radiation associated with this accretion itself would then heat the
ambient gas and evacuate the central dense region, self-regulating
their growth until their host halos grow much more massive
\citep{Alvarez+2009,Jeon+2012,Tanaka+2012}.
Finally, another obstacle to growth for low-mass BHs is their erratic motion 
around the central regions~\citep{Pfister+2019}; as a result, these BHs 
spend most of the time away from the dense core and accrete inefficiently
~\citep{Smith+2018}. 

In principle, mergers of Pop~III BHs could also drive their mass
growth.  However, the merged BHs experience strong recoil kicks with
typical velocities of $100~\kms$ owing to gravitational wave emission,
depending on the mass ratio and spin configuration of the merging pair
\citep{Herrmann+2007,Koppitz+2007,Campanelli+2007}.  Since this
typical recoil velocity is well above the escape velocity from
minihalos, merged BHs will typically be ejected from their parent
halos to the intergalactic medium, where they cannot accrete at
high rates \citep{Haiman2004}.  This strongly limits the role of
mergers in the early growth of most BHs.  However a few ultra-early,
rare BHs can avoid this fate by not experiencing mergers until they
grow significantly in mass; subsequent mergers will then be at very
unequal masses ($\lsim 1:100$), where recoil speeds diminish below a
few~$\kms$~(leading to a ``rich-get-richer'' runaway; see
\citealt{VolonteriRees2006,TanakaHaiman2009}).

In summary, {\it efficient, sustained growth of Pop~III remnant BHs
  likely has to wait until these BHs end up in much more massive DM
  halos, whose potential is deep enough to gravitationally bind a
  sufficient amount of gas and recoiled BHs}.

\subsection{Rapid Growth of Seed BHs in High-Redshift Protogalaxies}
\label{subsec:BHaccretion-apps}

We here discuss under what circumstances early seed BHs may realize 
super- or hyper-Eddington accretion, sustained over several orders of 
magnitude growth in mass.  This corresponds to the pathway marked by 
the purple line in \textbf{Figure~\ref{fig:rees}}.

The natural place where such rapid BH growth may occur is in the
``atomic cooling halos'' (ACHs) introduced in \S\ref{sec:intro}, which are DM
halos whose virial temperature is just above the atomic-cooling
threshold ($T_{\rm vir}\approx 8000~\K$).  In a typical ACH, cooling
and collapse of the gas will be dictated by heavy elements produced in
prior episodes of star-formation, as well as ${\rm H_2}$ molecules
formed directly during the collapse. However, in rare cases, prior
star-formation, as well as ${\rm H_2}$ cooling, can be suppressed by
several effects: including intense external H$_2$-dissociating LW
irradiation \citep{Haiman+1997,Machacek+2001,WiseAbel2007b}, 
streaming motions between DM and baryons \citep{Fialkov+2012,TanakaLi2014}, dynamical
heating associated with halo mergers~\citep{Yoshida+2003,Wise+2019},
or some combination of these effects.  The chemistry and
thermodynamics of gas in these halos will be discussed in detail in
\S~\ref{subsec:warmgas} below, in the context of a similar set of
requirements for forming a massive $M_\bullet \sim 10^5~\msun$ seed
BH via a supermassive star.

Just before the atomic-cooling regime ($T_{\rm vir}\gsim 8000~\K$),
the pristine gas remains essentially adiabatic, and settles into a
hydrostatic-equilibrium profile in the DM halo's gravitational
potential, having a core with $\sim 0.1~R_{\rm vir}$ and an envelope
following $\rho \propto r^{-2}$ \citep{Visbal+2014_nogo}, where
$R_{\rm vir}\approx 470~\pc ~(T_{\rm vir} /8000~\K)^{1/2}
[(1+z)/16]^{-3/2}$ is the halo virial radius.  Importantly, the mass
of the primordial gas in the core region is as high as $M_{\rm
  core}\approx 2\times 10^5~\msun ~(T_{\rm vir}/8000~\K)^{3/2}
[(1+z)/16]^{-3/2}$ \citep{Inayoshi+2015}, which serves as a rough
upper limit on the mass budget available to grow a massive BH.  Once
the halo crosses the atomic-cooling threshold, atomic-line cooling (primarily
Ly$\alpha$ transition) begins to operate, and gravitational
contraction of the gas in the core is triggered. The core region
develops a density profile as steep as $\rho \propto r^{-2}$, as seen
in cosmological simulations of high-$z$ protogalaxies without prior
star formation \citep{Wise+2008, Regan+2014}.

The situation envisioned here is that one of the minihalo progenitors
of the ACH did manage to form a Pop~III star, but its host minihalo
remained chemically pristine, because this Pop~III star quenched
subsequent star formation by its UV radiation \citep{OmukaiNishi1999}
and then collapsed into a BH without exploding as a SNe or ejecting
any metals \citep{HegerWoosley2002}.  In fact, this fate may be
typical for massive Pop~III stars, and therefore several of the
minihalo progenitors of the ACH (rather than just one) could have an
early star-formation episode, as long as stars in the Pop~III stellar
IMF which eject metals (low-mass stars and PISNe; see below) are not
sampled.  In this case, there can be multiple stellar-mass BHs,
initially spread spatially in the pristine ACH.  However, remnant BHs
would quickly decay their orbits due to dynamical friction on the DM
and the gas, depending on their initial orbital properties, and sink
to the dense central region.  \cite{Ryu+2016} simulated the orbital
motion of Pop~III remnant BHs embedded in gas-rich protogalaxies,
taking into account gas drag on the BHs, and found that most initial
BH configurations allow one BH (but no more than one) to sink to the
center and to grow rapidly.

As discussed in \S\ref{sec:lightseed}, Pop~III remnant BHs hardly grow
in low-mass minihalos.  However, the BH buried in the dense region in
the ACH ($\rho \propto r^{-2}$) can be fed at the full Bondi accretion
rate, unless BH radiative feedback prevents the inflowing matter.
When the ambient matter is self-gravitating, the accretion rate onto
the central object is simply given by $\dot{M}_{\rm B}\sim 20~c_{\rm
  s}^3/G$ \citep[see also][]{Becerra+2018}.  In fact, even if ${\rm
  H_2}$ formation is triggered after the halo crosses the atomic-cooling
threshold, the accretion rate onto the central BH may remain high,
since the Bondi radius is relatively large ($\sim 0.06$ pc; see
\textbf{Table~\ref{tab:phys}}), and the gas inside this radius is not
self-gravitating.  As a result, the hyper-Eddington condition, i.e.,
$\dot{M}_{\rm B}\gsim 500~\dot{M}_{\rm Edd}$, is satisfied until the
BH mass reaches $\approx 2\times 10^5~\msun$.  When the BH mass
exceeds this critical value, the ionized bubble created by the BH
expands outside the Bondi radius, where the expansion is further
accelerated because of $\frac{d\ln \rho}{d\ln r} < -1.5$
\citep[e.g.,][]{Mellema+2006}, and the inflowing gas is heated up and
would also likely be pushed outward by Ly$\alpha$ photons
\citep{Smith+2017}.  As a result, the rapid hyper-Eddington growth
phase of the remnant BH in the ACH is terminated at this mass,
independent of the initial seed mass \citep{Inayoshi+2016}.  By
coincidence, this critical mass is comparable to the core mass before
the rapid BH growth begins -- implying that the bulk of the gas in the
core region would be consumed in a hyper-Eddington phase.

In the above works, the collapsing gas had a quasi-spherical geometry.
\citet{Lupi+2016} considered a similar setup, but with stellar-mass
BHs orbiting in a circum-nuclear disk, in which clumpy structures form
by gravitational instability. They found that the orbiting BHs capture
and swallow massive clumps at super-Eddington rates.  Combining
merger-tree simulations, \cite{Pezzulli+2016} have discussed the
evolution of seed BHs at high redshifts, including AGN feedback with a
simplified model.  They concluded that seed BHs with $\sim
100-10^6~\msun$ would likely experience a rapid gas accretion phase in
gas-rich protogalaxies.  In particular, they find that $\sim
40\%~(10\%)$ of seeds can grow at hyper-Eddington accretion rates of
$\gsim 500~\dot{M}_{\rm Edd}$ at $z=15-20$ ($z=10-15$)\footnote{Note
  that because \cite{Pezzulli+2016} defined the Eddington rate as
  $\dot{M}_{\rm Edd}\equiv 16L_{\rm Edd}/c^2$, the hyper-Eddington
  criterion is set at $\dot{M}> 300~\dot{M}_{\rm Edd}$ in their Figure
  4.}.

{\em In summary, hyper-Eddington accretion can be sustained and
  quickly produce $10^{5-6}~\msun$ BHs by Pop~III remnant BHs that
  find themselves in special ACHs, with chemically
  pristine gas.}  Note that these conditions are somewhat less strict
than those necessary for the formation of massive seed BHs with
$M_\bullet \sim 10^5~\msun$ via a supermassive star (discussed in
\S\ref{subsec:warmgas} below), because in the hyper-accreting BH case,
some prior star-formation occurs in the ACH, and even ${\rm H_2}$
cooling may not prevent a brief hyper-Eddington phase once the atomic-cooling
threshold is crossed.  Nevertheless, pristine ACHs are required, which
are very rare at high-$z$. ACHs are more common at lower $z$, but are
more likely to be polluted by metals, yielding a ``sweet spot'' for
metal-free (or metal-poor) ACHs at $z=12-15$~\citep{Chon+2016}.  
These redshifts are still sufficiently high to permit further growth to $M_\bullet \approx
10^9~\msun$ by $z=6-7$ at the more leisurely Eddington rate.

\section{ANGULAR MOMENTUM TRANSPORT}
\label{sec:angmom}

As mentioned in \S~\ref{sec:growacc}, in order for infalling gas to
maneuver from galactic scales down to the nucleus and accrete onto a
compact central object (either a BH or a massive protostar), it needs
to shed its large angular momentum.  In this section, we discuss the
physics of angular momentum transport on multiple scales in the
general context of galaxy formation and AGN fueling
(\S~\ref{subsec:angmom1}), and then specialize to the analogous
problem in protogalaxies in the high-$z$ universe
(\S~\ref{subsec:angmom2}).

The angular momentum $J$ of an object of mass $M$ and radius $R$ can
be specified in terms of a dimensionless spin parameter $\lambda\equiv
J/\sqrt{2} M v_c R$, where $v_c=\sqrt{GM/R}$ is the Keplerian circular
velocity.  This parameter expresses the level of centrifugal support,
with $\lambda=0$ corresponding to no net rotation, and $\lambda\approx
1$ to full rotational support ($\lambda=1$ for an isothermal sphere).
Due to torques from nearby large-scale structures, DM halos
at the time of their virialization acquire a log-normal distribution
of $\lambda$, with a mean $\langle\lambda\rangle\approx 0.035$, weakly
dependent on either halo mass or collapse redshift, from galactic
halos in the local universe~\citep{BarnesEfstathiou1987,Bullock+2001},
down to the halo masses ($\approx 10^6~\msun$) and up to the redshifts
($z\approx15$) of interest for the formation of the first massive
BHs~\citep{DavisNatarajan2009}.

To illustrate the importance of angular momentum transport for the gas
component, it is useful to contrast two scales.  First, assuming that
the gas shares the halo's specific angular
momentum~\citep{vandenBosch+2002}, and that this specific angular
momentum is conserved as the gas cools and contracts inside the halo,
the centrifugal barrier would halt the collapse at $R_c\approx \lambda
R_{\rm vir}$~\citep{MoMaoWhite1998}.  This is a very large radius,
$R_c\approx 0.1~\kpc ~(M_{\rm vir}/10^8~\msun)^{1/3}[(1+z)/11]^{-1}
\approx 3\times 10^{20}~{\rm cm}$. By comparison, the Schwarzschild
radius of a stellar-mass BH of mass $M_\bullet$ is $R_{\rm Sch}=
3\times 10^{7}~{\rm cm}~(M_\bullet/100\,\msun)$, and even the radius
of a supermassive protostar (see \S\ref{subsec:journey-to-BH}) with
mass $M_\star$ is only $R_\star = 2.6\times 10^3~\rsun
~(M_\star/100~\msun)^{1/2} \approx 2\times 10^{14}~{\rm cm}~(M_\star
/100~\msun)^{1/2}$~\citep{Hosokawa+2012}.

{\em The conclusion is that the gas needs to reach distances $\sim 10^{6}$
times smaller than the centrifugal barrier, in order to be
incorporated into a central giant supermassive protostar, and needs to
move inward by a further factor of $\sim 10^7$ in order to accrete
onto a central stellar-mass BH.}

\subsection{Angular Momentum Transport in Galaxy Formation}
\label{subsec:angmom1}

It has indeed long been recognized that efficient angular momentum
transport is required to move gas inward by many orders of magnitude
in radius, from galactic ($\gsim$ kpc) scales down to the vicinity of
a central SMBH, to fuel active galactic nuclei~(see,
e.g. \citealt{Shlosman+1990} for a review).  In this broader context
of galaxy formation, several distinct processes have been understood
to play important roles, roughly staggered in three distinct spatial
scales (which can, however, overlap).

First, on the largest scales, both the gas, and the collisionless
components (DM and any pre-existing stars) develop non-axisymmetric
morphologies.  Such non-axisymmetries are inevitably produced in major
mergers~\citep{BarnesHernquist1991}, but can also develop as a result
of perturbations in minor mergers or tidal interactions, or even arise
in isolated galaxies that already have large self-gravitating disks.
Self-gravitating disks or flattened structures are known to be
globally unstable to a spontaneous loss of axisymmetry when the ratio
of their bulk kinetic energy to potential energy, $T$/$|$$W$$|$
exceeds a critical
value~\citep{OstrikerPeebles1973,Christodoulou+1995}.  The resulting
spiral waves and bar-like structures that develop are known, in turn,
to transport angular momentum outward and to facilitate mass
inflow~\citep{Lynden-BellKalnajs1972}.  On galactic scales, the
collisionless components (DM and stars) are also important, and
misalignments between non-axisymmetric structures in these components
relative to those in the gas provide extra torques that help gas
inflow~\citep{Shlosman+1989,BarnesHernquist1991}.

Second, once the gas has cooled and contracted, it eventually becomes
self-gravitating, and can develop its own non-axisymmetric bar-like
structures, allowing continued gas inflow~(``bars-in-bars'';
\citealt{Shlosman+1989,Shlosman+1990}). However, the physics of
angular momentum transport in this regime is complicated further by
the fact that the gas becomes prone to local Toomre
instability~\citep{Goodman2003}, which could produce fragmentation and
efficient star-formation~\citep{GoodmanTan2004}. This could consume
much of the gas, and prohibit the large majority of the gas from
crossing this ``minefield'' and ever reaching the innermost regions,
where the growing BH stabilizes the inner disk~\citep{Thompson+2005}.

One promising idea is that turbulence, which inevitably develops,
facilitates gas inflow.  Since the gas needs to cool below the host
halo's virial temperature, bulk gas speeds typically exceed the sound
speed, and turbulence becomes supersonic.  Several hydrodynamical
simulations have indeed converged on the following broad picture in
this regime~\citep{Escala2007,Mayer+2007,Levine+2008, Mayer+2010, 
HopkinsQuataert2010,HopkinsQuataert2011,Choi+2013,Choi+2015}.
Large-scale global instabilities generate gas inflow, and also
concurrent turbulence down to the smallest resolved spatial scales. This turbulence 
has been suggested to support the disk against
gravitational
fragmentation~(\citealt{Levine+2008,BegelmanShlosman2009,Choi+2013}; 
although these results may have not yet numerically converged -- see further 
discussion below).  As
a result, a compact nuclear self-gravitating disk forms, which remains
locally stable.  The system can reach a quasi-steady state on long
time-scales, in which global instability drives intermittent barlike
structures. These bar-like structures, as well as turbulence itself,
redistribute the angular momentum in the disk on a dynamical
timescale, and can sustain a large gas inflow rate.  In a suite of
$\sim 100$ simulations, surveying the parameter space of galaxy
properties, \citet{HopkinsQuataert2010} find a cascade of secondary
instabilities, but with a diverse range of non-axisymmetric
morphologies beyond ``bars'', which are intermittent, and produce a
large ($\sim \msunyr$) but correspondingly
time-variable accretion rate.

Up to this stage, the presence or absence of a central massive object
was immaterial.  However, finally, once the gas reaches well inside the 
sphere of influence of the central BH (if there is one), the disk is no longer
unstable to either bar-like modes, or to local
Toomre-instability. \citet{HopkinsQuataert2010} find a new
gravitationally-driven instability at the boundary of this regime, in
the form of a precessing lopsided disk (or one-armed spiral).
However, further inside this regime, within $\sim 10^{4-8}~R_{\rm Sch}$
of the central BH (depending on BH mass and
accretion rate; e.g. \citealt{Haiman+2009}), the disk is
gravitationally stable.  In this regime, viscosity is understood
to be provided by magnetic fields. Even a vanishingly small initial
seed field (though amplified by turbulence) is sufficient to generate
the magneto-rotational instability (MRI; \citealt{BalbusHawley1991}).
MRI, or related MHD effects can efficiently transport angular momentum
in this smallest-scale regime, even at super-Eddington accretion
rates~\citep[e.g.,][]{Jiang+2014}.

The major caveat to the above picture is that simulations rely on
sub-grid prescriptions for cooling, star-formation, and feedback on
unresolved scales.  In particular, simulations typically impose a
temperature or entropy floor, which could have a large effect on star
formation, angular momentum transport, and the global behavior. These effects can
be parameterized and modeled analytically~\citep{HopkinsQuataert2011}.

\subsection{Angular Momentum Transport in High-Redshift Protogalaxies}
\label{subsec:angmom2}

Much of the physics described above also applies in the context of
forming massive BHs in the first galaxies.  The key differences are
that the gas is metal-free or metal poor, and cools less efficiently;
furthermore, gas is unlikely to cool greatly below the halo's virial
temperature.  As a result, star-formation is likely less efficient (at
least initially), and the self-gravitating disks that form are likely
to be thicker and less prone to instabilities.

Even before discussing angular momentum transport, we note that one
way to ease the fueling problem is to start with gas with
lower-than-usual angular momentum.  Such gas could be found in halos
in the low-$\lambda$ tail of the halo spin
distribution~\citep{EisensteinLoeb1995}, and/or in the low-$j$ tail of
the specific angular momentum distribution of gas in individual
halos~\citep{Koushiappas+2004}.  A related idea is that in early DM
halos, which form at the knots of many filaments of the
proto-cosmic-web, the gas arrives from many directions along these
filaments, resulting in a significant cancelation of the net angular
momentum, allowing more efficient initial
contraction/inflow~\citep{Dubois+2012,Prieto+2015}.

While these effects can help, in order to reach the central massive
objects, all three of the above scenarios require significant further
outward transfer of angular momentum.  The probability distribution of
$\lambda$ is found to be approximately log-normal, but its extreme low
spin-tail, where $\lambda$ is orders of magnitude below the mean, has
not been determined from the limited number of halos followed in the
above simulations. Nevertheless, it would be unrealistic to appeal to
a near-perfect cancelation of large-scale structure torques, and then
to a conservation of the nearly vanishing specific angular momentum.
Further angular momentum transfer is therefore needed, and is likely
purely gravitational in origin initially (on the largest scales),
similar to the picture discussed in the previous subsection.  Gas
adiabatically condensed in the central regions of pristine ACHs, whose
temperature is close to the virial temperature, will remain locally
Toomre stable unless they spin exceptionally
slowly~\citep{OhHaiman2002}.  However, as long as gas in these early
halos can cool and form self-gravitating disks, it can become unstable
to global non-axisymmetric modes. These can lead to a redistribution
of angular momentum, and allow gas inflow to the central region that
can ultimately produce a
BH~\citep{Koushiappas+2004,LodatoNatarajan2006}, with a range of
different BH masses between different
halos~\citep{LodatoNatarajan2007}.

More specifically in this context, \citet{Begelman+2006} proposed that
a multi-stage cascade of gaseous bars may form and transport angular
momentum outward (with gas collapsing down to smaller scales and
eventually forming a `quasistar'; see \S~\ref{subsec:journey-to-BH} below).  
This is a follow-up on the
`bars-in-bars' scenario discussed in the previous section. The
original proposal consisted of two distinct stages: first a
collisionless (stellar) bar driving the gas inward, and then a single
gaseous bar forming in a self-gravitating
disk~\citep{Shlosman+1990}. In principle, however, a cascade of
several nested bars on increasingly smaller scales could arise, as
long as star-formation is avoided.  Adaptive mesh refinement (AMR)
simulations following the central collapse of pristine ACHs 
have indeed found such a cascade. \citet{Wise+2008} identified
four nested stages of barlike instabilities, each separated by a
factor of $\sim 100$ on successive scales ($10^{18}$, $10^{16}$,
$10^{14}$, $10^{11}$ cm), efficiently driving gas down to the inner
region as small as $10^9$ cm.  \citet{Choi+2013,Choi+2015}
have found similar results, confirming the importance of nested
gaseous bars.

These simulations also identified supersonic turbulence, which is
inevitably produced during the process of
virialization~\citep{WiseAbel2007a}, and highlighted its dynamical
importance.  Note that turbulence both suppresses and stimulates
fragmentation.  Since turbulence acts as a source of pressure, which
counteracts gravity, it tends to stabilize gas against fragmentation,
at least on large scales (larger than the size of the turbulent
eddies).  On the other hand, when supersonic turbulent eddies collide,
they produce shocks and compress the gas, which promotes fragmentation
on small scales (an effect absent in the case of thermal pressure).
The overall sign of the impact of turbulence depends on whether
fragments produced in the latter process can cool and collapse on a
timescale shorter than the eddy turnover time (i.e before they are
disrupted by another collision).  In the high-$z$ protogalaxies, where
cooling is inefficient, the net outcome appears to be that turbulence
helps stabilize the gas against fragmentation and
star-formation~\citep{Choi+2015}.  We note, however, that
fragmentation may not have been numerically resolved. Indeed, applying
the same argument as in the case of the local ISM, at the Mach numbers
$M\approx 3$ typical of turbulent inflows in high-$z$ galaxies, one
would expect $\sim1\%$ of the mass to reside in small self-gravitating
fragments, which may be difficult to resolve.  \citet{Regan+2014} find
that their highest-resolution simulations (with 26 levels of
refinement with {\it enzo}, reaching a resolution of $\sim1$ AU),
point to fragmentation on scales of order $\sim100$ AU.

We also note that these and similar studies
\citep{ReganHaehnelt2009,Johnson+2011,Latif+2013} studied ACHs without
H$_2$ chemistry or assuming sufficiently strong LW irradiation so that
${\rm H_2}$ has a negligible effect. While this could be justified in
a small subset of ACHs exposed to intense LW radiation
(e.g. \citealt{Shang+2010,Regan+2017}) and/or extreme dynamical
heating (e.g.~\citealt{Yoshida+2003,Wise+2019}), the large majority of
early halos will have prior episodes of ${\rm H_2}$--cooling induced
fragmentation and star-formation in the minihalo stage (see discussion
in the next section).  Rapid inflow to a central supermassive object
is likely inhibited in this case; simulations find a near-Keplerian
compact disk fragmenting into dozens of stars growing at sub-critical
rates (e.g. \citealt{Greif+2012}; see next section).

We end this section by noting that stars are not necessarily only a
hindrance for getting gas down to the central BH.  In a scenario in
which a dense cluster of star surrounds a central seed BH, the stars
can help with the angular momentum
problem. \citet{AlexanderNatarajan2014} considered the usual Bondi
accretion problem, but with angular momentum, and with the inclusion
of the acceleration of the BH that would be expected in the presence
of a dense star cluster. The gravitational force of the stars results
in a `jitter' in the location of the BH, which, as a result, will see,
in its own frame of reference, the angular momentum of some of the
infalling gas canceled to zero (or sufficiently small for the gas to
fall radially inside the Schwarzschild radius).  Provided that the gas
density is high, so that the Bondi accretion rate is well in excess of
the Eddington rate, \citet{AlexanderNatarajan2014} showed that this
mechanism can solve the angular momentum problem and permit extended
periods of super-Eddington accretion, producing $\approx 10^4~\msun$
BHs (limited by the need for the star cluster to outweigh the BH).

\section{THE (INITIAL) MASS FUNCTION OF EARLY BLACK HOLES}
\label{sec:BHIMF}

In the previous two sections, we reviewed the formation of
stellar-mass seed BHs (e.g., Pop~III remnants) and general issues
regarding their growth via gas accretion, radiative/mechanical feedback, 
and angular momentum transport of inflowing gas.  In this section, we
discuss the prompt formation of ``{\em massive seed BHs}'', which, for
concreteness, we define as {\em any BH heavier than the typical
Pop~III stellar mass of $\sim 100~\msun$}.
Motivated largely by the discovery of $\sim 10^9~\msun$ SMBHs at $z\sim 6$,
numerous pathways have been proposed to form such massive seeds, (see
\textbf{Figure~\ref{fig:rees}}).  We emphasize that these SMBHs are
unusually bright, massive, and very rare objects, having a comoving
number density of $\sim 1~\gpc ^{-3}$ \citep{Willott+2010}, but the
physics of these pathways is of interest, even if many of these seeds
do not actually grow to the masses required to power the high-$z$
quasars.

In this review, we therefore discuss the nature of massive seeds in
each scenario, and the corresponding ``initial mass function (IMF)''
of massive BHs over the range $100\lsim M_\bullet/\msun \lsim 10^6$ at
high redshifts, rather than judge which models can successfully form
high-$z$ SMBHs.  As emphasized in \S\ref{sec:iceberg}, investigation
of the underlying BH population that do not grow to extreme SMBHs is
also crucially important to constrain their BH seeding and growth models
with ongoing and future observations (see \S~\ref{sec:obs}) and to
better understand the transition between low-$z$ and high-$z$ quasar
populations.

In the following subsections, we first review the basic requirements
of massive seed formation (\S~\ref{subsec:massiveBHrequirements}),
and give a motivation to focus on gravitational collapse of chemically pristine, warm 
gas in ACHs with virial temperatures of $T_{\rm vir}\gsim 8000~\K$ 
(see the lower branch of \textbf{Figure~\ref{fig:rees}}).
Then, we discuss the possibilities of keeping gas in ACHs warm via suppressed 
H$_2$ cooling or enhanced heating (\S~\ref{subsec:warmgas}), the resulting 
emergence of massive seed BHs with $M_\bullet \sim 10^{5-6}~\msun$ 
via supermassive stars (\S~\ref{subsec:journey-to-BH}), or with
$M_\bullet \sim 10^{3-4}~\msun$ via runaway mergers in a dense star
cluster~(\S~\ref{sec:densecluster}), and the subsequent evolution
of the population of these BHs in the context of hierarchical galaxy
evolution~(\S~\ref{subsec:BHevolution}).

\subsection{Prompt Formation of Massive Black Holes}
\label{subsec:massiveBHrequirements}

Before going into details, we first describe two general requirements,
which are shared by all proposed pathways for massive seed BH formation.

First, monolithic collapse of a massive gas cloud is required to form
a single massive object, avoiding major episodes of gas fragmentation
before the gas reaches very high density.  The efficiency of
fragmentation depends crucially on the equation of state of the
collapsing gas, characterized by the effective heat index $\gamma_{\rm
  eff} \equiv d\ln p/ d\ln \rho$ \citep[][and references
  therein]{KlessenGlover2016}.  In rapid cooling phases, where
$\gamma_{\rm eff}<1$, pressure-free and pancake-like collapse of the
overdense regions tends to develop a highly flattened sheet-like
configuration or filamentary structure.  When efficient cooling
terminates ($\gamma_{\rm eff}\approx 1$), those filaments re-fragment
into dense cores, each of which tends to collapse in a quasi-spherical
way and not experience further hierarchical fragmentation
\citep{Larson1985,InutsukaMiyama1997}.

\begin{figure}[t]
  \includegraphics[clip,width=4.0in]{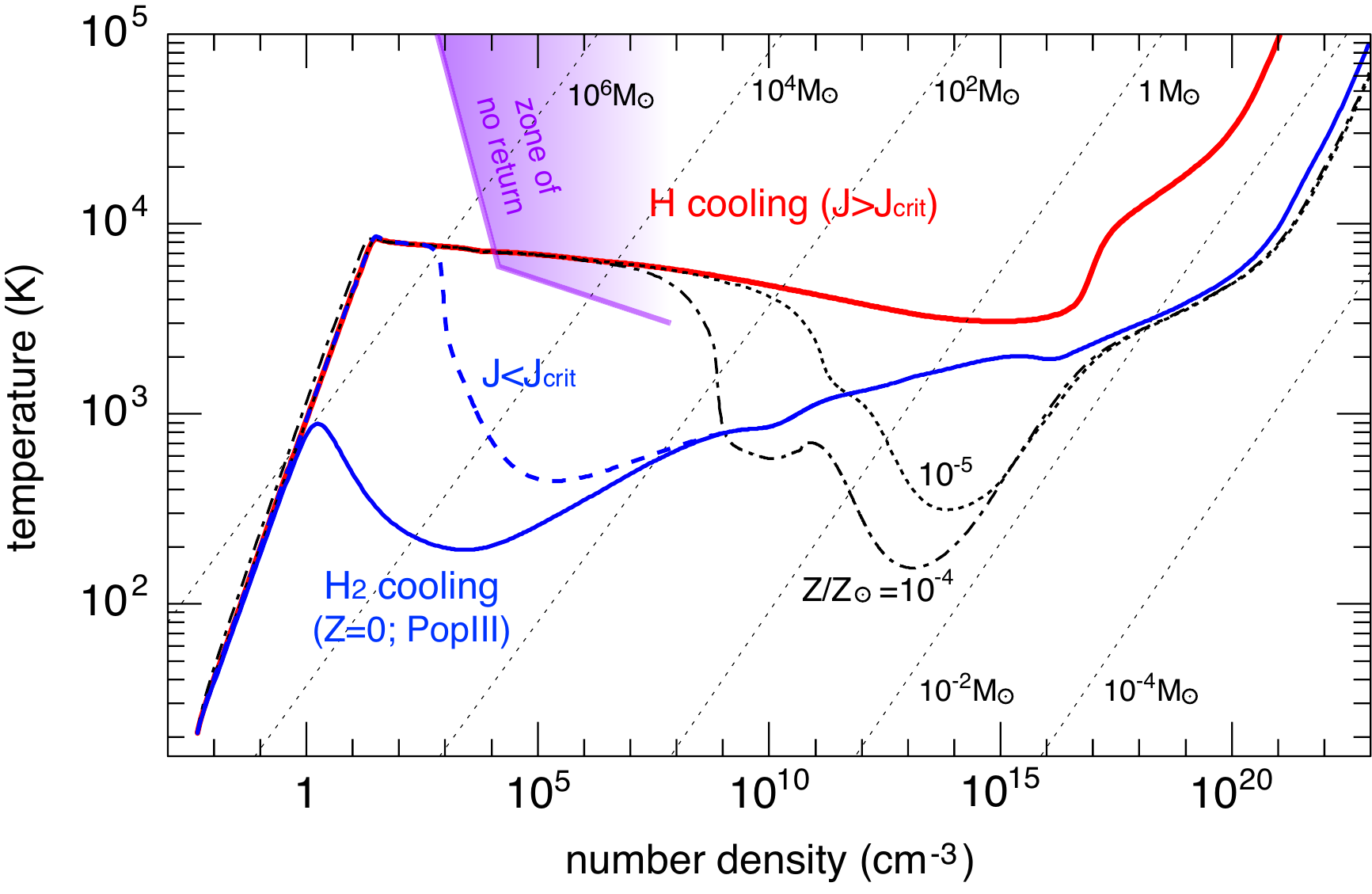}
  \vspace{-1\baselineskip}
\caption{Temperature evolution of a metal-free gas cloud, irradiated
  by LW radiation with three different intensities (blue and red
  curves), based on a spherical collapse one-zone model assuming a
  free-fall density evolution.  With a weak LW intensity ($J_{\rm
    LW}<J_{\rm crit}$), which dissociates H$_2$ only at lower
  densities, gravitational collapse of the cloud is led by H$_2$
  cooling toward higher densities.  With a sufficiently high LW
  intensity ($J_{\rm LW}>J_{\rm crit}$), which keeps H$_2$ dissociated
  until the gas enters into a dense and hot region (`zone of no
  return'; purple line), the cloud collapses nearly isothermally at
  $T\approx 5000~\K$ without rapid cooling phases (red curve).  Black
  curves show the evolution of metal- and dust-polluted gas with
  $Z/\zsun=10^{-4}$ and $10^{-5}$ leading to a rapid temperature drop
  due to dust thermal emission.  In such a rapid cooling phase, the
  gas is likely to fragment into small clumps, whose masses are
  approximately the value of the Jeans mass (dotted diagonal lines) at
  the temperature minimum.  
  This figure is based on data from \cite{Omukai+2008}.}
\label{fig:nt}
\end{figure}

\textbf{Figure~\ref{fig:nt}} illustrates the $n-T$ phase diagram of a
gas cloud collapsing under its self-gravity, obtained in one-zone
models under several different conditions.  Initially (at the lowest
densities) cooling is inefficient, and the gas is heated by
compression.  For metal-free gas with weak or no LW irradiation (blue
curves) and for slightly metal- and dust-polluted gas (black curves),
the collapsing gas eventually experiences a rapid temperature drop,
caused by cooling via H$_2$-line or thermal dust continuum emission
(see \S\ref{subsec:warmgas}), leading to vigorous fragmentation 
\citep[e.g.,][]{Clark+2008}.
The fragment mass is approximated by the Jeans mass at the temperature
loitering point (where cooling becomes less efficient and $\gamma_{\rm
  eff}\approx 1$). As indicated in the figure, $M_{\rm J}\sim
10^3~\msun$ for the H$_2$ cooling track and $\sim 0.1~\msun$ for the
dust-cooling track at this point.  Without rapid cooling phases
(i.e. due to intense LW irradiation; red curve), the thermal evolution
is quite different, with no clear single temperature minimum or
loitering point.  In fact, the collapse is nearly isothermal
($0.9<\gamma_{\rm eff}<1.1$) for over $\sim 16$ orders of magnitude in
density (from $\sim 10$ to $\sim 10^{17}{\rm cm^{-3}}$). Even in this
regime, however, the collapsing central core is unstable against
non-spherical perturbations.  The collapsing central region elongates
slowly with increasing central density, but the amplitude of this
distortion may not be large enough to produce fragmentation during the
extended isothermal
phase~\citep{Lai2000,HanawaMatsumoto2000,Sugimura+2017b}.

Second, rapid gas accumulation is required to avoid the formation of a
normal massive Pop~III star, and to instead form a supermassive star
with mass up to $M_\star \sim 10^6~\msun$.  The critical accretion
rate is $\dot{M} \approx 0.01-0.1~\msunyr$ (see
\S~\ref{subsec:journey-to-BH}), which needs to be sustained at the
center of the massive gas cloud collapsing under its self-gravity.
The gas accretion rate onto a new-born central protostar is
approximately given by $\approx A c_{\rm s}^3/G$
(assuming that the gas was initially in quasi-hydrostatic equilibrium;
\citealt{Larson1969,Penston1969}; see also equation \ref{eq:mdot_sg}),
with the numerical factor ranging between $A\approx 1-47$ depending on
the boundary conditions.  3D simulations typically find $A\approx 20$
\citep[e.g.,][]{Inayoshi+2014}.  The implication is that avoiding
normal Pop~III star formation requires the gas temperature to remain
as high as $\sim 10^4~\K$ during the collapse phase.  This requirement
could be satisfied if atomic cooling processes (e.g., Ly$\alpha$
emission) induce gravitational collapse of gas with a suppressed
H$_2$-abundance (red curve in \textbf{Figure~\ref{fig:nt}}).

As discussed below, these two conditions are fully satisfied only in
rare special environments, such as ACHs exposed to
strong LW irradiation by close neighbors, heating through rapid halo
mergers, or located in regions with an unusually high baryon streaming
velocity.  This rarity is qualitatively consistent with the fact that
SMBHs observed at high redshifts are hosted in rare, very massive
galaxies.  On the other hand, even if the requirements are not
achieved perfectly, relatively lower-mass but still massive seeds
would form in the IMBH range ($10^2-10^4~\msun$).  These IMBH seeds
could be abundant and have a significant contribution to the overall
mass density of the high-$z$ BH population.  In addition, the host
halos of the high-$z$ quasars formed in highly unrepresentative
regions of the universe, requiring $\sim5\sigma$ fluctuations on
$10^{12-13}\msun$ scales. These special environments may conspire to
meet the requirements of massive seed formation {\em and} also enable
their subsequent growth to the SMBHs regime.  Indeed, we expect that
only a small minority of massive seeds are born in regions which
evolve to massive $\sim 10^{12}~\msun$ galaxies by $z\approx 6$ -- the
others follow minor branches of the galaxy merger history and remain
lower-mass BHs in lower-mass galaxies or in satellite galaxies
\citep{Valiante+2016}.

In summary, {\it massive seed BHs are expected to result from the
  near-isothermal collapse of a gravitationally unstable massive gas
  cloud at a temperature of $T\approx 10^4~\K$, which produces a high
  mass accretion rate onto the central object and avoids major
  episodes of gas fragmentation.} This likely occurs only in rare
special environments, as we next discuss.    

It is, however, worth keeping in mind a caveat to these requirements. 
In sufficiently massive galaxies, rapid gas inflow could efficiently fuel a central BH,
and result in rapid growth, even if the above criteria are not met.
Indeed, this must occur for bright quasars at low redshift, which
are fueled at rates of $\gsim 10~\msunyr$, despite the gas in the nuclei
of their hosts being cold and highly metal-enriched. This contradicts 
the naive expectation that gas should fragment efficiently in the
outer regions of their accretion disks, before reaching the innermost
regions. High-$z$ galaxies may behave similarly, 
and assemble massive BHs rapidly, despite efficient cooling from metals and/or H$_2$,
once they reach a critical mass.  We expect this critical mass is well
above the ACH limit, but this is currently not understood.

\subsection{Keeping the Gas Warm: Suppressing H$_2$ Cooling and Enhancing Heating}
\label{subsec:warmgas}

It has long been recognized that the key physics governing the
formation of the first stars (and BHs) is the abundance of H$_2$
molecules. This is because collisional excitation of H$_2$ is
the only way for primordial gas to cool radiatively and collapse to
high density~\citep{SaslawZipoy1967}.  Early works have constructed
complete gas--phase reaction networks, and identified the two possible
ways of forming H$_2$ in primordial gas: via the intermediaries
${\rm H_2^+}$ or ${\rm H^-}$, of which the latter is relevant in the
collapse of high--redshift objects \citep{PeeblesDicke1968,Hirasawa1969,Matsuda+1969}.
As discussed in \S~\ref{sec:lightseed} above, the masses of the Pop~III stars, 
formed via ${\rm H_2}$-cooling, is tied to the
accretion rate of the protostellar core, which is determined primarily
by the gas temperature -- the cooler the gas, the lower the accretion
rate and the stellar mass.  Therefore, a key requirement to avoid
forming a ``normal'' massive Pop~III star, and to instead form a
much more massive SMS is the suppression of H$_2$ formation and
cooling.  Provided that H$_2$ is sufficiently suppressed, and the host
halo is sufficiently massive (near the atomic-cooling threshold)
to raise the gas temperature to several $1000$ K, the SMS can reach
masses of up to $\sim 10^{5-6}~\msun$ (and eventually collapse into a
massive seed BH with the same mass; see \S~\ref{subsec:journey-to-BH} below).

Various H$_2$-dissociating processes, focusing especially on ACHs, 
have been investigated by several authors
\citep{Omukai2001, OhHaiman2002, BrommLoeb2003, Shang+2010,
  Schleicher+2010b, Wolcott-Green+2011, InayoshiOmukai2012,
  Agarwal+2012, Regan+2014b, Sugimura+2014, Becerra+2015, Latif+2016,
  Wolcott-Green+2017, Regan+2018}.  In particular, three distinct
mechanisms could keep the H$_2$ fraction at low levels:
photodissociation of H$_2$ by LW radiation, photodetachment of H$^-$,
and collisional H$_2$ dissociation in dense and hot gas.  We next
discuss these mechanisms, along with other possible ways to keep the
primordial gas warm.

\subsubsection{Dissociating H$_2$ by Lyman-Werner radiation}
\label{sec:H2LWdis}

H$_2$ can be photodissociated by irradiation by soft UV photons, in a
two-step process.  Photons in the $\approx 11-15$ eV range are
resonantly absorbed in the Lyman and Werner lines of H$_2$. These are
transitions between the ground and excited electronic states of H$_2$,
analogous to the Lyman series of hydrogen atoms (but split into many
rotational and vibrational levels; see below).  Roughly $\sim 10\%$ of the
excited ${\rm H_2^{(*)}}$ decays radiatively into the split state of
two H atoms (rather than cascading back to the electronic ground state
of H$_2$), resulting in H$_2$ dissociation.  The
significance of this two-step process was first highlighted in the
context of the local ISM by Solomon in 1965 (see \citealt{Field+1966})
and subsequently studied by \citet{StecherWilliams1967}.

The earliest generation of stars and remnant BHs emitted LW
radiation at redshifts well before cosmic reionization.  Because of
the long mean free path of photons with energies $h\nu < 13.6$ eV to
the intergalactic medium, a LW background was built up in the early
universe, capable of suppressing H$_2$ formation in low-mass DM halos
\citep{Haiman+1997,Ciardi+2000}.  H$_2$ formation in primordial gas
occurs mainly through the H$^-$ channel:
\begin{equation}
{\rm H} + {\rm e}^- \rightarrow {\rm H}^- + \gamma,
\end{equation}
\begin{equation}
{\rm H}^- + {\rm H} \rightarrow {\rm H}_2 + {\rm e}^-,
\end{equation}
The critical value $J_{\rm crit}$ of the LW intensity for suppressing
the H$_2$ abundance follows from balancing the dissociation rate
($\propto J_{\rm LW}\cdot n_{\rm H}$) with the formation rate ($\propto n_{\rm H}^2$),
and therefore depends on the density $n_{\rm H}$ (linearly, in the
optically thin limit).  In minihalos, in the absence of H$_2$ cooling
the gas contracts adiabatically\footnote{Compton cooling on the cosmic
  microwave background is important at high redshift $z>10$ but only
  during the early stages of collapse, since the Compton-cooling time
  is independent of density.}, and the relevant density in this case
is the maximum value set by the entropy floor of the primordial gas
($n_{\rm H}\sim 0.1-10~{\rm cm^{-3}}$; e.g.,
\citealt{Visbal+2014_nogo}).  Depending on halo mass and redshift, the
resulting critical intensity is $J_{\rm crit} \approx
(0.01-1)~J_{21}$, where $J_{21}\equiv 10^{-21}~{\rm erg}~{\rm
  s}^{-1}~{\rm cm}^{-2}~{\rm Hz}^{-1}~{\rm
  sr}^{-1}$~\citep{Haiman+2000,Machacek+2001,WiseAbel2007b,
  OSheaNorman2008,LatifKhochfar2019}.
This is a relatively low value: for reference, the minimum flux
required to reionize the universe, i.e., one ionizing photon per
hydrogen atom (neglecting recombinations and the opacity of the IGM),
corresponds to a mean background UV flux of $\langle J\rangle\approx 3
[(1+z)/11]^3~J_{21}$.  As a result, a large fraction of the earliest
minihalos may be `sterilized' of H$_2$, and not form any
stars~\citep{OmukaiNishi1999,Haiman+2000,Ciardi+2000,Ricotti+2001,Ricotti+2002,Mesinger+2006}.
An important caveat here is that $J_{\rm crit}$ has not been reliably
computed in the most massive minihalos, just below the ACH.  
There are, however, hints that in these ``sub-atomic'' halos,
$J_{\rm crit}$ rises significantly~(to $\gsim 100~J_{21}$;
\citealt{Regan+2017}), especially since H$_2$ self-shielding can
become important (see below).  This caveat is important, since massive
BH seed formation requires that efficient star formation is avoided
all the way to the ACH stage.

The situation is dramatically different for ACHs.  As
the halo increases its mass and virial temperature to $T_{\rm vir}
\approx 8000~\K$, Ly$\alpha$ cooling kicks in and gas is able to cool
and collapse by collisional excitation of atomic hydrogen, even in the absence of
H$_2$ \citep{Omukai2001,OhHaiman2002}.  As the density increases,
$J_{\rm crit}$ rises.  Additionally, in these halos the column density
of H$_2$ reaches $\sim 10^{14}~{\rm cm^{-2}}$, 
at which point the LW lines start to become optically thick~
\citep{DraineBertoldi1996}:
H$_2$ is therefore self-shielded, and the LW flux in the core of
the halo is attenuated.  The relevant density here is the critical
density, of H$_2$ for local thermal equilibrium (LTE), because above
this density (i) the rovibrational states of H$_2$ are kept in
equilibrium via collisions and radiative cooling becomes ineffective,
and (ii) collisional dissociation from the excited rovibrational
levels of H$_2$ reduces the H$_2$
fraction~\citep{Omukai2001,Shang+2010,InayoshiOmukai2011}.  The
critical density for the most important transitions is $n_{\rm H}\approx
10^4~{\rm cm^{-3}}$~\citep{Wolcott-GreenHaiman2019}, which is several
orders of magnitude higher than the density in minihalos.  
The value of $J_{\rm crit}$ in the ACHs is therefore correspondingly several orders of
magnitude higher.

The thermal evolution of the gas sharply bifurcates, depending on
whether the LW intensity is below or above the threshold, as
illustrated in \textbf{Figure~\ref{fig:nt}}.  When $J<J_{\rm crit}$
(dashed blue curve), a rapid temperature drop is caused by radiative
cooling of self-shielded H$_2$ after a brief isothermal phase.  In
this case, the temperature track converges toward the one without LW
radiation (solid blue).  As a result of the rapid cooling phase
($\gamma_{\rm eff}<1$), this gas is expected to fragment into small
clumps with $M_{\rm J}\sim 10^3~\msun$~\citep{Regan+2018,
  Kulkarni+2019}.  When $J>J_{\rm crit}$ (solid red curve), the
temperature evolution is qualitatively altered.  The nearly isothermal
collapse at $T\approx 8000~\K$ continues until very high density ($\sim
10^{16}~\cc$) without being affected by H$_2$ cooling.  In this case,
fragmentation may be absent, or is at least strongly
suppressed~\citep{Regan+2018}.

There is now a very large literature on the value of $J_{\rm crit}$,
which depends sensitively on the detailed calculation of the
optically--thick H$_2$--photodissociation rate.  In general, this rate
must be computed by summing over the rate of resonant absorption into
thousands of LW lines, multiplied, in each line, by the probability of
eventual decay into the split atomic state.  Even in one-zone models,
this calculation is challenging: there are 301 rovibrational states of
the ground electronic state, and over half a million allowed
electronic transitions in total.  In each LW line, the rate
calculation must take into account both the shape of the incident flux
(including absorption lines in realistic galaxy spectra overlapping
with H$_2$ lines), self-shielding by the H$_2$ line itself (depending
on the H$_2$ column density) as well as possible shielding by the
damping wings of neighboring atomic Lyman lines (depending on the
${\rm H}$ column density).  Finally, the H$_2$ ro-vibrational levels
in the electronic ground state are not all in general in LTE, which
significantly affects the effective shielding, the resulting
dissociation rate, as well as the radiative cooling.  In the
three-dimensional case, additional complications arise from bulk
motions (which Doppler shift line frequencies), temperature
variations, and the basic fact that self-shielding is not a local
quantity, but rather depends on the direction-dependent column density
across the protogalaxy.

Overall, $J_{\rm crit}$ in ACHs has been found to range from $J_{\rm
  crit}\sim (10^3-10^5) J_{21}$. In one-zone models, the most complete
calculations, which adopt the most up-to-date chemical network, and
take into account realistic incident spectral shapes from
low-metallicity galaxies, give $J_{\rm crit} \approx
(1000-1400)~J_{21}$~(e.g.~\citealt{Sugimura+2014} and references therein).
When combined with calculations of self-shielding~\citep{Wolcott-Green+2017} 
and non-LTE effects~\citep{Wolcott-GreenHaiman2019}, the lower end of this range
is generally favored (because self-shielding is somewhat weaker when
the H$_2$ is spread over many different ro-vibrational states).
\citet{Latif+2015} find the exceptionally low value of 400~$J_{21}$ (with a 
somewhat different chemistry network).
The exact value of $J_{\rm crit}$ indeed depends on the chemical 
reaction networks and/or reaction rate coefficients adopted in the literature.
Uncertainties in the reaction rates translate to a factor of $\approx$ two
uncertainty in $J_{\rm crit}$~\citep{Glover2015a,Glover2015b}.

Three-dimensional simulations of ACHs find that the collapse dynamics
affects the thermal evolution of the gas, and can impact $J_{\rm
  crit}$ significantly.  \cite{Shang+2010} have noted that turbulent
shocks occur at various densities, and cause a $\sim 10-20\%$ scatter
in the temperature.  If the temperature is lower (high temperature
fluctuations are radiated away by Ly$\alpha$ cooling quickly), the
collisional dissociation rate is also lower, which requires a higher
$J_{\rm crit}$ to compensate. Overall, they found $J_{\rm
  crit}=(10^4-10^5)~J_{21}$. However, these values are highly sensitive
to the treatment of the self-shielding.  

Simulations typically adopt a
parameter, $f_{\rm shield}$, which depends only on the gas temperature, 
as well as a local estimate of an effective ${\rm H_2}$ column density.
to take into account self-shielding in the LW lines.  
\citet{Wolcott-Green+2011} have shown that $J_{\rm crit}$
is reduced by an order of magnitude when a more accurate shielding
factor (including excited ro-vibrational states), as well as a more
accurate local column density estimate is used.  Adopting the
shielding factors from \citet{Wolcott-Green+2011}, \citet{Latif+2015}
have found $J_{\rm crit}\approx (2-5)\times 10^4~J_{21}$.
\cite{Hartwig+2015} have implemented a new method to capture the gas
geometry and velocity field enabling a proper determination of the
direction-dependent H$_2$ self-shielding factor, reducing the critical
flux by a factor of two.  Overall, the values of $J_{\rm crit}$ in 3D
simulations tend to be a factor of few higher than in one-zone models
when they use the same input spectra and shielding treatment.
Importantly, 3D simulations can also include the fact that the LW flux 
seen from a neighboring galaxy is highly anisotropic~\citep{Regan+2016}, 
and have also revealed a strong halo-to-halo variation in $J_{\rm crit}$, 
by at least a factor of several~\citep{Shang+2010,Latif+2014a}.

Stellar populations with a significant binary fraction alter the
radiation spectra of source galaxies and increase the critical
intensity \citep{Agarwal+2017}.  A unique effect of a significant
binary population is X-ray irradiation associated with the LW emitting
galaxies.  Effective ionization by soft X-rays ($\approx 1$ keV)
enhances the electron fraction and thus activates H$_2$ formation
through the electron-catalyzed reactions \citep{Haiman+1996b}.  As a
result, the critical flux is boosted by one order of magnitude
\citep{InayoshiOmukai2011,InayoshiTanaka2015,Latif+2015}.

\begin{figure}[t]
  \includegraphics[clip,width=3.3in]{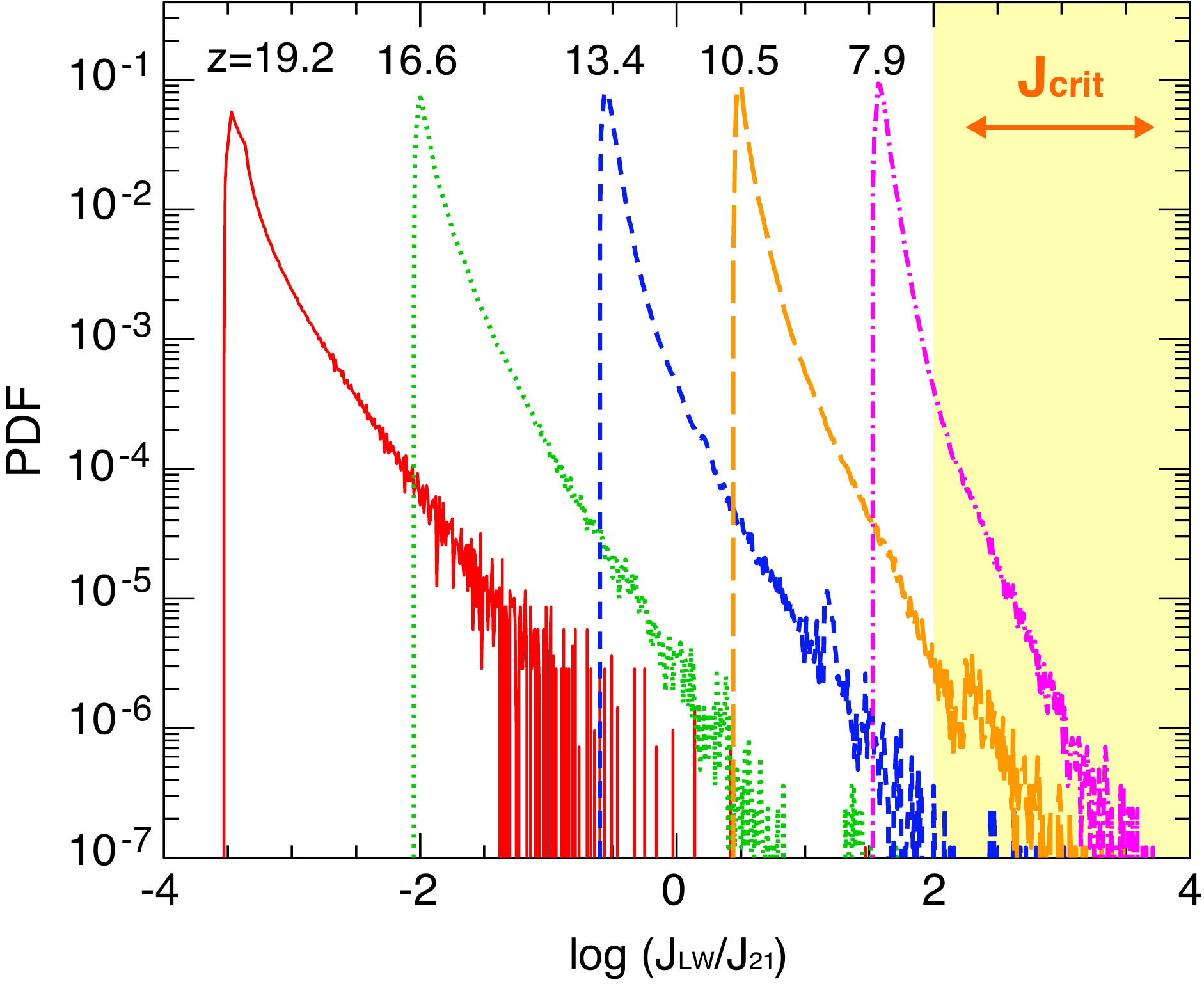}
  \vspace{-1\baselineskip}
\caption{Distribution function (PDF) of the LW radiation background
  intensity ($J_{\rm LW}$, in units of $J_{21}$) at different
  redshifts between $7.9<z<19.2$, obtained in a cosmological
  simulation~\citep{Ahn+2009}.  The shaded region on the right marks
  the expected range of the critical LW intensity $J_{\rm crit}$ for
  H$_2$ suppression in ACHs.  Since the critical LW flux $J_{\rm
    crit}$ lies in the far tail of the steep PDF, an exceedingly
  small, but non-zero fraction ($\lsim 10^{-5}-10^{-6}$) of ACHs at
  $z\gsim 10$ are exposed to sufficiently intense LW radiation and
  potentially form a massive seed BH.  Radiation from subhalos and
  low-mass minihalos, which were not resolved in the simulation and
  therefore not included in this figure, can significantly increase
  this fraction (see text).}
\label{fig:Jcrit-pdf}
\end{figure}

The correct knowledge of $J_{\rm crit}$ is crucial in estimating the
number of ACHs where massive seed BHs could form
\citep{Dijkstra+2008,Ahn+2009,Agarwal+2012,Johnson+2013,Dijkstra+2014}.
This is because $J_{\rm crit}$ is substantially higher than the
expected level of the LW background at high redshifts, well before
reionization
\citep{Haiman+1997,WiseAbel2007b,OSheaNorman2008}. Nevertheless, a
small fraction of ACHs, which are located in an extraordinary
overdense region and have bright galaxies nearby, could be irradiated
by a sufficiently high flux~\citep{Dijkstra+2008}.  The fraction of
ACHs that potentially form massive seeds directly reflects the
individual properties of a few bright, nearby source galaxies (e.g.,
stellar mass, star formation rate, LW luminosity), and their distances
from the ACH of interest.  The range of required distances is, in
fact, quite narrow~\citep{Visbal+2014_sync,Regan+2017}.  If the
neighbor(s) are too far, they must have unrealistically high star
formation efficiencies to produce a sufficiently high LW flux.  If
they are too close, then the gas in the ACH of interest tends to be either 
stripped by ram pressure or tidally disrupted by massive source 
galaxies \citep{Chon+2016} and/or photoevaporated by intense 
ionizing photons \citep{Johnson+2014,Regan+2016} and polluted 
by metal winds  produced from these neighboring sources \citep{Dijkstra+2014}.

Overall, the fraction of ACHs exposed to sufficiently strong LW
radiation ($J_{\rm LW}\geq J_{\rm crit}$) sharply decreases with
$J_{\rm crit}$, because the PDF of the background flux $J_{\rm LW}$ is
very steep \citep{Dijkstra+2008,Ahn+2009,Agarwal+2012,Johnson+2013,
  Dijkstra+2014,InayoshiTanaka2015, Chon+2016,Habouzit+2016}.
\textbf{Figure~\ref{fig:Jcrit-pdf}} shows an example of this PDF from
\citet{Ahn+2009}.  The probability of ACHs having $J_{\rm LW}>J_{\rm
  crit}$ at $z=10$ is as small as $\approx 10^{-6}-10^{-7}$.  Naively
multiplying this by the (comoving) number density of ACHs $n_{\rm
  ACH}\approx $ few $\mpc^{-3}$ at $z=10$, the expected number density
of massive seed BHs is $n_{\rm seed}\approx$ few ${\gpc^{-3}}~(J_{\rm
  crit}/10^3)^{-\beta}$, where $\beta \approx 5$
\citep{InayoshiTanaka2015}.  On the other hand, this neglects any
radiation from subhalos, which are unresolved in cosmological
simulations of the background.  Recent N-body~\citep{Visbal+2014_sync}
and hydrodynamical~\citep{Chon+2016} simulations which resolve
sub-halos, as well as lower-mass minihalos, finds a significantly
flatter flux PDF, and many orders of magnitude higher probability for
$J_{\rm LW}>J_{\rm crit}$, due to these extra radiation sources.  Many
of these halos may fail to collapse, because of tidal disruption
and/or ram-pressure stripping, as noted above. However,
\citet{Visbal+2014_sync} still find an abundance of massive seed
BH-forming halos as high as $\approx 10^{-4}~{\rm Mpc}^{-3}$.

\cite{Visbal+2014_sync} have further introduced the timing as an
important aspect, and considered only pairs of pristine ACHs which
form nearly synchronously (crossing the atomic-cooling threshold
within a few Myr of each other), and with a small spatial separation
(within a $\sim$kpc).  \cite{Regan+2017} have further investigated
this ``synchronized pairs'' scenario, using cosmological
hydrodynamical simulations, and found that a massive seed BH could
form {\it only if} (1) the separation of two halos is within $0.2-0.3$
kpc and (2) the time for the irradiated ACH to collapse and form a
seed BH is within $\sim 4$ Myr, to avoid the deleterious effects of
X-ray irradiation, photo-evaporation or metal pollution.  However,
since these conclusions depend on a number of uncertain parameters
(e.g., initial mass function, star formation efficiency, clumping
factor of the intergalactic medium, metal wind velocity), further
studies of this scenario are required to assess how frequently it
may ultimately produce massive BHs.

Finally, we note the alternative possibility that the first star(s)
within a halo provide the LW radiation that can suppress further
fragmentation in the same halo~\citep{Susa2007}.  Such internal ${\rm
  H_2}$ suppression likely would have to involve significant
fine-tuning, since both photoionization feedback and metal enrichment,
concurrently with the LW radiation, needs to be avoided, while a
strong LW flux must be present at the ACH stage. \citet{Dunn+2018}
recently investigated this scenario in hydrodynamical simulations (BHs
are treated similarly to \cite{Tremmel+2017}, but with LW radiation
included), and found massive BHs to be a common outcome, including
multiple massive BHs forming in the same halo (above the ACH limit),
although their resolution was insufficient to determine the history of
ACHs or to follow the formation of a massive seed BH.  In order to
determine whether massive BHs can form under these circumstances,
higher-resolution simulations are required.

\subsubsection{Other H$_2$ suppression mechanisms}
\label{sec:H2coldis}

We next mention two additional mechanisms which could suppress the H$_2$ abundance.

{\em $H^-$ photodetachment.} Low-energy ($\lsim 11$eV) near-IR and
optical photons can not efficiently dissociate H$_2$ via LW line absorption, 
due to the absence of strong LW lines, but
can indirectly suppress H$_2$ formation via H$^-$ photo-detachment
(${\rm H}^- + \gamma \rightarrow {\rm H} + {\rm e}^-$).  The photon
energy threshold for this detachment is $h\nu \geq 0.76$ eV.  Several
papers computed and quoted a critical flux, $J_{\rm crit}$, assuming
that the incident spectrum is represented by a black-body shape, with
a temperature of either $T_{\star}\approx 10^5~\K$ (where the peak
frequency is $h\nu_{\rm max}\approx 24$ eV), or a temperature of
$T_\star \sim 10^4~\K$ ($h\nu_{\rm max}\approx 2.4$ eV).  In the latter
case, $J_{\rm crit}$ is found to be as low as $30~J_{21}$, as a result
of efficient H$^-$ photo-detachment
(e.g. \citealt{Omukai2001,BrommLoeb2003,Shang+2010}).  
Unfortunately, this low value has caused significant confusion in the literature, 
and created the illusion that Pop~II galaxies with softer spectra can
suppress the H$_2$ abundance in early galaxies more easily. 
In reality, the opposite is true: suppressing the H$_2$ abundance with softer spectra is more difficult.

While technically correct, the low $J_{\rm crit}$ values quoted in the
steep Wien tail of soft black-body spectra are misleading. First,
since the H$_2$ suppression is caused by $\sim 2$eV photons,
what matters is this IR intensity, and not the LW flux.  A spectrum as
soft as $T_\star \sim 10^4~\K$ would be created by low-mass stars,
which {\em do not emit much UV radiation}.  
As shown by \citet{Wolcott-Green+2012}, in order to produce the low 
$J_{\rm LW}=30~J_{21}$ with low-mass stars requires 
a factor of few {\em more} mass in stars than to achieve the higher $J_{\rm LW}(\approx 10^3)$ 
with massive Pop~III stars.  
Second, in practice, realistic composite galaxy spectra are not as
soft as a $T_\star \sim 10^4~\K$ black-body, unless they are devoid of
$\gsim 1\msun$ stars.  Using one-zone models,
\citet{Wolcott-Green+2017} pointed out that independent of the
spectral shape, there is a critical curve in the ($k_{\rm LW},k_{\rm
  H-}$) plane, where $k_{\rm LW}$ and $k_{\rm H-}$ are the H$_2$
dissociation rates by LW and IR photons, which determines whether an
illuminated protogalaxy can cool efficiently via H$_2$.  Using
population synthesis models for Pop~III and Pop~II galaxy spectra, the
conclusion is, however, that unless the Pop~II IMF is even softer than
the low-$z$ Salpeter distribution, and consists predominantly of $\sim
1~\msun$ stars, the direct LW dissociation dominates and ${\rm H^-}$
detachment plays only a minor role (see also \citealt{Sugimura+2014}
and \citealt{AgarwalKhochfar2015}).

Even if ${\rm H^-}$-detachment by direct {\em external} illumination
is not important, the Ly$\alpha$ radiation generated inside the gas
cooling in an ACH will be highly trapped, building up a large {\em
  internal} Ly$\alpha$ photon density.  In toy models,
\citet{JohnsonDijkstra2017} found that these trapped Ly$\alpha$
photons can detach ${\rm H^-}$, and reduce the required $J_{\rm
  crit}$ for the external UV flux. Recent 3D simulations confirm this
conclusion and find reductions of $\sim20\%$ in $J_{\rm
  crit}$~\citep{Wolcott-Green+2019}.

\vspace{\baselineskip}

{\em Collisional $H_2$ dissociation.}  Alternatively, H$_2$
collisional dissociation (${\rm H} + {\rm H}_2 \rightarrow 3{\rm H}$)
potentially plays an important role in the formation of massive seed
BHs in dense ($n_{\rm H}\gsim 10^4~\cc$) and hot ($T\approx 8000~\K$)
shocked regions \citep{InayoshiOmukai2012}.  Such shocked regions may
result from colliding inflows at the centers of protogalaxies in their
assembly and/or via violent collisions of galaxies themselves
\citep{Mayer+2010,Inayoshi+2015,Mayer+2015}.  In the primordial case,
if the temperature and density of the metal-free post-shock gas is
high enough for H$_2$ rovibrational levels to reach LTE, the gas never
cools down below $\approx 3000~\K$ because of the lack of H$_2$
cooling due to collisional dissociation. This `zone of no return' is
marked in \textbf{Figure~\ref{fig:nt}}.  The shocked layer fragments
and forms a massive cloud with $\sim 10^5~\msun$, which collapses
near-isothermally via atomic cooling as in the LW-aided scenario.
\cite{Fernandez+2014} investigated this shock-aided scenario with 3D
cosmological simulations, and confirmed the basic idea of the `zone of
no return'.  However, they found that for ACHs with virial
temperatures near the atomic-cooling threshold ($\approx 8000~\K$),
cold gas flows accrete into the halo but experience shocks before
reaching the central region, at densities that are too low.  This, on
the other hand, may happen because the radiative cooling time-scale is
not short enough for hot gas heated by virial shocks to collapse well
inside the halo (\citealt{Visbal+2014_nogo} for sub-ACHs and
\cite{BirnboimDekel2003,DekelBirnboim2006} for more massive DM halos
with $\sim 10^{12}~\msun$ at lower redshifts).  In more massive
haloes, with $T_{\rm vir}> 10^4~\K$, where the shock-dissipated energy
is more quickly carried away by radiative cooling (the cooling rate is
a steep function of temperature, $\Lambda \propto T^\beta$ with $\beta
\sim 8$ at $8000~\K<T < 2\times 10^4~\K$, and the cooling timescale of
primordial gas has the minimum value at $T\simeq 2\times 10^4~\K$),
the colliding inflows may still produce the required high-density,
high-temperature shocked gas in the core.

\subsubsection{Baryonic streaming motions}
\label{sec:BSM}

An alternative way to avoid star-formation in early galaxies is
provided by the large relative velocities of baryons with respect to
DM, which develop in the wake of cosmological recombination at
$z\approx 1100$ \citep{TseliakhovichHirata2010}. These ``streaming
motions'' are coherent on $\sim$Mpc scales with typical
(root-mean-square) magnitudes of $\approx 30~\kms~[(1+z)/1100]$, and
can delay the collapse of gas into early DM
minihalos~\citep{Greif+2011,Stacy+2011,Fialkov+2012}.  The resulting
lack of star-formation can keep the halo gas pristine when it finally
collapses into more massive halos, near the atomic-cooling threshold
of $T_{\rm vir}\approx 8000~\K$ \citep{TanakaLi2014,Schauer+2017}.
Furthermore, the value of the streaming velocity has a Gaussian
distribution, and therefore in rare, high-velocity patches of the
universe, the onset of gas collapse is further delayed, until the DM
halos become as massive as $\approx 10^8~\msun$ ($T_{\rm vir} \approx
2\times 10^4~\K$ at $15<z<20$), which is a factor of $\sim 10-30$
above the atomic-cooling threshold~\citep{Hirano+2017}. In this regime, dynamical
effects due to frequent mergers of gaseous halos violently disturb the
gaseous cores of the interacting galaxies, and further prevent star
formation \citep{Hirano+2018}.  \cite{Inayoshi+2018} used Monte Carlo
merger trees to simulate the assembly history of DM halos with
streaming velocities at twice the r.m.s. value.  The fraction and
absolute number density of pristine halos with $T_{\rm vir} \approx
2\times 10^4~\K$ are estimated as $\sim 3\times 10^{-5}$ and $\approx
10^{-5}-10^{-4}~\mpc^{-3}$ at $15<z<20$.  In such massive halos, well
above the atomic-cooling threshold, efficient Ly$\alpha$ cooling could drive
cold, pristine streams penetrating deep inside the halo and directly
feeding a dense central galactic disk, where massive seed BHs might
form from the dense and warm shocked gas, surrounded by a massive disk
of Pop~III stars.

\subsubsection{Rapid galaxy assembly}
\label{sec:DH}

Finally, a yet different way to avoid star-formation in early galaxies
is through unusually rapid galaxy assembly (note that the existence of
streaming motion is not required here, though high streaming
velocities would likely trigger violent mergers of gaseous halos as
discussed in \S\ref{sec:BSM}).  If mergers are sufficiently frequent,
they may continuously interrupt H$_2$ cooling and heat the gas back up
to the virial temperature~\citep{Fernandez+2014}, and/or counteract
any cooling via enhanced compressional
heating~\citep{Yoshida+2003,Chon+2016,Wise+2019}.  In particular,
\cite{Chon+2016} found that frequent, relatively minor mergers of ACHs
generally decrease the gas density at the core via dynamical heating,
and prevent its gravitational collapse.  Collapse of the cloud is
induced only in rare cases of frequent major-merger events where the
core mass is boosted by a factor of $\gsim 10$ within one dynamical
time-scale.  Recent numerical simulations by \cite{Hirano+2017} and
\cite{Wise+2019} have indeed suggested that the combination of
dynamical heating and a weak LW radiation background with $J_{\rm
  LW}\ll J_{\rm crit}$ increases the gas temperature on $\sim 10~\pc$
scales, enhancing the mass inflow rate toward the center.
\citet{Mayer+2015} also proposed that short-cutting their prior
history, shock heating during Milky-Way size galaxy mergers at
$z\approx 6$ plays a major role in balancing cooling, keeping the
temperature of the core at $T\approx 8000~\K$, and producing
supersonic infall at rates as high as $>10^4~\msunyr$, even if the gas
is heavily metal-polluted.  However such rapid mass accretion likely
leads to the formation of a gravitationally unstable nuclear disk, and
accretion then might be quenched by efficient
fragmentation~\citep{Ferrara+2013}.

\subsection{Massive Black Holes via a Supermassive Star}
\label{subsec:journey-to-BH}

Assuming that H$_2$ suppression, extra heating, streaming motions, or
some combination of these kept the gas warm in an ACH, the next
question is whether this gas fragments into clumps, or collapses
monolithically into one single object. Here we discuss the formation
of the protostar, its growth into an SMS, and its eventual collapse
into a massive BH.

\subsubsection{Pre-stellar collapse: the birth of a protostar}
\label{subssubsec:prestellar}

Three-dimensional hydrodynamical simulations have found that an
H$_2$-free cloud exposed to intense LW radiation could avoid
fragmentation and continue to collapse monotonically
\citep{BrommLoeb2003,Wise+2008,ReganHaehnelt2009,Shang+2010,Choi+2013,Latif+2013}.
However, the issue is not settled, because most of these cosmological
simulations have utilized simplifying assumptions in studying the
fragmentation process (such as turning off H$_2$ cooling by hand, or
adopting an optically thin treatment of Ly$\alpha$ cooling) and had
limited spatial resolution~(AU-resolution runs by \citealt{Regan+2014}
point to possible fragmentation on scales of $\sim100$ AU).
\cite{Inayoshi+2014} have studied gravitational collapse of a warm
primordial gas cloud with $T\sim 8000~\K$ up to densities high enough
for the gas to become optically thick and form a protostar ($n_{\rm H}
\gsim 10^{16}~\cc$ and $r \lsim 0.1$ AU), using a 3D
simulation which includes all the relevant cooling processes of both
H$_2$ and H, but adopting idealized, non-cosmological initial conditions with a weakly
turbulent field.
\cite{VanBorm+2014} have performed similar
simulations with somewhat different initial conditions and simplified
treatments of the chemical reaction network and radiative cooling.

There are several crucial findings: {\it the central core collapses
  almost isothermally} ($T\approx 5000-8000~K$) {\it until $\sim
  10^{16}~\cc$}, {\it forms one single object without major episodes
  of fragmentation, and accretes onto the central protostar at a high
  rate of $\sim 1~\msunyr (\approx 20~c_{\rm s}^3/G)$.}

\begin{figure}[t]
\includegraphics[clip,width=4.0in]{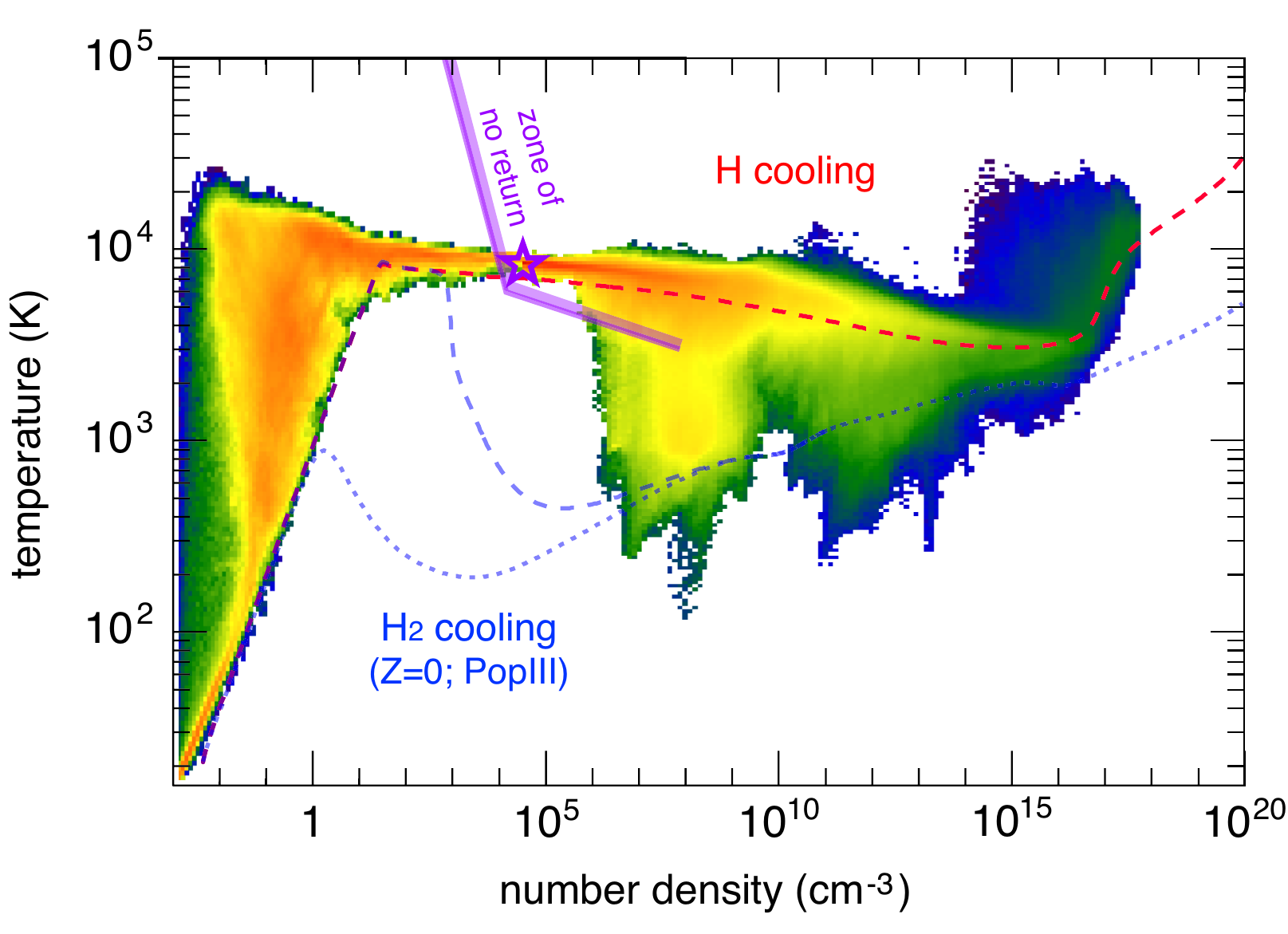}
  \vspace{-1\baselineskip}
\caption{3D simulation result showing the density-temperature phase
  diagram of a massive collapsing cloud in an ACH.  If the gas in the
  core can avoid H$_2$ cooling by some mechanisms (e.g., strong LW
  irradiation in this case) until it enters the ``zone of no return''
  (purple shaded line), then subsequent H$_2$ cooling can be naturally
  averted, and most of the collapsing gas resides in the hot component
  ($T\approx 5000~\K$), which ultimately forms a rapidly accreting
  protostar at the center.  The overall behavior is essentially
  consistent with the one-zone models shown in
  \textbf{Figure~\ref{fig:nt}} (note that if H$_2$ cooling is
  efficient, then the simulation results are also consistent with the
  H$_2$ cooling track shown in \textbf{Figure~\ref{fig:nt}}).  The
  simulation data are taken from \cite{Fernandez+2014} at $n_{\rm
    H}<10^4~\cc$, and \cite{Inayoshi+2014} at $n_{\rm H}>10^4~\cc$.
  The two results are connected at the boundary of the zone of no
  return (star symbol).  }
\label{fig:3D_1}
\end{figure}


\begin{figure}[t]
\includegraphics[clip,width=4.7in]{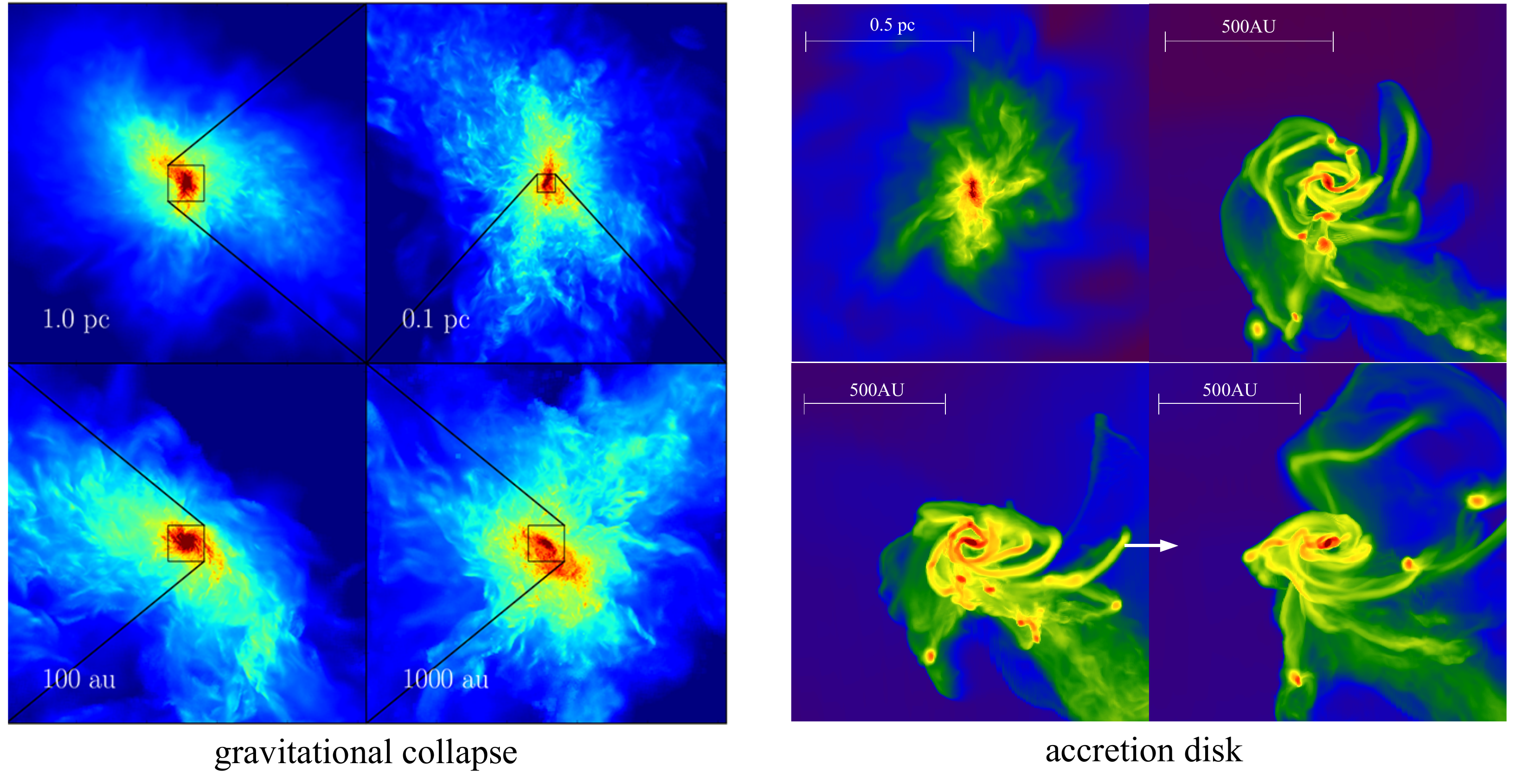}
  \vspace{-1\baselineskip}
\caption{Distribution of gas density in an ACH in
  which ${\rm H_2}$ cooling is suppressed by strong LW radiation with
  $J_{\rm LW}\gsim J_{\rm crit}$.  The results on the left
  \citep{Becerra+2015} and right~\citep{Regan+2014} panels are from
  two different simulations, which overall find a similar
  behavior. The central collapsing region does not undergo a major
  episode of fragmentation despite the complex flow structure caused
  by turbulence (left panels, down the few 1000 AU scales, and panel
  labeled ``0.5pc'' on the right).  However, eventually a disk forms
  (panels labeled ``100 au'' on the left and ``500AU'' on the right),
  which accretes at a high rate, becomes gravitationally unstable and
  soon fragments into clumps (best seen in the right panels).  The
  fragments are found to migrate quickly toward the center and
  are expected to coagulate to form a single supermassive star (see text).  }
\label{fig:3D_2}
\end{figure}

\textbf{Figure~\ref{fig:3D_1}} presents thermal evolution of a pristine, massive
collapsing gas cloud where H$_2$ is initially dissociated by intense LW radiation, 
obtained from a 3D hydrodynamical simulation.  
In the low-density regime ($n_{\rm H}\lsim 10^{4}~\cc$), the cooling 
is mainly via Ly$\alpha$ emission, as assumed in most previous work.  
In fact, Ly$\alpha$ cooling becomes less efficient and continuum cooling 
via two-photon emission leads to
further collapse until $n_{\rm H}\sim 10^8~\cc$. At higher density,
the dominant cooling process shifts to the H$^-$ free-bound emission
(${\rm H}+{\rm e}^- \rightarrow {\rm H^-} +\gamma$).  For $n_{\rm
  H}>10^{15}~\cc$, photons from the H$^-$ free-bound emission are
self-absorbed, as well as Rayleigh scattered by neutral H.  The gas
collapse proceeds further by H$^-$ free-free emission (${\rm H}+{\rm
  e}^- \rightarrow {\rm H} + {\rm e}^- +\gamma$) until $\sim 10^{16}~\cc$
  with the temperature decreasing gradually to $\sim 3000~\K$.  
  Finally, at this stage, 
  the cloud becomes completely opaque to all continuum emission and 
  forms an adiabatic core, i.e., a central protostar, with a mass of 
  $\sim 0.2 ~\msun$.

At the beginning of the cloud collapse ($n_{\rm H}\gsim 10^4~\cc$ in
the zone of no return), the H$_2$ fraction is at its equilibrium value
($x_{\rm H_2}\approx 10^{-8}$ almost independent of density),
balancing formation through ${\rm H^-}$ and collisional dissociation.
This H$_2$ fraction is too small to cool the gas via line emission.
As the density reaches $\sim 10^{11}~\cc$, the H$_2$ fraction jumps up
to $x_{\rm H_2} \sim 0.1$ by the three-body reaction ($3{\rm H}
\rightarrow {\rm H}_2 + {\rm H}$) in the inner region ($\lsim
10^3$~AU).  However, neither the H$_2$-line nor
collision-induced-emission (CIE) cooling plays a significant role in
the thermal evolution: H$_2$ lines are optically thick at $n_{\rm
  H}>10^{14}~\cc$ and other continuum cooling is more important than
the H$_2$ CIE cooling.  Combined with adiabatic cooling due to
turbulent expansion, radiative cooling related to H$_2$ induces
thermal instability, producing cold gas with $T<10^3~\K$ over the wide
density range of $10^6~\cc < n_{\rm H}< 10^{13}~\cc$ where the
temperature evolution deviates significantly from the one-zone model
results (blue curves) as shown in \textbf{Figure~\ref{fig:3D_1}}.
Since the cold components are not massive enough to be gravitationally
bound, the evolution of the central collapsing region is ultimately
not affected \citep[see Figures 3 and 4 in][]{Inayoshi+2014}.

Once the dense core becomes optically thick, the effects of the
trapped radiation become important. \citet{Luo+2018} performed AMR
simulations with {\it enzo}, in which the effects of the radiation
were included via the flux-limited diffusion (FLD) approximation.
These simulations exclude ${\rm H_2}$ and adopt tabulated opacities
for equilibrium abundances (i.e. do not follow the non-equilibrium
H$^-$ fraction), but they reach resolutions of $0.01-0.1$AU, and
resolve the complex shape and time-dependence of the photosphere.
They find qualitatively similar results to the above in their
"adiabatic" control run. On the other hand, when radiation via FLD is
included, they find near-Eddington luminosities in the core, with the
radiation escaping in directions of the steepest density and
temperature gradients.  This leads to intermittent outflows,
originating near the photosphere, while the core inside the
photosphere is quasi-spherical (rather than disky) and has very little
rotation.

The pre-stellar collapse of H$_2$-suppressed gas has also been
investigated in high-resolution cosmological simulations
\citep{Regan+2014,Becerra+2015,Latif+2016} which included H$_2$
chemistry and cooling and have converged on similar results, as
illustrated in \textbf{Figure~\ref{fig:3D_2}}. The early stages of
collapse essentially agree with previous non-cosmological simulations.
This is because the dynamics of the collapsing gas obeys a
self-similar solution, where the initial and boundary conditions have
been forgotten \citep{Larson2003}.  Namely, the density profile
consists of the central core and accreting envelope with the $\rho
\propto r^{-2}$ law \citep{Larson1969}, and the rotational velocity is
as large as half the Keplerian velocity
\citep{Narita+1984,SaigoHanawa1998}, which agrees with the expected
universal value \citep{Abel+2002,Yoshida+2008}. This stage is
illustrated in the left panels of \textbf{Figure~\ref{fig:3D_2}}, down
to $\sim 1000$ AU, as well as the first panel (labeled $0.5$ pc)
on the right.  Subsequently, when the bulk of the envelope mass is
accreting onto the central region and protostar, they are reminded of
their initial and boundary conditions.  The infalling matter forms
a compact accretion disk, surrounding a central embryonic protostar,
but first still without undergoing a major episode of fragmentation
(see the bottom row in the left panel of
\textbf{Figure~\ref{fig:3D_2}}).  Because of the high accretion rate
($>0.1~\msunyr$), the disk, however, soon becomes massive enough to be
unstable under its self-gravity, and is likely to fragment into
smaller clumps even when H$_2$ cooling does not play an important role
(see the three panels labeled ``500 AU'' on the right in
\textbf{Figure~\ref{fig:3D_2}}).

Numerical limitations have precluded following the subsequent
evolution for longer than $\sim$$200$ years after this
stage~\citep{Regan+2014}.  \citet{InayoshiHaiman2014} discussed the
evolution of clumps in the disk with an analytical model, taking into
account the growth of clumps via accretion and inward migration.  The
clumps can rapidly migrate inward on a timescale of $\sim 10^{4-5}$
yr, which is shorter than the internal Kelvin-Helmholtz timescale in
the clumps.  Therefore, most of the clumps can merge with the central
protostar before forming massive stars.  The clumpy structure of the
disk at a high accretion rate provides episodic burst-like accretion,
affecting the protostellar evolution \citep{Sakurai+2016b}.

As noted above, an important caveat is that the trapped radiation can
become dynamically important inside the optically thick core, and can
have a strong impact on the nature of the central object (i.e. its
angular momentum, shape, density, and accretion rate).
\citet{Ardaneh+2018} performed cosmological versions of the AMR
simulations by \citet{Luo+2018} and found qualitatively similar
results.  In particular, the core structure is irregular, and strongly
shaped by recurrent outflows driven by both radiation and thermal
pressure.  Such a rapidly accreting and thus spinning-up protostar
would also lead to non-axisymmetric deformation to a bar-like shape
that enables efficient angular momentum transfer to the surrounding
medium \citep{Lin+2011}.  These outflows mix with the inflow and are
ultimately trapped, but they help outward transfer of angular momentum
and result in a rapidly accreting, quasi-spherical central object
without significant rotation.

\subsubsection{Growth of a rapidly accreting protostar}
\label{sec:SMS}

What is the fate of the protostar surrounded by unlimited amounts of
gas?  Theoretically, such a rapidly accreting protostar is expected to
evolve into an SMS, for which the entropy input by rapid accretion and
energy generation by nuclear burning support the entire stellar
structure, before it ultimately collapses to~a~BH.

\begin{figure}[t]
\includegraphics[clip,width=3.7in]{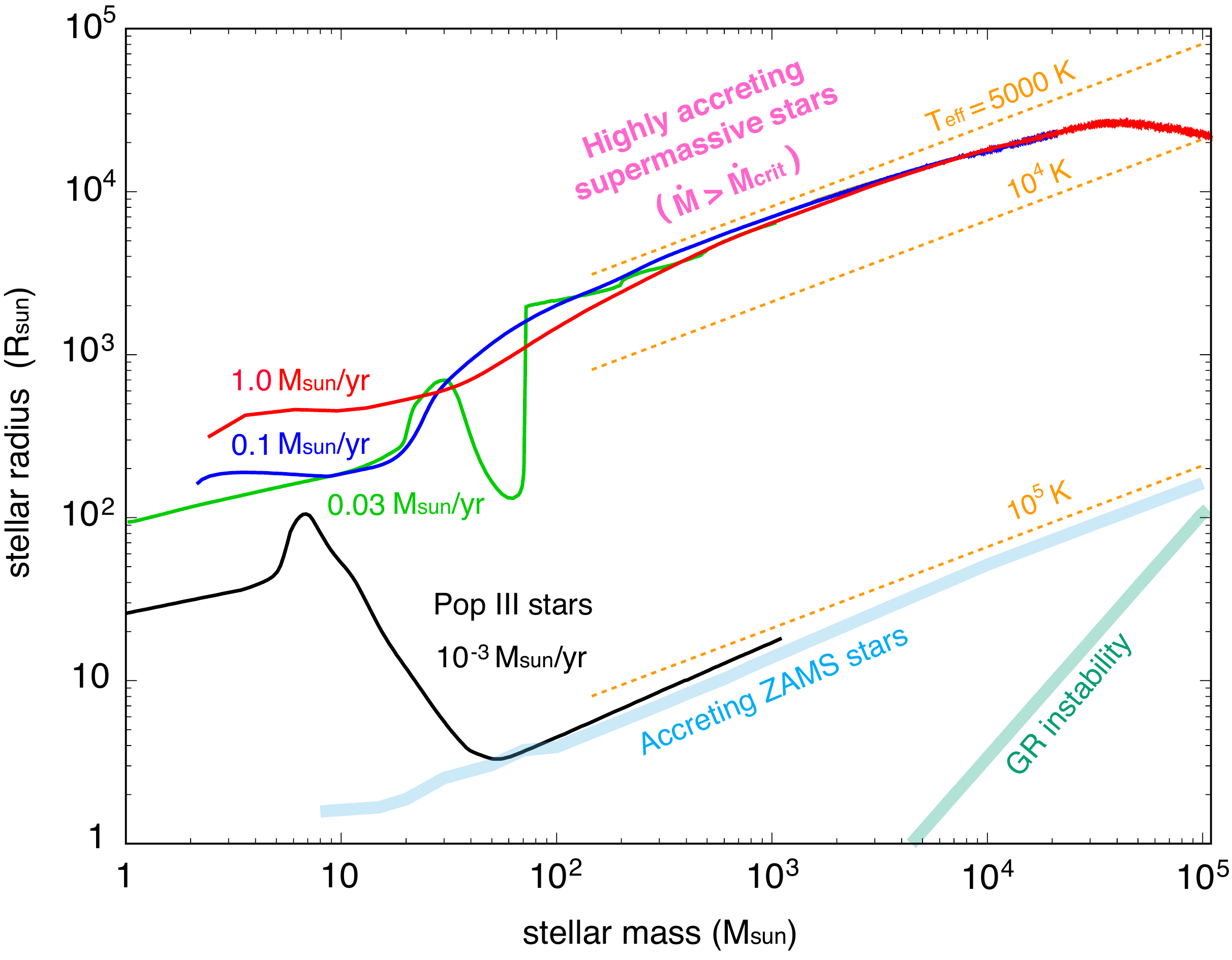}
  \vspace{-1\baselineskip}
\caption{Evolution of the protostellar radius for several different
  accretion rates in the range between $10^{-3}\leq \dot{M}_\star
  /\msunyr \leq 1.0$.  For the lowest accretion rate (ordinary massive
  Pop~III stars; black curve), the stellar structure contracts and
  settles down to that of a ZAMS star with a high effective
  temperature ($\approx 10^5~\K$).  For higher accretion rates
  ($\dot{M}_\star > \dot{M}_{\rm crit}$), the protostar continues to
  expand until its mass reaches $M_\star \sim 10^5~\msun$, above which
  the core region (which contains most of the mass) enters the GR instability regime.  The bloated
  envelope has a low temperature of $T_\star \approx 5000~\K$, for
  which UV stellar feedback does not halt gas accretion
  (the effective temperature indicated by orange lines is estimated
  assuming $L_\star=L_{\rm Edd}$, which is a good assumption for massive stars
  with $M_\star>10^2~\msun$).  
  The data are taken from \citet{Hosokawa+2012, Hosokawa+2013}.}
\label{SMS_R}
\end{figure}

\textbf{Figure~\ref{SMS_R}} shows the evolution of the radii of
protostars accreting at different rates, based on spherical
stellar-evolution models by \citet{Hosokawa+2012,Hosokawa+2013}.  In
the ordinary Pop~III star case, where $\dot{M} \approx
10^{-3}~\msunyr$ is set by H$_2$ cooling, the protostar initially
expands as it gains mass, due to adiabatic heat input by accretion. At
$M_\star \sim 10~\msun$, it begins to contract by cooling via
radiative diffusion (the so-called Kelvin-Helmholtz contraction phase)
until nuclear ignition occurs at the center
\citep{Stahler+1986,OmukaiPalla2001,OmukaiPalla2003}.  On the other
hand, at higher accretion rates of $\dot{M}\gsim 0.03~\msunyr$, the
growing protostar continues to expand without any Kelvin-Helmholtz
(KH) contraction \citep{Hosokawa+2012}.  In fact, the interior
material contracts and increases the central temperature to the onset
of nuclear burning, while the outermost layers significantly swell up,
resembling a red giant star.  This is because the outermost envelope
absorbs a part of the outward heat flux, and gains energy from the
accreted material.

Importantly, the effective temperature of the bloated atmosphere is
almost constant at $T_\star \sim 5000~\K$, regardless of the initial
mass, due to the strong temperature dependence of H$^-$ bound-free
absorption opacity \citep{Hayashi1961}.  As a result, the ionizing
flux is reduced by $\gsim 8$ orders of magnitude compared to a
zero-age main sequence (ZAMS) Pop~III star with the same mass.  Thus,
the UV feedback which could limit the stellar masses at lower
accretion rates to at most a few 100 $\msun$ (\citealt{Hirano+2014},
see also \citealt{McKeeTan2008,Hosokawa+2011}) never operates until
the mass reaches $M_\star \gsim 10^5~\msun$ where GR instability
induces collapse \citep{Hosokawa+2013}.  The accreting protostar with
its bloated envelope is pulsationally unstable, similar to red giants,
due to the $\kappa$ mechanism excited in the He$^+$ ionization layer
in the envelope \citep{Inayoshi+2013}.  However, the mass-loss rate is
significantly lower than the mass accretion rate.  In summary, the
growth of an accreting SMS is not prevented either by UV feedback or
by pulsational instability.

Gas accretion, in reality, likely proceeds through an accretion disk.
Since the mass inflow rate onto the disk from the parent cloud is
high, the disk becomes gravitationally unstable and is likely to feed
the central protostar via episodic mass accretion
(\citealt{Regan+2014,InayoshiHaiman2014, Becerra+2015,Latif+2015b,
  Latif+2016}; see also \textbf{Figure~\ref{fig:3D_2}}).  If the
intervals of episodic accretion due to clump migration through the
disk is sufficiently long, the bloated envelope of the SMS would
contract and emit strong UV radiation.  \cite{Sakurai+2016b} have
performed simulations of a circumstellar disk and the structure of the
central SMS self-consistently by reflecting burst-like accretion
episodes.  Since the typical interval of episodic accretion is shorter than 
the local KH-contraction timescale in the protostellar surface layer 
(where gas opacity is high due to H$^-$ free-bound absorption, and the 
KH-contraction timescale is  an order of magnitude longer than averaged 
over the entire star), the stellar UV feedback never inhibits gas accretion.

The bifurcation in the evolution is determined by the accretion rate.
The interior stellar structure within an inflating envelope was
discussed by \cite{OmukaiPalla2001,OmukaiPalla2003}, who found a
critical rate of $\dot{M}\approx 4\times 10^{-3}~\msunyr$, above which
the total luminosity (interior + accretion) during the KH contraction
marginally exceeds the Eddington luminosity.
\cite{Hosokawa+2012,Hosokawa+2013} found that the inflated structure
appears and stably exists only at higher accretion rates of $\gsim
3\times 10^{-2}~\msunyr$.  Recently, \cite{Haemmerle+2018} found that
at somewhat lower $\dot{M}\approx 10^{-2}~\msunyr$, the stellar
envelope is still bloated by rapid entropy input at $M_\star \gsim
10^3~\msun$ unless UV feedback and pulsational instability prevent
mass accretion.  These accretion rates may, in fact, be realized in
some of the massive minihalos, and/or in ACHs in which H$_2$ cooling
is not fully suppressed and the temperature is as low as a few
$10^3~\K$, potentially leading to many IMBHs with $10^2\lsim M_\bullet
/\msun \lsim 10^5$, as remnants of SMSs with intermediate accretion
rates in these halos.

\subsubsection{The final fates of growing supermassive stars}

According to the classical
argument~\citep{Chandrasekhar1964,ZeldovichNovikov1971,
  ShapiroTeukolsky1983}, an SMS exceeding a critical mass of $M_{\rm
  GR}$ is thought to directly collapse to massive BHs via a GR
instability\footnote{ In the interior of a very massive star,
  radiation pressure dominates gas pressure, leading to the adiabatic
  index $\Gamma_{\rm ad} \approx 4/3+\beta/6$, where $\beta \equiv
  P_{\rm gas}/(P_{\rm rad}+P_{\rm gas})\approx
  0.027~(M_\star/10^5~\msun)^{-1/2}$.  On the other hand, in the
  relativistic regime, the critical index against a small radial
  perturbation is $\Gamma_{\rm crit}\approx 4/3 + 1.12~(R_{\rm
    Sch}/R)$, where the second term comes from the GR effect.  Note
  that in the Newtonian limit for less massive stars, $\Gamma_{\rm
    ad}(= 5/3)$ is larger than $\Gamma_{\rm crit}(= 4/3)$, where the
  stellar structure is stable.  The green line in
  \textbf{Figure~\ref{SMS_R}} represents the critical radii of the GR
  stability ($\Gamma_{\rm crit}>\Gamma_{\rm ad}$;
  \citealt{Fricke1973}).}.  The critical mass is on the order of $\sim
10^5-10^6~\msun$, depending on the detailed properties of the stellar
rotation and radial structure.  \cite{ShibataShapiro2002} investigated
the gravitational collapse of a rotating SMS in full GR simulations,
and found that most of the stellar mass is eventually swallowed by the
newly-born BH, ejecting only $\sim 10\%$ of the mass.  Note that even
if the star initially has fast differential rotation, the angular
momentum would be quickly transported by turbulent viscosity driven by
MRI, but a higher fraction of the initial rest mass of the star forms
a disk instead of directly collapsing into the BH \citep{Sun+2018}.
If the star is rotating sufficiently fast at the beginning of
gravitational collapse, the SMS collapses and turns into an
intermediate stage where a close binary BH forms
\citep{Reisswig+2013}.  Some authors proposed a different picture, in
which only the central part of the SMS collapses to form a smaller,
$\sim 100~\msun$ BH, and the outer envelope is still inflated by
energy input from the accreting BH (``quasi-star";
\citealt{Begelman+2006, Begelman+2008}).  However, these results are
based on a simplified treatment of the equation of state; namely a
fully convective star with a homogeneous entropy distribution is
assumed.  Recent stellar evolution calculations suggest that rapid
mass accretion onto SMSs drastically changes their stellar structure,
causing significant bloating with a positive entropy gradient
\citep{Hosokawa+2013,Haemmerle+2018}.  As the homogeneous convective
core gradually extends after the ignition of hydrogen burning, not the
entire star but only its inner core might collapse due to the GR
instability.

Although runaway nuclear fusion might cause a very energetic supernova
(SN) explosion for an SMS with high metallicities, this is less likely
at zero or low metallicity \citep{Montero+2012}.  \citet{Umeda+2016}
extended the stellar structure calculations to the onset of the GR
instability, and found the critical mass $M_{\rm GR}$ increasing
monotonically with stellar accretion rate (see also
\citealt{Woods+2017}): (1) at $\dot{M}<0.1~\msunyr$, the stellar
collapse begins to occur when the nuclear fuel is exhausted,
(2) at $\dot{M}\approx 0.3-1.0~\msunyr$, the
star becomes GR unstable during the helium-burning stage at $M_{\rm
  GR}\approx 2-3.5~ \times 10^5~\msun$, and (3) in an extreme case with
$\dot{M}\approx 10~\msunyr$, the star collapses during the
hydrogen-burning stage at $M_{\rm GR}\approx 8.0 \times 10^5~\msun$\footnote{
For even more extreme cases with $\dot{M}\gg 10~\msunyr$, which could be 
achieved in the merger of massive galaxies \citep{Mayer+2010} rather than in an ACH, 
the central region could directly collapse into a BH without forming a hydrostatic equilibrium structure such as an SMS 
\citep{MayerBonoli2019,Haemmerle2019}.}. 
Note that in regime (3), there is a several orders of magnitude discrepancy in final stellar mass 
found by different groups \citep[see Figure 10 in][]{Woods+2018}.
The higher values of $M_{\rm GR}$ are caused by the positive entropy
gradient in the realistic stellar structure.  
At the end of the stellar evolution, $60-80\%$ of the total stellar mass is 
enclosed with the GR instability regime (Figure 7 of \citealt{Hosokawa+2013}).
Since the unstable region is as hot as $T\sim 10^7~\K$, the mass accretion
rate after the onset of collapse is as high as $\sim 5000~\msunyr$, which 
corresponds to a hyper-Eddington value of 
$\dot{M}/\dot{M}_{\rm Edd}>10^6~(M_\bullet/10^5~\msun)^{-1}$.

\subsection{Massive Black Holes via Runaway Mergers in a Dense Star Cluster}
\label{sec:densecluster}

The previous subsection outlined the scenario in which primordial gas,
cooling and contracting in an ACH, produces a massive seed BH via
monolithic collapse into a supermassive star.  While this is the
conventional wisdom for metal-free gas, the issue of whether the cloud
may fragment at very high densities is not entirely
settled. Furthermore, as mentioned in the introduction (see
\textbf{Figure~\ref{fig:rees}}) and in
\S~\ref{subsec:massiveBHrequirements}, when the gas collapsing in such
a halo has some modest level of pre-enrichment by metals and/or dust,
fragmentation is expected to occur.  This fragmentation represents
another possible pathway to massive seed BHs, or at least to IMBHs,
which has received less attention to date than it deserves.

In particular, following the arguments outlined in
\S~\ref{subsec:massiveBHrequirements}, this fragmentation may occur at
extreme densities -- as high as $n_{\rm H} \sim 10^{11-13}{\rm cm^{-3}}$ for
very metal-poor clouds with $Z \gsim 10^{-5}~\zsun$
(corresponding to $\gamma\approx 0.8$, just prior to reaching the temperature minima of the dotted and dot-dashed
curves in \textbf{Figure~\ref{fig:nt}}), while the fragments have
masses as low as $\sim 0.5~\msun$, at least initially (corresponding
to the Jeans mass at this density, shown in the same figure).  The
natural interpretation is therefore that an ultra-dense cluster of
low-mass stars may result~\citep{Omukai+2008,DevecchiVolonteri2009}.
Because of the extremely high density ($\sim 10^{9-11}~\msun~{\rm
  pc^{-3}}$, i.e. orders of magnitude higher than the densities at the
  resolution limit of local stellar clusters; see below), such star
clusters can undergo efficient runaway core collapse, which may
consume a non-negligible fraction of the stars, leading to a central
IMBH with a mass of up to $\sim 10^4~\msun$.  

The precise way a runaway collapse unfolds could take different
shapes.  Direct collisions between stars can occur over a timescale
that is shorter than the lifetime of even massive stars
\citep{Katz+2015,YajimaKhochfar2016,Sakurai+2017}, resulting in the formation of a single
massive star. Note that this is a runaway process, because once stars
collide, their masses and radii increase, accelerating the rate of
subsequent mergers.  Indeed, the basic features of this runaway have
been elucidated in several studies addressing possible IMBH formation
in the local universe~\citep{PortegiesZwart+2004,Gultekin+2004,Freitag+2006, Stone+2017}.
We note, however, that all of these works addressed star clusters with
much lower densities than the densities possible in the high-$z$
universe.  The bottom line is that stellar mergers could build a
single SMS, essentially reproducing the pathway of the previous
section, where such an SMS grew via accretion.

In the high-$z$ context, a ``hybrid'' scenario is also possible. In
particular, the newly born dense star cluster may be still embedded
in a dense gas cloud.
The dynamics of the runaway collapse itself may be aided by sudden 
significant gas inflows into the cluster~\citep{Davies+2011}.
In this case, some of the protostars may also be
accreting at high rates and have bloated envelopes, significantly
increasing their geometric cross-section; gas dynamical friction also
facilitates more efficient mergers~\citep{Boekholt+2018,Reinoso+2018}.
Additionally, the star-star mergers themselves may provide enough
heating to keep a central star bloated and on track to the SMS
regime~\citep{Tagawa+2019}. In these cases, a significant fraction
($\gsim 10\%$) of the stars may end up merging into a single SMS, and
the resulting IMBH could reach masses as high as $10^4-10^5~\msun$.

Finally, a variation on this possibility is that the cluster is
somewhat less dense, and the stars are massive and short lived, and
produce remnant stellar-mass BHs before they merge.  There will then
be a population of stellar-mass BHs embedded in a dense gas cloud.
Mergers between these BHs, aided by gas dynamical friction, could also
build a more massive $10^{3-4}~\msun$ IMBH~\citep{Tagawa+2016}.
\cite{Ryu+2016} have addressed a similar setup, 
except assuming an even
lower space density of the initial BH distribution (spreading the BHs
over $\sim 100$ pc, rather than over $<10$ pc). They simulated the
orbital motion of the BHs, taking into account gas drag on the BHs,
and found that most initial BH configurations lead to the formation of
a single massive BH in the center of the protogalaxy, reaching a mass
of $10^{3-5}~\msun$ through hyper-Eddington growth, rather than via
BH-BH mergers.

For reference, a typical density in the core of a globular cluster is
$\sim 10^{3}~\msun~{\rm pc^{-3}}$, although many are much denser, and
the densest known star cluster in the Milky Way, the Arches cluster,
has a central density of $\approx 2\times 10^{5}~\msun~{\rm pc^{-3}}$.
These densities are $4-5$ orders of magnitude lower than that of the
fragmenting clusters hypothesized at high redshifts, but are measured
on larger scales~\citep{Nguyen+2018}.  Nuclear star
clusters~\citep{Walcher+2005} and even an isolated local ultra-compact
dwarf galaxy~\citep{Strader+2013} reach surface densities at their
half-light radii which are similar to the densest globular clusters.
However, their core densities can be much higher. For example, M32 has
a central density, resolved by HST at $\sim 0.2$pc, in excess of
$>10^{7}~\msun~{\rm pc^{-3}}$~\citep{Lauer+1998}. M32 has a near-solar
metalicity, and it also has a central SMBH with an estimated mass of
$3\times10^6~\msun$ which could have generated the central
Bahcall-Wolfe-like cusp~\citep{MerrittSzell2006}.  More generally, the
above suggests that it may be possible to look for relics of the
high-$z$ star clusters in the local universe.  Such relics, assuming
that they preserved their identities, would consist of the IMBH and a
dense inner core of low-mass, extremely metal-poor stars, reaching
$\sim 10^{7}~\msun~{\rm pc^{-3}}$ at $\sim 10^3$ AU (where the cluster
did not have time to disperse in a Hubble time;
\citealt{Tagawa+2019}).  Extrapolation of the stellar density in the
nucleus of the Milky Way down to this scale is consistent with this
value~\citep{Genzel+2010}.

In summary, {\em runaway mergers between stars and/or their remnant BHs,
  facilitated by the extreme densities and the presence of gas
  enveloping the young star cluster, represents a viable pathway to
  forming IMBHs, possibly via a stage of an SMS, but with masses
  expected to be somewhat below the most extreme massive seed BH
  produced in the other pathways.}

\subsection{Subsequent Growth and Cosmological Evolution}
\label{subsec:BHevolution}

\begin{figure}[t]
\includegraphics[clip,width=5.0in]{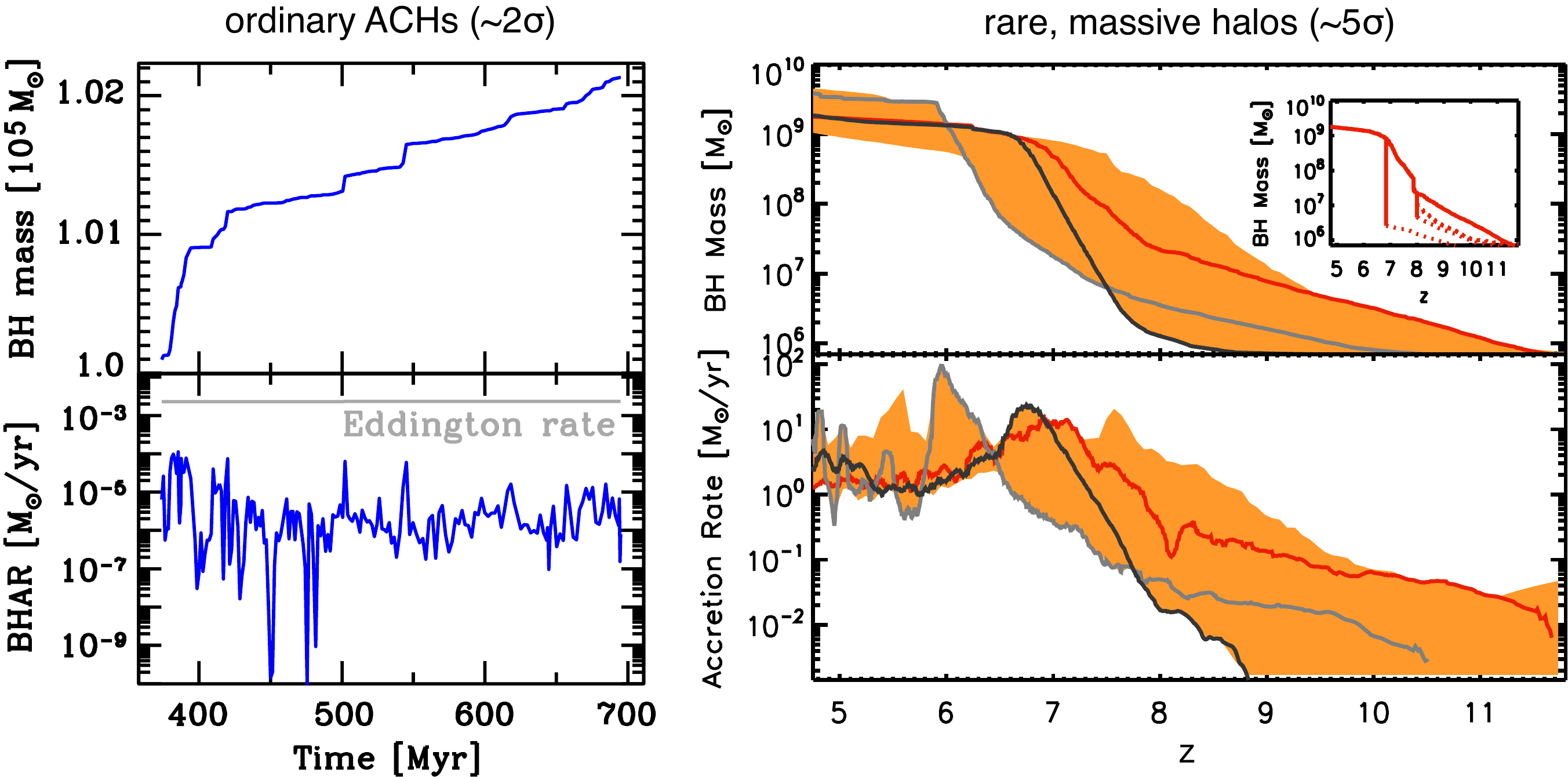}
  \vspace{-0.5\baselineskip}
\caption{Black hole mass and the accretion rate vs. redshift in two
  different simulations. The left panel shows the growth of a
  $10^5~\msun$ seed BH in a typical ACH halo, between $z=12$ and
  $z=7.5$ \citep{Latif+2018}.  The host galaxy reaches $3\times
  10^{10}~\msun$ at $z=7.5$. The right panel shows the evolution of
  seed BHs with the same mass, but placed in the progenitors of more
  massive galaxies that reach $\sim 10^{12}~\msun$ by $z\approx 6$
  \citep{DiMatteo+2012}.  The three lines represent three examples,
  and the orange band the full range in their simulated sample.  In
  the left panel, the seed BH never grows because of energetic SN and
  AGN feedback.  In the right panel, the seed BH is fed with cold gas
  streams supplied from large-scale structure at high accretion rates,
  and thus the BHs can grow to $M_\bullet \approx 10^9~\msun$ by
  $z\sim 7$.  We caution that the numerical resolution and sub-grid
  model treatments for feedback also differ in these simulations, and
  could impact the overall BH growth efficiency.}
\label{fig:BH_coldacc}
\end{figure}

We have reviewed several possible formation pathways for seed BHs with
masses of $10^2\lsim M_\bullet/\msun \lsim 10^6$ in ACHs (or
sub-ACHs).  In this section, we briefly discuss the
subsequent growth of these seeds in the cosmological context of
hierarchical structure formation.  Two approaches have been utilized
to model the population of growing BHs.  The first is semi-analytical
modeling, in which BH seed formation, gas accretion, and mergers of
BHs are modeled in a simplified way.  This is an effective method for
examining the statistical properties of BH populations and making
theoretical predictions that can be directly compared with
observations such as the quasar LF (see \S\ref{sec:obs}).  The second
approach is to use cosmological simulations of early galaxy formation.
These simulations do not have adequate dynamical range for accurate
global statistical predictions while resolving minihalos, but can
capture much more detailed properties of large-scale structure
formation, the growth process of individual seeds into SMBHs, feedback
associated with BH accretion, stellar radiation, and SN explosions.

Previous works employing semi-analytical methods have elucidated the
ingredients (e.g.~seeding mechanisms/conditions and BH accretion
physics) required to model the early BH population 
\citep[e.g.,][and references therein]{Volonteri+2003, TanakaHaiman2009, 
Agarwal+2013,Valiante+2016,RicarteNatarajan2018}.
Based on two seed formation scenarios (Pop III
remnants vs. massive seeds with $M_\bullet \approx 10^5~\msun$),
\cite{TanakaHaiman2009} explored possible channels to explain the
extremely rare high-redshift SMBH population. They varied key physical
parameters such as the BH seeding fraction and the accretion duty
cycle and found that models with the optimistic assumptions required
to explain the SDSS SMBHs ``overproduce" the mass density in lower
mass BHs by a factor of $10^2-10^3$. They find that this
overproduction can be avoided if seed formation stops or BHs accrete
at lower duty cycles at $z \lsim 20-30$ (such suppression is expected
due to the negative global feedback from the X-rays emitted by these
BHs themselves; \citealt{Tanaka+2012}).  More recently,
\cite{Valiante+2016} developed semi-analytical models that include
additional physics related to BH seeding mechanisms including LW
irradiation, chemical enrichment of halos, and cosmic reionization.
They claim that a contribution of massive seeds with $M_\bullet
\approx 10^5~\msun$ is needed to explain the most extreme SMBH
population unless the typical mass of Pop III remnants extends to
$\sim 10^3~\msun$.  Interestingly, this implies that Pop III remnant
BHs could play an important role in the formation of extremely massive
BHs as well as the less-massive BH population.

In recent decades, large-scale cosmological simulations have been
exploring the evolution of SMBHs and the coevolution of their host
galaxies.  Various feedback processes including SNe and AGN activity
have been examined.  Due to numerical limitations, cosmological
simulations with a spatial resolution of $\sim O({\kpc})$ must treat
feedback effects with sub-grid models instead of directly resolving
physical processes in the nuclear region.  There are many different
simulation studies adopting different sub-grid feedback models. These
simulations have generally found that most (massive) seed BHs formed
in protogalaxies hardly grow in mass via gas accretion because dense,
cold gas is expelled by energetic SN feedback associated with star
formation.  Radiative/mechanical feedback caused by BH accretion is
expected to produce a more modest effect~\citep{Dubois+2013, Dubois+2015, PrietoEscala2016, Smidt+2018,Smith+2018,Latif+2018}.  
For example, \cite{Latif+2018} focused on the growth of a $10^5~\msun$ BH at the
center of an ACH, marginally resolving the Bondi radius for cold gas
with $T\approx 8000~\K$, and found that the accretion is completely
shut off by SN feedback all the way down to $z\sim 6$ (see left panel
of \textbf{Figure~\ref{fig:BH_coldacc}}).  \cite{Habouzit+2017}
investigated the growth of seed BHs using large-scale cosmological
simulations and found that most seeds cannot grow due to SN feedback
until lower-redshifts when those seeds settle in the centers of more
massive galaxies.  As a result of this SN feedback, they find most
seed BHs stop their growth at $M_\bullet \approx 3\times 10^3~\msun$.
Their most massive BH reaches $M_\bullet \approx 3\times
10^4~(10^7)~\msun$ for their strong (weak) SN feedback model by
$z\approx 6$.

It is very important to note that most simulations where SN feedback
quenches BH growth have focused on typical ACHs in $\approx 2\sigma$
regions of the universe that will grow to $\lsim 10^{11}~\msun$ by
$z\approx 6$. However, the first SMBHs are likely hosted by rare
$10^{12-13}~\msun$ halos (corresponding to 4-5$\sigma$ fluctuations on
these scales).  In such massive halos, a sufficient gas supply would
be maintained by cold gas streams from large-scale structure at
$\dot{M}\sim 300~\msunyr (V_{\rm circ}/200~\kms)^3$ {\it if all
  feedback processes are ignored}.  Cold gas streams feeding the
center of the halo can exist only when thermal energy generated by
shocks associated with virialization is quickly radiated away,
otherwise a hot and diffuse medium fills the halo
\citep{ReesOstriker1977, Silk1977,
  BirnboimDekel2003,Keres2005,DekelBirnboim2006}.  Large-scale
cosmological simulations demonstrate that cold flows are not very
susceptible to feedback from growing BHs and a high accretion rate is
maintained until the mass of the galaxy reaches $\approx
10^{12}~\msun$, when the cold mode of accretion turns into the usual
hot virialization mode and the gas supply to the nuclear region is
strongly quenched \citep{DiMatteo+2012,Khandai+2012}.  As a result, a
seed BH with $M_\bullet =10^5~\msun$ is able to grow up to $\gsim
10^9~\msun$ by $z\approx 6$ (see right panel of
\textbf{Figure~\ref{fig:BH_coldacc}}), assuming a simple BH-feeding
prescription of $\dot{M}_\bullet ={\rm min}(\dot{M}_{\rm
  Edd},\dot{M}_{\rm B})$.  This overall picture seems consistent with
a number of different mechanisms for seeding BHs of $\sim
10^{5-6}~\msun$ including intense LW radiation (\S\ref{sec:H2LWdis}),
formation of dense shocked gas by colliding cold accretion flows
(\S\ref{sec:H2coldis}), strong streaming velocities (\S\ref{sec:BSM}),
and dynamical heating due to rapid halo mergers (\S\ref{sec:DH}).
Similar conclusions have been reached in other studies, including by
\citet{Li+2007}, \citet{Sijacki+2009}, and by \citet{Lupi+2019}, who find that SNe initially slow the growth, but once the galaxy is sufficiently massive, the BH is able to grow and reach $10^9~\msun$ (but see \citealt{Dubois+2013},
where depending on the AGN feedback model, the BH growth is severely
self-regulated even in very massive halos.).  {\it Therefore, further
  cosmological simulations where the early growth of seed BHs takes
  place in peculiar environments self-consistent with their seeding
  models and the rarity of high-$z$ SMBHs, as well as various feedback
  effects caused by SNe and AGN activities, are required.}

There are several important limitations of current numerical
simulations.  First, most large-scale cosmological simulations do
resolve the dynamics of DM/gas/stars on galactic scales at $\sim
O(\kpc)$, but the BH sphere of influence $R_{\rm B}$ is not resolved.
Although some studies mentioned above marginally resolve the BH
gravitational sphere of influence for neutral warm gas with $T\approx
8000~\K$, the accuracy of their prescriptions for energy and/or
momentum feedback injected in unresolved regions still remain
uncertain.  Additionally, the threshold density above which gas turns
into stars is typically set to $n_{\rm H}\approx 1-100~\cc$. This is
orders of magnitude lower than the density in star forming regions in
the local universe.  The efficiency of stellar and SN feedback must
strongly depend on this star formation density parameter.  For
example, if the threshold is set to a very low value, dense and cold
gas clouds are disrupted because gas consumption and feedback strength
are artificially overestimated.  This could lead to the very
inefficient growth of BHs seen in many cosmological simulation
studies, and could suppress possible short-duration super- or
hyper-Eddington accretion phases.  Another important piece of sub-grid
physics is how the orbits of massive BHs are handled during the
mergers of their host galaxies.  A common approach is to simply assume
that once two BHs are sufficiently close, they merge instantly, and
the merger remnant is moved to the center of mass of the host galaxy.
In reality, the unresolved inspiral of massive BHs can be inefficient
and take a significant fraction of the Hubble time; during this
inspiral their growth will be suppressed by the lower ambient density
and high orbital speed (see, e.g. \citealt{Gabor+2016} and references
therein).
Cosmological simulations that self-consistently connect with much
higher-resolution non-cosmological simulations, and able to resolve
the orbits of massive BHs to within the GW-emitting regime, will
likely be needed in the future for more accurate predictions.

{\it In summary, most massive seed BHs formed in early protogalaxies
  at $z\gsim 10$ would hardly grow via gas accretion because energetic
  SN feedback quickly evacuates gas from the nuclear region.  However,
  a small minority of seeds that were born in highly biased regions of
  the universe, could be fed with intense cold accretion streams
  through large-scale cosmic filaments.}  Additional theoretical work
on the link between the environments required for BH seeding and the
assembly history of SMBH host galaxies is needed.

\section{ALTERNATIVE BLACK HOLE FORMATION CHANNELS}
\label{sec:exotica}

In the absence of a definitive conclusion that early BHs formed via
one or more of the astrophysical scenarios above, it is prudent to
keep an open mind to alternative possibilities. 
Also, {\em JWST} and other instruments will have the observational capability 
of detecting more distant and more massive SMBHs, which can only be produced 
in exotic scenarios, should such SMBHs exist. Here we briefly review ideas 
invoked to form (perhaps some of the) early massive BHs.

\subsection{Primordial Black Holes}

The notion that primordial black holes (PBHs) may have formed in the
early universe was suggested over 50 years ago
\citep{ZeldovichNovikov1967,Hawking1971}.  Many specific PBH formation
mechanisms have been developed, 
based on large density fluctuations which can
decouple from the cosmic expansion and collapse into BHs.
Such overdensities could be produced in many different ways, including
in phase transitions, a temporary softening in the equation of state
(reducing pressure), quantum fluctuations, or specific ``designer''
models of inflation leading to a narrow peak in the fluctuation power
spectrum \citep[see, e.g., a review by][]{Carr2005}.  Equating the
cosmological background density at cosmic time $t$ with the density of
a Schwarzschild BH yields a characteristic PBH mass tracking the
increasing horizon mass,
\beq
M_{\rm PBH} \approx \frac{c^3 t}{G} \approx 10^5 \left(\frac{t}{\rm 1s}\right)\msun.
\label{eq:Mpbh}
\eeq

Interest in PBHs has also come from many contexts, such as the various
consequences of the Hawking radiation from small PBHs ($\lsim
10^{15}$ g) that have and are currently evaporating \citep[resulting in, e.g.,
  a $\gamma$ ray background;][]{PageHawking1976}, the possibility that PBHs make
up most, or even all of the DM \citep[e.g.][]{Carr+2016}, or that they
may account for recent LIGO detections of mergers between stellar-mass
BHs in the local universe~\citep{Bird+2016}.

Most interesting for this review is the possibility that PBHs as
massive as $10^5~\msun$ exist, and take on the role of massive seeds at
high redshifts~\citep[e.g.][]{BeanMagueijo2002,Dolgov2018}.  Such
massive PBHs form relatively late (i.e. at $t\sim 1$ s;
eq.~\ref{eq:Mpbh}), near the epoch of electroweak decoupling, and
therefore could potentially disturb the successful predictions of big
bang nucleosynthesis, introducing small (perhaps as large as $\sim 1\%$) 
inhomegeneities in the He abundance~\citep{CarrSilk2018}.  The
strongest constraints on such massive PBHs, however, come from
the angular power spectrum of cosmic microwave background (CMB) temperature and polarization anisotropies.
Massive PBHs begin to accrete gas efficiently after cosmic 
recombination \citep{Miller2000}. The corresponding radiation
pre-ionizes and heats the IGM~\citep{Ricotti+2008}. The accompanying
impact on the CMB anisotropies is subject to uncertainties about the
nature of the accretion (e.g. typical angular momentum, and
corresponding mode of accretion) and the emerging radiation (overall
radiative efficiency and spectrum).  The effect is strongest when the
radiation is assumed to arise from a disk, and yields a limit of
\beq
\rho_{\rm PBH} \lsim 1.1\times 10^{3}\left(\frac{10^5~\msun}{M_{\rm PBH}}\right)^{1.6} 
\left(\frac{0.01}{\lambda}\right)^{1.6} ~\msun\,{\rm Mpc}^{-3}
\label{eq:rhopbh}
\eeq
on the comoving mass density of these PBHs.  This limit was derived by
\citet{Poulin+2017}, based on radiatively inefficient
advection-dominated accretion flow (``ADAF'') models, and includes an
uncertain fudge factor $\lambda~(\approx 0.01)$ by which the disk
accretion rate is reduced (by winds and outflows) compared to the
Bondi-Hoyle-Lyttleton rate.  A model adopting radiatively even less
efficient spherical accretion \citep{HaimoudKamionkowski2017} finds a
$\sim100$ times weaker limit.  Equation~\ref{eq:rhopbh} represents a
tiny fraction ($\approx 3\times10^{-8}$) of the comoving DM density,
and only 0.3\% of the total present-day SMBH mass
density~\citep[e.g.][]{YuTremaine2002}.  Nevertheless, it corresponds
to only a weak upper limit of $\rho_{\rm PBH}/M_{\rm PBH}\approx
0.01~{\rm Mpc}^{-3}$ on the comoving number density of such BHs, which
is comparable to the present-day galaxy number density, and is $\sim
10^{6-7}$ times larger than the abundance of bright quasars at
$z\approx 6-7$.  Therefore, at least by their abundance, massive PBHs
remain viable as seeds of rare early quasars.
Equation~\ref{eq:rhopbh} also shows that the abundance of smaller PBHs
allowed by CMB constraints increases rapidly (since these smaller BHs
accrete much less efficiently from the IGM).  In the $1~\msun \lsim
M_{\rm PBH}\lsim 100~\msun$ range (where CMB constraints are weak), a
limit is provided by the (lack of) weak gravitational lensing of Type
Ia Supernovae~\citep{ZumalaSeljak2018}. This limit is still weak, and
allows up to $\sim 30\%$ of the DM to be composed of such stellar-mass
PBHs. The strongest current limit comes from the number of LIGO
events, which suggests that $1~\msun \lsim M_{\rm PBH}\lsim 300~\msun$
BHs can not make up more than $\sim 1\%$ of the DM \citep{Yacine2017}.

\subsection{Dark-Matter Powered Stars}

Another possibility is that some first-generation stars in the
universe were so-called ``dark stars'' \citep{Douglas2008}. Dark stars
are similar to normal metal-poor stars, except that they are powered
by dark matter annihilation \citep[see a recent review
  by][]{Freese+2016}.  The energy release from DM self-annihilation
can replace nuclear fusion as the dominant energy source, and maintain
a hydrostatic structure as long as the DM fuel lasts.  Annihilation
converts the majority of the rest-mass of the DM particles to energy
that can heat the star, compared to the 0.7\% efficiency of baryonic
fusion.  As a result, DM typically makes up a small fraction ($\lsim
10^{-3}$) of the total mass of the dark star.

The conditions for a dark star to exist are that (i) the DM density within
the star is sufficiently high for annihilation to dominate
over other heating processes (nuclear fusion, gravitational
contraction) and (ii) the annihilation products must deposit their
energy in the stellar envelope, where it must thermalize, rather than
escape the star.  One of the most popular (and best motivated) DM
candidates is a weakly interacting massive particle (WIMP). In many
models the WIMP is its own antiparticle, and self-annihilates with a
cross-section $\langle\sigma v\rangle\approx 10^{-26}~{\rm
  cm^3~s^{-1}}$ determined by the strength of weak interactions, which
yields the observed DM density~\citep{Jungman+1996}.  A natural place
where dark stars may be expected to form is in regions of high DM
density in the early universe, at the cores of DM minihalos.  Assuming
an NFW profile in these early halos, consisting of WIMPs, satisfies
condition (i) above. Whether condition (ii) is satisfied depends on
the specific types of annihilation products (e.g. photons,
electrons/positrons, or neutrinos) and their energy spectrum.  For
typical WIMP models, model stellar atmospheres have been found to be
opaque to these annihilation products (or their secondary products as
they cascade through a sequence of photon / electron / positron
conversions; see \citealt{Freese+2016}). More generally, the requirements
for dark star existence have been shown to be met generically for a wide
range of  WIMP properties and dark matter halo density profiles \citep{Freese2009}.

The main difference between normal metal-poor massive stars and dark
stars is that the latter are much puffier (several AU in radius), and
therefore due to their large surface area and correspondingly low
surface temperature ($\lsim 10^4$K), they emit primarily infrared
radiation.  Similarly to the case of the rapidly accreting protostars
discussed in \S~\ref{sec:SMS} above, this avoids UV feedback shutting
down their accretion and limiting their masses: dark stars can
continue to grow. However, for dark stars, the external gas accretion
rate need not be very high to maintain their large size.  For example,
at the accretion rate of $10^{-3}~\msunyr$,
DM annihilation can keep the star bloated and its effective
temperature near $\sim 10^4$ K, so that in $10^8$ yrs, a
$\sim10^5~\msun$ supermassive star is built
\citep[e.g.][]{Rindler-Daller+2015}.  In principle, dark stars can
continue to grow to even higher masses (up to $10^7~\msun$, given a
sufficiently high ${\dot M}$, as long as the original DM fuel lasts,
and/or if the DM within the star is replenished by the capture of
new DM particles.  In this case, they may reach luminosities high
enough to be detectable by NIRCam with the {\em James Webb Space
  Telescope (JWST)} in Msec exposures \citep[e.g.][]{Ilie+2012}.
Ultimately, once their DM fuel is exhausted, the most massive dark
stars will collapse to BHs via a general relativistic instability,
without ever going through a phase of nuclear fusion.

\subsection{Heating by a (Primordial) Magnetic Field}

As argued elsewhere in this review, the key ingredient of massive BH
seed formation is for the collapsing gas to avoid fragmentation, which
in turn requires avoiding efficient H$_2$ cooling and remaining
at temperatures near $T\sim 10^4$ K.  This condition can be satisfied
either by reducing cooling, or, alternatively, by invoking an extra
source of heating, beyond the compressional and shock heating that is
typically included in studies of gas collapse in protogalaxies.  In
principle, a sufficiently strong ($\sim$nano-Gauss) primordial
magnetic field (PMF) could help provide the required
heating~\citep{Schleicher+2009,Sethi+2010}.

Several mechanisms have been proposed to produce a global PMF, with a
comoving field strength of order 1 nG (quoted here and throughout this
subsection on the scale of 1 Mpc), during inflation and/or during
various phase transitions in the early universe \citep[see,
  e.g. extended reviews by][and references
  therein]{Widrow2002,Kandu2016}.  If present, the weak seed PMF can
be amplified by flux--freezing inside a collapsing primordial
gas~\citep{MakiSusa2007,Sethi+2008,Schleicher+2009}.  Dynamo effects
should also help amplify small initial seed
fields~\citep{Schleicher+2010}; simulations resolving turbulence have
confirmed this and suggested that magnetic fields can become
dynamically important~\citep[e.g.][]{Turk+2012}.

Several studies have pointed out that a seed PMF can therefore affect
the fragmentation properties of the first
protogalaxies~\citep[e.g.][]{MachidaDoi2013,Latif+2014}.  First,
$\sim$ nG seed magnetic fields can elevate the Jeans mass, and delay
collapse until past the atomic-cooling threshold
~\citep{Schleicher+2009,Sethi+2010}.  Second, heating by
ambipolar diffusion or decaying turbulence can dominate H$_2$
cooling and keep the gas warm, and the gas accretion rate high.  In
one zone models, \citet{Sethi+2010} find that ambipolar diffusion
dominates, and the critical PMF value to keep the gas at $T\sim 10^4$ K is $\sim3$ nG.

This critical magnetic field strength is somewhat higher than recent
upper limits from the CMB.  PMFs leave several distinct signatures in
the CMB anisotropies.  From the angular power spectra, Planck finds a
limit of 2 nG on a nearly scale-invariant
PMF~\citep{Planck-Bfield2016}.  This limit is strengthened by a factor
of $\sim$2 by including the impact of the PMF on the ionization
history~\citep{KunzeKomatsu2015,Paoletti+2019} or by combining Planck
and SPT data to use the small-scale B-mode polarization power
spectrum, which is sourced by the PMF and survives damping well past
the Silk scale~\citep{Zucca+2017}.  Ultra-faint dwarfs, whose
formation at high redshift would be suppressed for large PMFs, give
an even stronger limit of $\sim$0.5 nG~\citep{SafarzadehLoeb2019}.

Overall, this suggests that for a PMF near its maximum allowed
amplitude of $\sim 1$ nG, H$_2$ cooling can be fully suppressed
in rare $\gsim 3\sigma$ regions of the spatially fluctuating
$B$--field.  However, it is worth noting that even weaker fields can
be important, because they can reduce the LW flux ($J_{\rm crit}$)
required to disable H$_2$ cooling; \citet{VanBormSpaans2013}
finds a factor of 10 reduction for $B=2$ nG.

\subsection{Massive black Holes from Collisional Dark Matter}

A handful of well-documented issues with galaxy formation in the
$\Lambda$CDM model (e.g. the so-called cusp/core, missing satellite,
and too-big-to-fail problems; reviewed recently by
\citealt{BullockBoylan-Kolchin2017}) have motivated several proposals
to change the particle properties of CDM.  In general, these
modifications are designed to reduce the fluctuations on small scales
(below $\sim 1$ Mpc, or $\sim 10^{11}~\msun$).  Such reductions can be
dramatically important for early SMBH formation.  In the CDM paradigm,
the smallest objects collapse first, and they subsequently merge
together to form larger objects. It then follows that the loss of
small-scale power modifies structure formation most severely at the
highest redshifts; in particular, the number of self-gravitating
objects at high redshift is reduced.  For example, in the context of
warm dark matter (WDM) models, \citet{BHO2001} have shown that a WDM
particle mass significantly below 1 keV would make it difficult to
form any DM halos at $z>7$ to host a high-redshift quasar.
Likewise, the highest-redshift $z=10$ galaxies detected in the {\it
  Hubble} deep fields behind strong lensing clusters would be hard to
explain in WDM models with this particle mass~\citep{Pacucci+2013}.
On the other hand, a particle mass near~1~keV would erase the
population of high-$z$ minihalos and delay the onset of structure
formation to commence with ACHs~\citep{Dayal+2017}. This would help create a large number of
metal-free ACHs, conducive to massive BH formation.

In the case of self-interacting DM
(SIDM;~\citealt{SpergelSteinhardt2000}), two other interesting effects
arise.  First, due to the self-interactions (compared to much weaker
purely gravitational interactions), the `gravothermal catastrophe'
sets in on a much shorter timescale, and can result in the
relativistic collapse of the core of a DM halo into a BH (see,
e.g. \citealt{KodaShapiro2011} for a detailed discussion).
\citet{BalbergShapiro2002} found that $\sim 10^6~\msun$ BHs in the
cores of massive ($\sim 10^{12}~\msun$) halos can form by $z\sim 10$ as
a result, although they used a cross section $\sigma=5~{\rm cm^2
  g^{-1}}$ which is now ruled out.  The interaction cross-section is
limited to $\sigma<0.6~{\rm cm^2 g^{-1}}$; this comes from the
displacement of the gas with respect to the centroid of the total mass
in the merging subcluster component of the Bullet Cluster, which shows
that unlike the baryons, the DM component has not been slowed down by
collisions~\citep{Randall+2008}.  A twist on the SIDM idea, however,
is that the bulk of DM is normal CDM, but a small fraction is SIDM
with a large cross-section.  This hybrid model avoids essentially any
constraint from the overall behavior of DM, and forming $\sim
10^6~\msun$ BHs by $z\sim 10$ appears feasible in such strong-SIDM
sub-component models~\citep{Pollack+2015,Choquette+2018}.

A second feature of SIDM is that a pre-existing BH can accrete
efficiently. This is because SIDM is presumed not to radiate, so the
Eddington limit does not apply, and because scattering allows a rapid
diffusive refilling of the loss-cone, and thus efficient accretion continues
even when the SIDM mean free path
is larger than the Bondi radius~\citep{Ostriker2000}.  This could help
growing massive BHs, starting from stellar-mass seed BHs formed in the
usual way (i.e. a remnant of a massive star), located at the dense
core of its DM halo.

\section{FUTURE OBSERVATIONAL DIAGNOSTICS}
\label{sec:obsdiag}
Existing $z\approx 6-7$ quasar observations only probe the most
massive SMBHs, representing the tip of the iceberg of the
high-redshift BH population. In order to better understand early BH
formation and growth, it will be important to characterize a much
wider mass range, and to go to higher redshifts. Fortunately, a
variety of planned or proposed instruments, such as \emph{JWST},
\emph{LISA}, and \emph{Lynx} will make this possible\footnote{see
  www.jwst.nasa.gov, www.lisamission.org and www.lynxobservatory.com.}.
In this section, we discuss the most promising observational probes
including both direct detections and indirect methods at high
redshift, as well as fossil evidence from BHs in the
local universe.

\subsection{Direct Observations} 

Current observations of SMBHs at $z\approx 6-7$ are unlikely to
contain information on SMBH seeding.  This is because, even for
massive seeds, SMBHs have grown many orders of magnitude in mass,
erasing the memory of seed formation and early accretion.
Furthermore, whatever combination of seeding and accretion produced
these rare SMBHs is likely to be a small unrepresentative fraction of
all the BHs born at much higher redshifts.  To distinguish seed
models, and to diagnose the IMF of early BHs at their birth and their
subsequent growth, it will be important to observe BHs near (or below)
the masses predicted for massive seeds, $\sim 10^5~\msun$, at
redshifts $z\gsim 10$ where they are expected to form. For reference,
we note that the flux from a $10^5~\msun$ BH at $z=10$, assuming that
it shines at the Eddington limit, and has the same spectral shape as
typical lower-$z$ quasar, would be $\sim$0.5 nJy in the near-IR
(observed at 1$\mu$m) and $\sim 3\times 10^{-19}~{\rm
  erg~s^{-1}~cm^{-2}}$ in the soft X-ray (observed at 1 keV) bands.
Directly detecting these BHs will be challenging, but should be
possible in Msec exposures with {\it JWST's} NIRCam,
and in the soft X-ray bands with a new, sensitive instrument such as
the planned {\it Lynx} telescope, which can reach a sensitivity of
$\sim 10^{-19}~{\rm erg~s^{-1}~cm^{-2}}$ in 4 Msec.  Searching areas
of strong gravitational lensing behind massive clusters could improve
these detection limits by an order of magnitude, although only in small
solid angles.
Once these
high-$z$ point sources are detected, the next question is whether
there is any clear signature of their origin.  Three distinctive
features of massive seed BHs born from pristine gas in ACHs are (i)
obscuration by an unusually large gas column, placing the bulk of
their emergent flux in the IR and X-ray bands, (ii) the lack of
any (or very little) metals in their spectra, and (iii) the lack of
an appreciable host galaxy.

With the sensitivities above, the IR and X-ray luminosity functions of
accreting high-redshift BHs at $z=10$ should be possible to measure
over 2-3 orders of magnitude in flux, which will help characterize the
BH population and constrain seeding and evolution models.
Stellar-mass seeds are generally thought to be much more abundant than
massive seeds: cosmological hydrodynamical simulations and
semi-analytic calculations find a cumulative Pop~III stellar density
of $\approx 3\times 10^5~\msun ~\mpc^{-3}$ forming by $z\approx 6$
\citep[see e.g.][]{Wise+2012, Visbal+2015}.  If the Pop~III IMF is
top-heavy as predicted by \cite{Hirano+2014}, a significant fraction
of this total Pop~III mass density will end up in BHs.  While the
abundance of massive seeds is highly uncertain, they are expected to
be much rarer.  For massive seeds created through strong LW feedback
from a neighboring galaxy, an extremely optimistic value of the
critical LW flux ($J_{\rm crit}\approx 100~J_{21}$), leads to a number
density of $\sim 10^{-5}~\mpc^{-3}$ \citep{Dijkstra+2014}.  Including
radiation from satellite galaxies in the same halo, and/or including
other effects mentioned in \S~\ref{subsec:warmgas}, the predictions
rise as high as $\sim 10^{-3}~\mpc^{-3}$.  This value is reached under
optimistic assumptions about (the lack of) metal enrichment in the
``synchronized halo pairs'' scenario \citep{Visbal+2014_sync}.
\citet{Wise+2019} find a similar value due to dynamical heating of
rare ACHs, although in their simulations this value only applies in
highly biased regions (overdense on large scales) and the global
average abundance is $\sim (10^{-6}-10^{-7})~\mpc^{-3}$.  In order to
reach even higher values requires unusual or ad-hoc assumptions.  For
example, if $J_{\rm crit}$ were 1-2 orders of magnitude lower than
expected, then the abundance could reach up to $\sim
10^{-1}~\mpc^{-3}$, corresponding to a large fraction ($\sim 10\%)$ of
{\it all} ACHs at $z\approx 10$ (see \citealt{Habouzit+2016} and
references therein).  These values, however, are still orders of
magnitude below the abundance of Pop III seeds.  For reference, we note
that the space density of $10^{-3}~\mpc^{-3}$ at $z\approx 10$ would
correspond to $\approx 1~{\rm arcmin^{-2}}$ per unit redshift, which
is sufficient to collect a statistical sample with IR/X-ray surveys.

Recent semi-analytic models \citep{RicarteNatarajan2018} made
predictions for the high-redshift luminosity functions for stellar and
massive seeds. At sufficiently high-redshift ($z \gsim 10$), the
luminosity functions diverge, and depend strongly on the seed model
implemented. In particular, the bright end is strongly impacted by the
seed mass: stellar BHs whose growth is capped both by the $\sim$
Eddington rate and by radiative feedback (assumed to maintain the
$M_\bullet-\sigma$ relation) cannot produce as massive and luminous
SMBHs by $z=10$ as the brightest massive seeds.  On the other hand,
the faint end is strongly affected by the abundance of seeds (with
stellar seeds producing a much higher number of quasars at, e.g., {\it
  Lynx}'s detection threshold). However, it must be kept in mind that
the assumed accretion properties of high-redshift BHs strongly impact
the LF. \cite{TanakaHaiman2009} find that the BH mass function depends
on a combination of accretion, duty cycle, and seeding fraction. These
degeneracies will likely make it difficult to characterize the
population of high-redshift BHs from the luminosity functions alone.

 \subsubsection{Gravitational Waves and Other Transients}

\begin{figure}[t]
\includegraphics[width=3in]{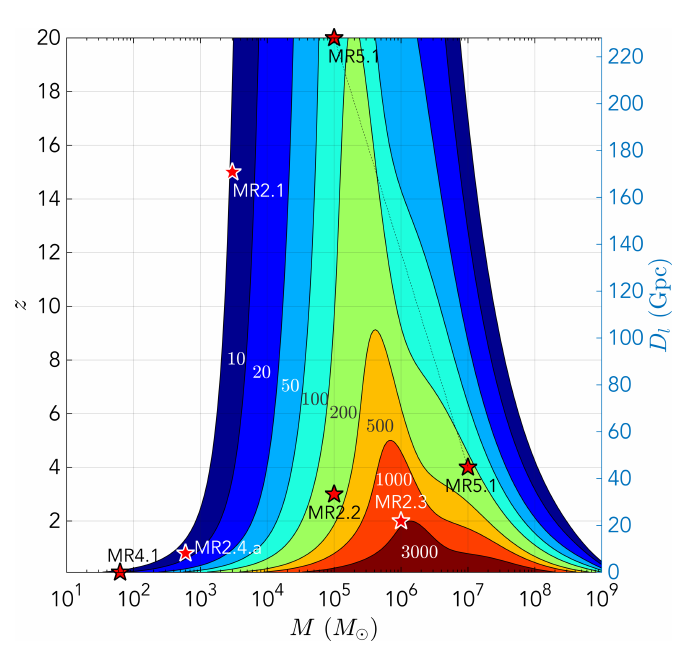}
\vspace{-2\baselineskip}
\caption{Contours of constant signal-to-noise for detection of
  gravitational waves from massive BH mergers by \emph{LISA}, as a
  function of redshift and binary mass.  Stars indicate reference
  binaries used to define Mission Requirements, which include the
  detection of $10^3~\msun$ binaries to $z=15$ (MR2.1) and the accurate
  characterization of the post-merger waveform from $10^5~\msun$
  binaries to $z=20$ (MR5.1).  Figure adapted from the LISA mission
  proposal approved by ESA \citep{LISA}.}
\label{fig:lisa}
\end{figure}

The IR and X-ray luminosity functions of BHs will inform us mainly
about gas accretion onto high-redshift BHs. This will be strongly
complemented by future gravitational wave measurements with the
space-based detector \emph{LISA}, planned for launch in
2034. \emph{LISA} will have the sensitivity to detect BH mergers for
masses from $\sim 10^{4-7}~\msun$ out to redshifts beyond
$z\approx20$ (see \textbf{Figure~\ref{fig:lisa}}). \emph{Simultaneously
  probing BH mergers and accretion will give us a complete picture of
  BH growth in the early Universe.}

Theoretical predictions of \emph{LISA} event rates for merging massive
BHs have been derived from numerous semi-analytical models, as well as
more recently from cosmological galaxy evolution simulations.  The
predictions span a wide range, from no detectable events in the most
pessimistic case, up to several tens of events per year per unit
redshift at $z=10$, and a few events per year even at $z\gsim 15$ in
the most optimistic models (for recent predictions, see,
e.g. \citealt{Klein+2016,Hartwig+2018,RicarteNatarajan2018}; for
earlier studies that included the $z\gsim 10$ universe, see,
e.g. \citealt{Sesana+2007,TanakaHaiman2009} and references therein).
The most pessimistic scenario is for stellar-mass BHs to stay below
{\it LISA}'s detection threshold, and for massive seed BHs to form
inefficiently or to be unable to promptly merge when their hosts merge.
The largest rates arise when abundant stellar-mass BH seeds are
assumed to grow above {\it LISA}'s $\gsim 10^{3-4}~\msun$ detection
threshold by $z\sim 10$, or in models where massive seeds are assumed
to form efficiently in a large fraction of ACHs.  Aside from the total
event rates, the mass spectrum is expected to depend on the seeding,
and should help disentangle models.  It has been shown that
\emph{LISA} data can indeed distinguish between a wide variety of
models which include different prescriptions for seed formation,
feedback, accretion efficiency and accretion geometry
\citep{Sesana+2011}.  \emph{LISA} may also be able to probe SMS star
formation itself, by measuring mergers of the remnants of binary SMSs
\citep{Hartwig+2018} or directly detecting their formation process,
either via the ``burst'' accompanying the collapse of a rotating SMS
into a singe massive BH~\citep{SaijoHawke2009,Shibata+2016}, or the
breakup of such an SMS producing two massive BHs, which subsequently
inspiral and merge~\citep{Reisswig+2013}.  

While \emph{LISA} will
probe a BH mass range important for probing massive seeds,
significantly smaller stellar seed mergers cannot be detected. The
proposed interferometry mission \emph{DECIGO}~\citep{Decigo}, 
as well as ongoing and future experiments based on quantum interferometry with cold atoms,
such as the terrestrial MAGIS-100~\citep{MAGIS-100} or the AEDGE satellite~\citep{AEDGE}
would fill in the gap between LIGO and \emph{LISA}.  This is warranted especially
since the LIGO discoveries have already uncovered stellar-mass BHs more massive than had
been expected (currently up to $85~\msun$; \citealt{GW170729}), raising the possibility that the
mass function extends to even higher masses.

In addition to gravitational waves, there are several other transient
observables which will be important for constraining the high-redshift
BH population. Surveys with \emph{JWST} or \emph{WFIRST} will have the
ability to see the potentially extremely luminous pair-instability
supernovae (PISNe) from Pop~III stars with masses of $M_\star \sim 140
- 260~\msun$ to redshifts as high as $z\sim 30$~\citep{Kasen+2011}.
Less massive but potentially much more numerous PISNe from lower-mass
($M_\star \sim 90 - 140~\msun$) Pop~III stars could still be visible
to $z\sim 10$~\citep{Smidt+2015}.  Observing these explosions would
put constraints on the abundance and IMF of Pop~III stars (and thus
the prevalence of light BH seeds).  However, \cite{Hartwig+2018_JWST}
pointed out that the identification of PISNe and differentiation from
other sources that could have a similar photometric signature, such as
AGN or high-z galaxies, are very challenging; in fact, the optimal
strategy would require at least 50,000 different fields of view with
an exposure time of $\approx 600$ s for each field to detect one PISN
at $z<7.5$.  It may also be possible to directly detect the explosions
of SMSs, yielding information on massive seeds~\citep{Whalen+2013},
although these explosions may occur only for non-rotating SMSs in a
narrow mass-range near $55,000~\msun$~\citep{Chen+2014}.  Tidal
disruption events of stars formed in the dense accretion disk around a
newly born massive seed BH could occur, and the bright X-ray and radio
emission from the jets from these events could be detected out to
$z\sim 20$ and diagnosed by their long duration of $\sim
10^{5-6}(1+z)$ s~\citep{KashiyamaInayoshi2016}.

\subsubsection{Spectral Signatures}

Another approach is to identify accretion onto massive seeds shortly
after their formation from their spectra.  Due to the monolithic
collapse of hydrogen gas required to form massive seeds, a newly
formed massive BH seed should still be buried in a very large column
density of gas.  The impact of the corresponding obscuration has been
investigated in one-dimensional radiation-hydrodynamic simulations
\citep{Pacucci+2015a}. These simulations find strong X-ray (0.1-100
keV) and submm/IR (1-100$~\mu {\rm m}$) emission, with a large gap in
the spectrum at wavelengths below the Ly$\alpha$ line due to
absorption and reprocessing to lower-wavelength radiation.  Using
similar simulations, \cite{Pacucci+2016} and \cite{Natarajan+2017}
have predicted that $\emph{JWST}$ will have the sensitivity to observe
newly formed massive seeds and proposed a set of color-color cuts
which can be used to distinguish these objects from high-redshift
galaxies and low-redshift contaminants, based on their very red IR
colors.  Additionally, the combined spectra of growing BH seeds and
their host galaxies were predicted in \cite{Valiante+2018}. They find
that accreting high-redshift seeds will be detectable with
\emph{JWST}, but that it will be difficult to distinguish stellar
versus massive BH seeds.  We also note that theoretical predictions
for the spectra of SMS (before BH formation) suggest that they could
also be detected by $\emph{JWST}$~\citep{Hosokawa+2013,Surace2018}.

An additional spectral signature of massive seeds was identified by
\cite{Dijkstra+2016}.  The high column density of pristine gas and the
absence of dust lead to numerous Ly$\alpha$ scatterings. This creates
ideal conditions for pumping of the 2$p$ level of atomic hydrogen,
which results in 3 cm (rest frame) maser emission from the
$2p\rightarrow2s$ transition. This bright line could be observed by
the \emph{Square Kilometer Array (SKA)}.

Observing the characteristic spectrum of a massive BH seed buried by a
very high column of gas and/or producing 3 cm maser emission could
provide a smoking gun for the presence of a massive seed BH. However,
one major challenge is that these observations require the seed BH to
have formed very recently ($z\lsim 6-10$). 
The abundance of massive seeds therefore
needs to be close to the highest predictions, to have a chance of
catching their formation within a narrow time window.

A related spectral diagnostic is the Ly$\alpha$ line emitted from an
accreting massive seed BH. \cite{DijkstraGronkeSobral2016} model the
Ly$\alpha$ line profile in this scenario and find line offsets and
widths exceeding ${\sim}1000~{\rm km~s^{-1}}$. This is higher than
typical Ly$\alpha$ emitting galaxies and thus may be an interesting
observational discriminant.

Another interesting spectral signature is the presence of
strong HeII recombination lines (with the strongest line at rest-frame
wavelength of 1640\AA). This line is expected, and should be
detectable by {\it JWST} to $z\approx 10$, both from Pop~III stars and
accreting massive BHs~\citep{Tumlinson+2000,Oh+2001}. An important
difference is that the Pop~III stellar lines are expected to be
narrow, because of the absence of strong Wolf-Rayet type winds, while
lines produced in the vicinity of an accreting BH should generally be
much broader.  The claimed detection~\citep{Sobral+2015} in the bright
$z=6.6$ Ly$\alpha$ emitter CR7 of a strong HeII 1640\AA\ line without
corresponding metal lines was interpreted by several authors as
evidence either for Pop III stars or for a massive seed BH (although
the narrow observed width, which favors a stellar origin, received
very little attention).  Although a subsequent analysis found any HeII
line much weaker, and the metallicity higher than originally
believed~\citep{Shibuya+2018}, CR7 served as an intriguing case study.
Future detection of a similarly strong, but broad HeII line without
metals could signal an SMS that recently collapsed to a BH. A strong
and narrow HeII line would likewise indicate a significant population
of Pop~III stars, which are predicted to form concurrently with the
massive seed BH in models where the seed is forming in a galaxy
significantly above the atomic-cooling threshold ~\citep{Inayoshi+2018}.

\subsubsection{BH Host Galaxies}

Low-$z$ SMBHs follow tight correlations with their hosts, which are
widely believed to result from feedback
processes~\citep{KormendyHo2013}. Therefore seed BHs will not
generally be born on this relation, but will rather settle onto the
relation over time (typically "from above", with BH masses initially
above the relation).  The high-$z$ evolution of the low-mass end of
well-known BH-host relations should therefore contain information on
the earliest seeds and their
growth~(e.g. \citealt{Volonteri+2009,Pacucci+2018}).

More specifically, a discussed above, the standard formation scenario for massive seeds
requires ACHs with no prior star formation and a suppressed ${\rm
  H_2}$ abundance, to avoid metal cooling and gas fragmentation. Thus,
DM halos hosting massive seeds will initially have an ``obese-BH
galaxy" stage~\citep{Agarwal+2013}, where there is either no
appreciable host galaxy at all, or else BH mass strongly dominates
over the stellar mass, and the accretion onto the BH outshines the
stellar component of the host galaxy.  An important question is how
long a typical newborn massive BH will stay unusually obese, before
its host grows more massive.  This was investigated by
\cite{VisbalHaiman2018}, who tracked a large number of ACHs in a
cosmological N-body simulation.  It was found that before they can
grow by an order of magnitude in mass, essentially all massive seeds
have BH mass-to-stellar mass ratios significantly higher than
stellar-mass seeds that have grown to the same mass.  A promising
strategy to diagnose massive seeds will therefore be to measure their
accretion rates with next generation X-ray telescopes such as
\emph{Lynx} and compare their host galaxies' stellar properties using
either \emph{JWST} or thirty meter-class ground-based telescopes.
Strong X-ray sources which are not accompanied by any detectable
stellar host component will be strong massive seed candidates. Even in
a few rare cases where the ACH host of a massive seed promptly merges
with a nearby massive galaxy, \cite{VisbalHaiman2018} found that the
ACH remains offset from this galaxy at a distance (few kpc) that can
be resolved with \emph{JWST} or \emph{Lynx}.  {\em In short, the tell-tale
evidence for a massive seed BH is the absence of a host galaxy, or
a galaxy offset by a few kpc.}   In a variant of this picture, the X-rays from 
the growing massive seed BH trigger the formation of a small cluster of 
Pop~III stars, which could lead to characteristic spectral signatures~\citep{Barrow+2018}.

\subsection{Indirect Observations}

In addition to direct detection of individual objects, there are
promising indirect observations which can shed light on the
high-redshift BH population. One possible approach is to measure the
heating and partial ionization of the IGM caused by X-rays produced
during BH accretion. When an X-ray ionizes a hydrogen or helium atom,
a large amount of energy is imparted to the escaping electron. This
energetic electron then interacts with the surrounding gas leading to
heating and secondary ionizations. For a nearly neutral IGM, a
significant fraction of X-ray energy goes into both heating and
ionization \citep[for detailed calculations
  see][]{2010MNRAS.404.1869F}. As the ionization fraction increases, a
larger percentage of X-ray energy is deposited as heat, making X-rays
less efficient than UV photons at ionizing the IGM to an ionization
fraction close to unity.

One probe of such early ``preionization'' is via CMB temperature and
polarization anisotropies.  The optical depth to electron scattering,
$\tau_e$, provides an integral constraint on the ionization of the
IGM.  Due to the increased density of the Universe at high-redshift,
even a relatively small pre-reionization can contribute significantly
to $\tau_e$. Thus, CMB observations can be utilized to probe the total
accretion of seed BHs in the early Universe \citep{RicottiOstriker04,
  Madau+2004, Ricotti+2005}. Since $\tau_e$ is only one number, it is
not possible to disentangle the quantities impacting a BH-driven
pre-reionization (e.g. spectral properties, abundance, duty cycle,
radiative efficiency, etc), $\tau_e$ measurements only yield upper
limits on early preionization.  With reasonable (but uncertain)
assumptions, \cite{Visbal+2015} show that if most Pop~III remnant BHs
accrete radiatively efficiently near the Eddington limit, the
resulting pre-reionization would violate the electron optical depth
constraints from \emph{Planck}.

In principle, the shape of the large angular-scale CMB polarization
power spectrum contains information on the evolution of the ionized
fraction $x_e(z)$ that goes beyond $\tau_e$
\citep{Holder+2003,Kaplinghat+2003}.  The above studies have shown
that two different reionization histories could produce the same value
of $\tau_e$, but could predict different shapes of a polarization
``bump'', which could be distinguished at high significance in {\em
  Planck} data.  In particular, a long period of partial ionization
extending out to high redshift would shift power towards smaller
angular scales.  \citet{Heinrich+2017} and \citet{Miranda+2017} have
recently fit parametric reionization models to the public 2015 {\em
  Planck} LFI polarization data.  Interestingly, they concluded that
this analysis favors reionization histories with somewhat elevated
$\tau_e$ values ($\tau_e \approx 0.08$, compared to the value
$\tau_e\approx 0.05$ obtained by assuming prescribed, sharp ionization
histories; \citealt{Planck2018}), and a shape that mimics a tail of
partial ionization extending to high redshift (with $10-20\%$
ionization at $z=15-20$), but a relatively sudden full reionization at
$z=6-7$.  A similar feature, however, could also be produced by a
high-$z$ tail of partial ionization from early stars~\citep{Ahn+2012}.
Future, more sensitive CMB polarization experiments (aiming to detect
signatures of primordial B-modes from inflation), as well as the
analysis of {\it Planck}'s HFI data, can provide better measurements
of any high-$z$ preionization, e.g., the Stage-4 ground-based CMB
experiment (CMB-S4; \citealt{CMB-S4_2019}) and LiteBIRD
\citep{Matsumura+2014,Hazumi+2019}.

The ionization and thermal history of the high-redshift IGM can also
be probed by radio observations of the redshifted 21cm line of neutral
hydrogen \citep{Furlanetto+2006, PritchardLoeb2012}.  These
observations can break degeneracies between any preionization by
softer (UV) radiation from stars, vs. harder (X-ray) radiation from
accreting BHs. The latter is expected to produce a much smoother
spatial morphology, than the `swiss-cheese' ionization structure
produced by UV photons~\citep{Zhang+2007, MesingerFerraraSpergel2013}.
Interferometers such as \emph{HERA} are designed to measure spatial
fluctuations in the emission/absorption signal, while ``global''
experiments such as \emph{EDGES} seek to measure the signal averaged
over the entire sky. One of the main predicted features of the global
21cm signal is an upturn in brightness temperature corresponding to
early X-ray heating from accreting BHs or stellar remnants.
\cite{Tanaka+2016} have argued that if Pop~III stars serve as the
seeds of the first SMBHs, this should result in early X-ray heating of
the IGM observable in the 21cm signal at $z>20$. If this signal is not
observed it implies either that SMBHs formed later with massive seeds,
SMBHs only occur in a small fraction of galaxies at high redshift, or
accreting BH seeds emit significantly fewer or softer X-rays than
would be expected based on low-redshift observations \citep[the impact
  of a softer X-ray spectrum is computed
  in][]{FialkovBarkanaVisbal2014}.

Recently, the \emph{EDGES} experiment published the first detection of
the global signal \citep{Bowman+2018}, which shows a very strong
absorption feature at $z\approx20$ followed by rapid heating at
$z\approx16$. The large depth of this feature is puzzling, and to date
has no physically compelling interpretation without exotic
assumptions.  However, if additional observations confirm that the
feature (even if with a more realistic, reduced amplitude) is
cosmological, and not a residual from foreground or instrumental
modeling, this would contain interesting new information on the
formation of the first SMBHs.

Another indirect method to probe the first SMBHs is via the unresolved
X-ray background. Using again the Soltan-Paczy\'nski argument
\citep{Soltan1982}, it is possible to put constraints on the total
amount of high-redshift BH accretion. \citet{Dijkstra+2004} have shown
that this strongly limits the number density of faint, undetected BHs
at high redshift; in particular, these BHs cannot contribute
significantly to reionization without overproducing the unresolved
X-ray background from \emph{Chandra}. In a similar analysis,
\citet{Salvaterra+2012} derived limits of $\rho_\bullet \lsim 0.7
\times 10^{4}~\msun~\mpc^{-3}$ on the BH mass density at
$z>5$. However, these constraints can be significantly weakened by
varying assumptions about the spectral properties of the BH emission
\citep{Cappelluti+2017}. Due to the high abundance of BHs, seed models
based on Pop~III stars which accrete efficiently are the most likely
to be ruled out by X-ray background measurements
\citep{RicarteNatarajan2018}.  In a related analysis, one can look for
X-ray emission from BHs by stacking the X-ray observations of known
optically detected galaxies.  \citet{Treister+2013} has stacked {\it
  Chandra} data for galaxies found in the Hubble Deep Fields, and has
derived upper limits on the average X-ray luminosities of BHs in
$z=6-8$ galaxies.  These limits can be translated to a mass density
with the same Soltan-Paczynski approach as above, and yield
$\rho_\bullet \lsim 10^{3}~\msun~\mpc^{-3}$, implying (with further
assumptions about the frequency of occurrence and spectrum of BHs in
individual galaxies) an upper limit of $\sim 3\times 10^6~\msun$ on
the mass of the nuclear SMBH in a typical $z\approx 6$ galaxy.

\subsection{Fossil Evidence in the Local Universe} 

Local observations of IMBHs also have the potential to constrain
high-redshift BHs (see the review by Greene et al. in this issue). Due
to their relatively quiet merger and star formation histories, dwarf
galaxies may share similarities with galaxies in the high-redshift
universe, and retain information on BH seeding mechanisms.
Theoretical predictions based on Monte Carlo halo merger trees suggest
that the BH occupation fraction in these galaxies depends on the
high-$z$ seed properties~\citep{Volonteri+2008,vanWassenhove+2010}.
Because Pop~III stars are expected to be much more abundant than
massive seeds, models with stellar-mass BHs formed from these stars
tend to have higher occupation fractions. However, if these seeds have
not grown much beyond their initial masses, they will be very
difficult to observe.  It has also been suggested that the low-$z$
$M_\bullet - \sigma$ relation could be sensitive to SMBH seeding
\citep{Volonteri+2009}, with massive seeds leading to a flatter
low-mass end (i.e.~BHs with masses above relation found at large
$\sigma$).  However, subsequent work suggests that this relation
depends more strongly on the accretion model than the seeding
prescription \citep{RicarteNatarajan2018}.

BHs have been discovered in a number of local dwarf galaxies, many
with masses below $10^6~\msun$ \citep{Greene+2007,Barth+2008,
Reines+2013,Moran+2014,Baldassare+2018,Chilingarian+2018}.  
The lowest-mass confirmed BH resides in the galaxy RGG 118 and 
has an estimated mass of $\sim 5 \times10^{4}~\msun$ \citep{Baldassare+2015}.  
Additionally, there are a number of dynamical BH candidates in globular clusters with similar
masses \citep[e.g.][]{Lutzgendorf+2013}, but their lack of X-ray or
radio emission signatures makes it difficult to confirm their
existence \citep{Strader+2012,Wrobel+2015}. For a comprehensive review
on local observations of IMBHs see \cite{Mezcua+2017}.

While these observations are intriguing, the current BH occupation
fraction is not well enough constrained to draw strong conclusions
about the SMBH seeding mechanism.  The fact that IMBHs have only been
found with $M_\bullet > 10^4~\msun$ along with the observed flattening
of the $M_\bullet - \sigma$ relation at low masses \citep[e.g. see
  figure 10 in][]{Mezcua+2017} may hint at the importance of massive
seeds.  However, this inference should be taken with caution.  A
population of lower-mass IMBHs may have escaped detection to date,
or stellar-mass seeds formed early on in dwarf galaxies could have
grown substantially over time.

One potential challenge in studying massive seeds with BHs in dwarf
galaxies is that their abundance is expected to be exceedingly
low. As discussed above, if massive seed formation relies on full
photodissociation of H$_2$, the required LW flux is very
high, resulting in an abundance perhaps as low as $\approx
1~{\rm Gpc}^{-3}$, just matching that of the bright $z\approx
6$ quasars \citep{Dijkstra+2014}. In this case, we could not expect to
find such a seed in the local universe.  Other formation channels can
create many more massive BHs, but, as discussed above, the
overall number density is still quite low. Taking the
synchronized halo pairs scenario \citep{Visbal+2014_sync} as an
example, and assuming that massive seed formation is shut off by metal
pollution at relatively high redshift, the number density of $\approx
10^{-3}~{\rm Mpc^{-3}}$ would represent roughly the total abundance of
massive seeds created throughout the entire universe. For this  
density, the fraction of dwarf galaxies containing massive seed BH
fossils would be at most $\sim 10^{-2}$.  Thus, if such massive BHs
are found in dwarf galaxies, and their formation can be securely
placed to high-$z$ (e.g. from the old age of the host's stellar
population), then this will imply either that the abundance of these
seeds is much higher than expected, or that some
lower-mass BHs grew efficiently. A recent study did find much larger occupation fractions: the majority of halos somewhat above the ACH limit, followed in simulations by \citet{Bellovary+2019}, 
were found to form massive seed BHs at $z\sim 15-20$. These simulations included treatments of ${\rm H_2}$ chemistry and metal enrichment, but did not resolve the small-scale collapse dynamics or the history of gas (and stars) in the earlier stages of these halos (including minihalos). Higher-resolution simulations are required to assess whether massive BHs could indeed be produced so often.
      
 \section{CONCLUSIONS}
 \label{sec:conc}

 In an ideal world, this review would describe a robust prediction for
 the evolving BH mass function -- together with the corresponding LF
 -- as a function of redshift.  The state of the field is far from
 this goal.  However, there are well-developed ideas for the formation
 of BHs in the range from stellar masses ($\sim 10-100~\msun$) to BHs
 as massive as $\sim 10^6~\msun$, in early protogalaxies.  This
 review has therefore focused mostly on describing these ideas.

 In analogy with the stellar IMF, there is an IMF of BHs, describing
 the distribution of BH masses at their birth.  It is nearly certain
 that this IMF has a strong peak at stellar masses ($\sim
 10-100~\msun$), because the first generation of metal-poor stars were
 likely massive, and often (perhaps in the majority of cases) leaving
 behind BH remnants.  It is also very likely that BHs more massive
 than these were formed by the processes described in this review, and
 summarized in \textbf{Figure~\ref{fig:rees}}.  It is often useful to
 contrast these ``massive seeds'' with the ``light'' stellar-mass BHs,
 for the purposes of illustrating extreme scenarios.  However, the
 high-$z$ BH IMF more likely covers the full range of $10 \lsim
 M_\bullet / \msun \lsim 10^6$.

 The most massive BHs with $M_\bullet \approx 10^{5-6}~\msun$ require
 very special conditions -- pristine metal free gas, with a
 suppressed ${\rm H_2}$ abundance in relatively large ``atomic
 cooling'' halos with masses of $\gsim$ several $\times~10^7~\msun$ --
 producing rapid, monolithic infall of $10^6~\msun$ of gas.  This will
 be realized only in rare, highly biased regions of the universe,
 containing ACHs that are exposed to unusually intense Lyman-Werner
 radiations, have unusually and rapid assembly histories, reside in
 regions with unusually high baryonic streaming velocity
 -- or some combination of these three.  On the other hand, we expect
 that IMBHs ($\sim 10^2-10^4~\msun$) should form in larger abundance,
 in regions which are biased, but fail to fully meet the above
 conditions for ``massive seed BH'' formation. For example, IMBHs
 could arise in overdense regions where H$_2$ cooling is suppressed
 but not fully disabled (so that an embryonic supermassive star does
 not grow to the massive BH regime), or where gas condensing in ACHs
 had some modest prior star-formation and metal-enrichment (so that
 the gas undergoes some fragmentation rather than accreting onto a
 sole central protostar).  We generally expect that criteria to form
 increasingly massive BHs are increasingly harder to realize, and
 therefore BH IMF will span the full range of $\sim 10-10^6~\msun$,
 monotonically declining with mass in this range.

 The subsequent growth of these seed BHs due to accretion and mergers
 will determine the evolution of their mass-function over cosmic time.
 Some ``lucky'' BHs will find themselves in the dense cores of growing
 galaxies, and will be able to accrete efficiently, but the majority,
 especially at early times, will fail to grow.  This is because both
 gas and BHs can be relatively easily ejected from the shallow
 potential wells of the first ``microgalaxies'', via radiative
 processes and supernova explosions, and via merger-induced
 gravitational recoil, respectively.  The BHs hosted by the most
 massive and most rapidly growing host halos will have the best chance
 to grow efficiently with a high duty-cycle, and to evolve into the
 quasars observed at $z\sim 6-7$.  However, feedback from the SMBHs
 and accompanying stars is likely important in regulating their growth
 even in large ($10^{10-12}~\msun$) galaxies. This feedback, in the
 form of SNe, as well as radiation and mechanical energy from strong
 outflows driven by the SMBH itself,
 requires simulations spanning the full dynamical range
 from where this radiation and outflow is generated, to galactic
 scales, where feedback operates over cosmological timescales.
 Understanding the details of this is the challenging frontier in
 numerical simulations.

 Finally, the admittedly poor state of our ab-initio understanding of
 the early BH population is an opportunity for observations.
 Measuring the luminosity function of quasars in optical and X-ray
 bands, 1-2 orders of magnitude below the present limits, should be
 feasible, and should yield strong constraints on assembly models,
 especially at $z\gsim 10$.  These constraints will likely suffer from
 degeneracies between seeding and growth, but these degeneracies
 should be lifted by the direct GW detections of merging BHs in the
 $\sim$10$^4$--$10^6~\msun$ range by {\it LISA}, and by combinations
 of indirect probes of early BHs via their imprint on the cosmic 21cm
 signal and on large-scale CMB polarization anisotropies.

 In order to probe the specific seed models, it will be necessary to
 detect BHs with masses of $\lsim 10^5~\msun$ at redshifts $z\gsim
 10$, because the newly-born BHs will likely lose the memory of their
 birth by the time they grow well above $\sim 10^5~\msun$ and are
 incorporated into more massive host galaxies.  While challenging,
 this should be feasible in the future in ultra-deep observations with
 X-ray telescopes such as {\em Lynx} and in the optical/IR with {\em
   JWST} and next-generation, 30m-class optical/IR
 telescopes. Unusually massive seed BHs, if caught near (within $\sim
 100$Myr) the time of their formation, should also have smoking-gun
 signatures measurable with these instruments, due to the large
 obscuring column of gas still present in their host halos, as well as
 due to the exceedingly low mass and small size of their hosts.

\section*{DISCLOSURE STATEMENT}
The authors are not aware of any affiliations, memberships, funding,
or financial holdings that might be perceived as affecting the
objectivity of this review.

\section*{ACKNOWLEDGMENTS}
We thank Yacine Ali-Ha\"{\i}moud, Eduardo Ba{\~n}ados, Jillian
Bellovary, Mitch Begelman, Xiaohui Fan, Katie Freese, Jenny Greene,
Tilman Hartwig, Takashi Hosokawa, Yoshiki Matsuoka, Priya Natarajan,
Kazu Omukai, Fabio Pacucci, Jim Stone, Hiromichi Tagawa, Fabio Vito,
Marta Volonteri, Feige Wang, Jemma Wolcott-Green, Tyrone Woods, Jinyi
Yang and Naoki Yoshida for numerous useful comments on an earlier
version of this manuscript,
and Kyungjin Ahn, Tiziana di Matteo and Yoshiki Matsuoka for
permission to adopt figures from their published work.  KI
acknowledges recent support from the National Science Foundation of
China (11721303, 11950410493) and the National Key R\&D Program of
China (2016YFA0400702).  ZH acknowledges support from NASA through
grants NNX15AB19G and NNX17AL82G and from the National Science
Foundation through grant 1715661.

\bibliographystyle{ar-style2}

\section*{Supplemental Material}

\begin{center}
  \textcolor{cyan}{\bf List of the 203 quasars at $z\geq6$, shown in Figure 1 of the manuscript.}
  \begin{tabular}{@{}l|c|c|c|c@{}}
\hline
NAME                         & Redshift  &   ${\rm M}_{1450}$    &   Mass$^{(a)}$	 &     Reference\\
                             &           &                      &   ${\rm M_\odot}$ &          \\
\hline
ULAS J134208.10+092838.61	&   7.54     &      -26.76     &      7.80e+08     & 	 Banados et al. 2018$^{(b)}$     \\
ULAS J1120+0641			&   7.0842     &      -26.63     &      1.85e+09     & 	 Banados et al. 2016$^{(c)}$     \\
HSC J124353.93 +010038.5 	&   7.07     &      -24.13     &      1.85e+08     & 	 Matsuoka et al. 2019a     \\
DELS J003836.10-152723.6 	&   7.021     &      -27.1     &      2.85e+09     & 	 Wang et al. 2018b     \\
DES J025216.64-050331.8		&   7.02     &      -26.50     &      1.64e+09     & 	 Yang et al. 2019     \\
SHELLQ J235646.33+001747.3	&   7.01     &      -25.31     &      5.49e+08     & 	 Matsuoka et al. 2019b     \\
SHELLQ J160953.03+532821.0	&   6.92     &      -22.75     &      5.20e+07     & 	 Matsuoka et al. 2019b     \\
DELS J083946.88+390011.5 	&   6.905     &      -26.29     &      1.35e+09     & 	 Wang et al. 2018b     \\
VIK J2348-3054			&   6.9018     &      -25.80     &      8.63e+08     & 	 Banados et al. 2016     \\
SHELLQ J2210+0304		&   6.9     &      -24.44     &      2.46e+08     & 	 Matsuoka et al. 2018b     \\
DES J221100.60-632055.8		&   6.88     &      -25.10     &      4.53e+08     & 	 Yang et al. 2019     \\
DES J024655.90-521949.9		&   6.87     &      -25.35     &      5.70e+08     & 	 Yang et al. 2019     \\
HSC J1205-0000 			&   6.85     &      -24.98     &      4.05e+08     & 	 Banados et al. 2016     \\
VDES J0020-3653   		&   6.834     &      -26.92     &      2.42e+09     & 	 Reed et al. 2019     \\
DES J031941.66-100846.0		&   6.83     &      -25.71     &      7.94e+08     & 	 Yang et al. 2019     \\
SHELLQ J011257.84+011042.4	&   6.82     &      -24.07     &      1.75e+08     & 	 Matsuoka et al. 2019b     \\
DELS J041128.63-090749.8  	&   6.81     &      -26.61     &      1.82e+09     & 	 Wang et al. 2018b     \\
SHELLQ J1429-0104 		&   6.8     &      -23.00     &      6.55e+07     & 	 Matsuoka et al. 2018a     \\
VIK J0109-3047			&   6.7909     &      -25.64     &      7.45e+08     & 	 Banados et al. 2016     \\
SHELLQ J161207.12+555919.2	&   6.78     &      -23.02     &      6.67e+07     & 	 Matsuoka et al. 2019b     \\
DELS J082931.97+411740.4  	&   6.768     &      -26.36     &      1.44e+09     & 	 Wang et al. 2018b     \\
DELS J110421.59+213428.8  	&   6.74     &      -26.67     &      1.92e+09     & 	 Wang et al. 2018b     \\
VDES J0244-5008   		&   6.724     &      -26.72     &      2.01e+09     & 	 Reed et al. 2019     \\
DELS J091013.63+165629.8  	&   6.72     &      -25.57     &      6.98e+08     & 	 Wang et al. 2018b     \\
SHELLQ J134400.87+012827.8	&   6.72     &      -23.46     &      1.00e+08     & 	 Matsuoka et al. 2019b     \\
SHELLQ J0213-0626		&   6.72     &      -25.24     &      5.15e+08     & 	 Matsuoka et al. 2018b     \\
DELS J083737.84+492900.4  	&   6.710     &      -26.42     &      1.52e+09     & 	 Wang et al. 2018b     \\
SHELLQ J000142.54+000057.5	&   6.69     &      -24.49     &      2.58e+08     & 	 Matsuoka et al. 2019b     \\
SHELLQ J123137.77+005230.3	&   6.69     &      -24.39     &      2.35e+08     & 	 Matsuoka et al. 2019b     \\
PSO J338.2298+29.5089		&   6.658     &      -26.14     &      1.18e+09     & 	 Banados et al. 2016     \\
DELS J121627.58+451910.7  	&   6.654     &      -25.58     &      7.05e+08     & 	 Wang et al. 2018b     \\
VHS  J210219.22-145854.0  	&   6.648     &      -25.50     &      6.55e+08     & 	 Wang et al. 2018b     \\
DELS J091054.53-041406.8  	&   6.63     &      -26.36     &      1.44e+09     & 	 Wang et al. 2018b     \\
DELS J104819.09-010940.21	&   6.63     &      -25.95     &      9.91e+08     & 	 Wang et al. 2017     \\
PSO J006.1240+39.2219		&   6.621     &      -25.94     &      9.82e+08     & 	 Tang et al. 2017     \\
VIK J0305-3150			&   6.6145     &      -26.18     &      1.22e+09     & 	 Banados et al. 2016     \\
SHELLQ J0923+0402		&   6.6     &      -26.18     &      1.22e+09     & 	 Matsuoka et al. 2018b     \\
PSO J323.1382+12.2986  		&   6.5881     &      -27.06     &      2.75e+09     & 	 Mazzucchelli et al. 2017     \\
PSO J231.6576-20.8335		&   6.5864     &      -27.14     &      2.96e+09     & 	 Mazzucchelli et al. 2017     \\
DELS J070626.39+292105.5  	&   6.583     &      -27.51     &      4.17e+09     & 	 Wang et al. 2018b     \\
DELS J113508.93+501133.0  	&   6.583     &      -26.19     &      1.23e+09     & 	 Wang et al. 2018b     \\
SHELLQ J0921+0007		&   6.56     &      -24.79     &      3.40e+08     & 	 Matsuoka et al. 2018b     \\
PSO J036.5078+03.0498		&   6.5412     &      -27.33     &      3.53e+09     & 	 Banados et al. 2016     \\
VDES J0224-4711   		&   6.526     &      -26.94     &      2.46e+09     & 	 Reed et al. 2019     \\
J043947.08+163415.7		&   6.51     &      ---     &      4.29e+08     & 	 Fan et al. 2019$^{(d)}$     \\
PSO J167.6415-13.4960		&   6.508     &      -25.62     &      7.31e+08     & 	 Banados et al. 2016     \\
\hline
\end{tabular}
\end{center}

\newpage
\begin{center}
\begin{tabular}{@{}l|c|c|c|c@{}}
\hline
NAME                         & Redshift  &   ${\rm M}_{1450}$    &   Mass$^{(a)}$	 &     Reference\\
                             &           &                      &   ${\rm M_\odot}$ &          \\
\hline
SHELLQ J1545+4232		&   6.50     &      -24.15     &      1.88e+08     & 	 Matsuoka et al. 2018b     \\
SHELLQ J135012.04-002705.2	&   6.49     &      -24.38     &      2.33e+08     & 	 Matsuoka et al. 2019b     \\
PSO J247.2970+24.1277		&   6.476     &      -26.53     &      1.69e+09     & 	 Mazzucchelli et al. 2017     \\
DES J213110.29-435902.5		&   6.45     &      -24.93     &      3.87e+08     & 	 Yang et al. 2019     \\
PSO J261.0364+19.0286		&   6.44     &      -25.69     &      7.80e+08     & 	 Mazzucchelli et al. 2017     \\
PSO J183.1124+05.0926		&   6.4386     &      -26.99     &      2.58e+09     & 	 Mazzucchelli et al. 2017     \\
VIK J2318–3113			&   6.4435     &      -26.11     &      1.15e+09     & 	 Decarli et al. 2018            \\
CFHQS J0210-0456		&   6.4323     &      -24.53     &      2.68e+08     & 	 Banados et al. 2016     \\
PSO J011.3899+09.0325		&   6.42     &      -25.94     &      9.82e+08     & 	 Mazzucchelli et al. 2017     \\
SDSS J1148+5251			&   6.4189     &      -27.62     &      4.61e+09     & 	 Banados et al. 2016     \\
CFHQS J2329-0301		&   6.417     &      -25.25     &      5.20e+08     & 	 Banados et al. 2016     \\
DES J021638.85-522620.6		&   6.41     &      -25.19     &      4.92e+08     & 	 Yang et al. 2019     \\
SHELLQ J1004+0239		&   6.41     &      -24.52     &      2.65e+08     & 	 Matsuoka et al. 2018b     \\
DELS J153532.87+194320.1  	&   6.40     &      -27.01     &      2.63e+09     & 	 Wang et al. 2018b     \\
SHELLQ J113753.64+004509.7	&   6.40     &      -24.20     &      1.97e+08     & 	 Matsuoka et al. 2019b     \\
SHELLQ J084456.62+022640.5	&   6.40     &      -21.57     &      1.75e+07     & 	 Matsuoka et al. 2019b     \\
HSC J2236+0032 			&   6.40     &      -23.60     &      1.13e+08     & 	 Banados et al. 2016     \\
HSC J0859+0022			&   6.39     &      -23.62     &      1.15e+08     & 	 Banados et al. 2016     \\
PSO J159.2257-02.5438		&   6.38     &      -26.80     &      2.16e+09     & 	 Banados et al. 2016     \\
DELS J080305.42+313834.2  	&   6.37706     &      -26.51     &      1.66e+09     & 	 Wang et al. 2018b     \\
VIK J1152+0055			&   6.37     &      -25.13     &      4.65e+08     & 	 Banados et al. 2016     \\
SHELLQ J0211-0203		&   6.37     &      -23.36     &      9.12e+07     & 	 Matsuoka et al. 2018b     \\
SHELLQ J2304+0045		&   6.36     &      -24.28     &      2.12e+08     & 	 Matsuoka et al. 2018b     \\
DELS J131608.14+102832.8 	&   6.35     &      -25.73     &      8.09e+08     & 	 Wang et al. 2018b     \\
SHELLQ J0857+0056 		&   6.35     &      -23.01     &      6.61e+07     & 	 Matsuoka et al. 2018a     \\
SHELLQ J2255+0251		&   6.34     &      -23.87     &      1.45e+08     & 	 Matsuoka et al. 2018b     \\
VIK J2211–3206			&   6.3394     &      -26.71     &      2.00e+09     & 	 Decarli et al. 2018            \\
SHELLQ J1406-0116		&   6.33     &      -24.96     &      3.98e+08     & 	 Matsuoka et al. 2018b     \\
SDSS J010013.02+280225.8	&   6.3258     &      -29.14     &      1.24e+10     & 	 Banados et al. 2016$^{(b)}$     \\
ATLAS J332.8017-32.1036	        &   6.32     &      -26.79     &      2.14e+09     & 	 Chehade et al. 2018     \\
ULAS J1148+0702			&   6.32     &      -26.48     &      1.61e+09     & 	 Banados et al. 2016     \\
VST-ATLAS J0142-3327 		&   6.31     &      -27.80     &      5.45e+09     & 	 Carnall et al. 2015           \\
VST-ATLAS J025.6821-33.4627	&   6.31     &      -27.81     &      5.49e+09     & 	 Banados et al. 2016     \\
SDSS J1030+0524			&   6.308     &      -26.99     &      2.58e+09     & 	 Banados et al. 2016     \\
SHELLQ J1146-0005		&   6.30     &      -21.46     &      1.58e+07     & 	 Matsuoka et al. 2018b     \\
SHELLQ J152555.79+430324.0	&   6.27     &      -23.61     &      1.14e+08     & 	 Matsuoka et al. 2019b     \\
SHELLQ J0905+0300 		&   6.27     &      -22.55     &      4.32e+07     & 	 Matsuoka et al. 2018a     \\
SHELLQ J1146+0124		&   6.27     &      -23.71     &      1.25e+08     & 	 Matsuoka et al. 2018b     \\
SHELLQ J2239+0207 		&   6.26     &      -24.69     &      3.10e+08     & 	 Matsuoka et al. 2018a     \\
SDSS J1623+3112			&   6.26     &      -26.55     &      1.72e+09     & 	 Banados et al. 2016     \\
CFHQS J0050+3445		&   6.253     &      -26.70     &      1.97e+09     & 	 Banados et al. 2016     \\
VDES J0330-4025 		&   6.25     &      -26.42     &      1.52e+09     & 	 Reed et al. 2017     \\
VDES J0323-4701 		&   6.25     &      -26.02     &      1.05e+09     & 	 Reed et al. 2017     \\
VDES J0143-5545 	        &   6.25     &      -25.65     &      7.52e+08     & 	 Reed et al. 2017     \\
SHELLQ J0844-0052 		&   6.25     &      -23.74     &      1.29e+08     & 	 Matsuoka et al. 2018a     \\
PSO J308.0416-21.2339		&   6.24     &      -26.35     &      1.43e+09     & 	 Banados et al. 2016     \\
SDSS J1048+4637			&   6.2284     &      -27.24     &      3.25e+09     & 	 Banados et al. 2016     \\
VDES J0410-4414 		&   6.21     &      -26.14     &      1.18e+09     & 	 Reed et al. 2017     \\
\hline
\end{tabular}
\end{center}

\newpage
\begin{center}
\begin{tabular}{@{}l|c|c|c|c@{}}
\hline
NAME                         & Redshift  &   ${\rm M}_{1450}$    &   Mass$^{(a)}$	 &     Reference\\
                             &           &                      &   ${\rm M_\odot}$ &          \\
\hline
CFHQS J0136+0226		&   6.21     &      -24.66     &      3.02e+08     & 	 Banados et al. 2016     \\
SHELLQ J0217-0208 		&   6.20     &      -23.19     &      7.80e+07     & 	 Matsuoka et al. 2018a     \\
PSO J184.3389+01.5284		&   6.20     &      -25.37     &      5.81e+08     & 	 Banados et al. 2016     \\
CFHQS J0227-0605		&   6.20     &      -25.28     &      5.34e+08     & 	 Banados et al. 2016     \\
SHELLQ J1208-0200 		&   6.2     &      -24.73     &      3.22e+08     & 	 Matsuoka et al. 2018a     \\
SHELLQ J151248.71+442217.5	&   6.19     &      -22.07     &      2.78e+07     & 	 Matsuoka et al. 2019b     \\
SHELLQ J0918+0139		&   6.19     &      -23.71     &      1.25e+08     & 	 Matsuoka et al. 2018b     \\
CFHQS J1429+5447		&   6.1831     &      -26.10     &      1.13e+09     & 	 Banados et al. 2016     \\
SHELLQ J225520.78+050343.3	&   6.18     &      -24.43     &      2.44e+08     & 	 Matsuoka et al. 2019b     \\
SHELLQ J1425-0015 		&   6.18     &      -23.44     &      9.82e+07     & 	 Matsuoka et al. 2018a     \\
PSO J060.5529+24.8567		&   6.18     &      -26.95     &      2.49e+09     & 	 Banados et al. 2016     \\
SHELLQ J0844-0132		&   6.18     &      -23.97     &      1.60e+08     & 	 Matsuoka et al. 2018b     \\
HSC J2232+0012 			&   6.18     &      -22.76     &      5.25e+07     & 	 Banados et al. 2016     \\
DELS J121721.35+013142.52	&   6.17     &      -25.76     &      8.32e+08     & 	 Wang et al. 2017     \\
CFHQS J0221-0802		&   6.161     &      -24.70     &      3.13e+08     & 	 Banados et al. 2016     \\
SHELLQ J2201+0155 		&   6.16     &      -22.97     &      6.37e+07     & 	 Matsuoka et al. 2018a     \\
SHELLQ J1146-0154		&   6.16     &      -23.43     &      9.73e+07     & 	 Matsuoka et al. 2018b     \\
VIMOS2911001793			&   6.16     &      -22.60     &      4.53e+07     & 	 Banados et al. 2016     \\
HSC J2219+0102			&   6.156     &      -23.10     &      7.19e+07     & 	 Kashikawa et al. 2015        \\
CFHQS J2229+1457		&   6.1517     &      -24.78     &      3.37e+08     & 	 Banados et al. 2016     \\
SHELLQ J134733.69-015750.6	&   6.15     &      -24.73     &      3.22e+08     & 	 Matsuoka et al. 2019b     \\
PSO J359.1352-06.3831		&   6.15     &      -26.79     &      2.14e+09     & 	 Banados et al. 2016     \\
SDSS J1250+3130			&   6.15     &      -26.53     &      1.69e+09     & 	 Banados et al. 2016     \\
SHELLQ J0909+0440		&   6.15     &      -24.88     &      3.70e+08     & 	 Matsuoka et al. 2018b     \\
SHELLQ J0834+0211		&   6.15     &      -24.05     &      1.72e+08     & 	 Matsuoka et al. 2018b     \\
SDSS J235632.44–062259.2	&   6.15     &      -26.85     &      2.27e+09     & 	 Wang et al. 2016	     \\
VIK J2318–3029			&   6.1458     &      -26.21     &      1.26e+09     & 	 Decarli et al. 2018            \\
SHELLQ J144823.33+433305.9	&   6.14     &      -24.36     &      2.29e+08     & 	 Matsuoka et al. 2019b     \\
PSO J065.4085-26.9543		&   6.14     &      -27.25     &      3.28e+09     & 	 Banados et al. 2016     \\
PSO J210.4472+27.8263		&   6.14     &      -26.54     &      1.70e+09     & 	 Banados et al. 2016     \\
ULAS J1609+3041			&   6.14     &      -26.38     &      1.47e+09     & 	 Banados et al. 2016     \\
PSO J217.9185-07.4120		&   6.14     &      -26.35     &      1.43e+09     & 	 Banados et al. 2016     \\
ULAS J1319+0950			&   6.133     &      -27.05     &      2.73e+09     & 	 Banados et al. 2016     \\
SHELLQ J151657.87+422852.9	&   6.13     &      -24.35     &      2.27e+08     & 	 Matsuoka et al. 2019b     \\
SHELLQ J000133.30+000605.4	&   6.13     &      -23.72     &      1.27e+08     & 	 Matsuoka et al. 2019b     \\
SHELLQ J125437.08-001410.7	&   6.13     &      -20.91     &      9.55e+06     & 	 Matsuoka et al. 2019b     \\
SHELLQ J1440-0107 		&   6.13     &      -22.59     &      4.49e+07     & 	 Matsuoka et al. 2018a     \\
SHELLQ J1423-0018 		&   6.13     &      -21.88     &      2.33e+07     & 	 Matsuoka et al. 2018a     \\
CFHQS J0033-0125		&   6.13     &      -25.14     &      4.70e+08     & 	 Banados et al. 2016     \\
CFHQS J1509-1749		&   6.121     &      -27.14     &      2.96e+09     & 	 Banados et al. 2016     \\
PSO J065.5041-19.4579		&   6.12     &      -26.62     &      1.83e+09     & 	 Banados et al. 2016     \\
FIRST J1427+3312		&   6.12     &      -26.10     &      1.13e+09     & 	 Banados et al. 2016     \\
SDSS J2315-0023			&   6.12     &      -25.66     &      7.59e+08     & 	 Banados et al. 2016     \\
SHELLQ J2252+0225		&   6.12     &      -22.74     &      5.15e+07     & 	 Matsuoka et al. 2018b     \\
PSO J239.7124-07.4026		&   6.11     &      -27.46     &      3.98e+09     & 	 Banados et al. 2016     \\
PSO J217.0891-16.0453		&   6.11     &      -26.93     &      2.44e+09     & 	 Banados et al. 2016     \\
SHELLQ J000445.81-004944.3	&   6.10     &      -23.90     &      1.50e+08     & 	 Matsuoka et al. 2019b     \\
VDES J0454-4448a 		&   6.10     &      -26.36     &      1.44e+09     & 	 Reed et al. 2017     \\
\hline
\end{tabular}
\end{center}

\newpage
\begin{center}
\begin{tabular}{@{}l|c|c|c|c@{}}
\hline
NAME                         & Redshift  &   ${\rm M}_{1450}$    &   Mass$^{(a)}$	 &     Reference\\
                             &           &                      &   ${\rm M_\odot}$ &          \\
\hline
PSO J002.3786+32.8702		&   6.10     &      -25.65     &      7.52e+08     & 	 Banados et al. 2016     \\
HSC J2216-0016			&   6.10     &      -23.62     &      1.15e+08     & 	 Banados et al. 2016     \\
SHELLQ J1406-0144		&   6.10     &      -23.37     &      9.20e+07     & 	 Matsuoka et al. 2018b     \\
SHELLQ J0235-0532 		&   6.09     &      -23.01     &      6.61e+07     & 	 Matsuoka et al. 2018a     \\
SDSS J1602+4228			&   6.09     &      -26.94     &      2.46e+09     & 	 Banados et al. 2016     \\
DES J0454-4448			&   6.09     &      -26.47     &      1.60e+09     & 	 Banados et al. 2016     \\
CFHQS J2100-1715		&   6.087     &      -25.55     &      6.85e+08     & 	 Banados et al. 2016     \\
SHELLQ J093543.32-011033.3	&   6.08     &      -21.97     &      2.53e+07     & 	 Matsuoka et al. 2019b     \\
SHELLQ J2228+0152 		&   6.08     &      -24.00     &      1.64e+08     & 	 Matsuoka et al. 2018a     \\
PSO J293.0317+71.6523		&   6.08     &      -26.92     &      2.42e+09     & 	 Banados et al. 2016     \\
SDSS J0303-0019			&   6.078     &      -25.56     &      6.92e+08     & 	 Banados et al. 2016     \\
SDSS J0353+0104			&   6.072     &      -26.43     &      1.54e+09     & 	 Banados et al. 2016     \\
ATLAS J158.6938-14.4211 	&   6.07     &      -27.23     &      3.22e+09     & 	 Chehade et al. 2018     \\
DELS J155909.09+221214.43	&   6.07     &      -25.83     &      8.87e+08     & 	 Wang et al. 2017     \\
VDES J0420-4453 		&   6.07     &      -26.25     &      1.30e+09     & 	 Reed et al. 2017     \\
SHELLQ J0911+0152 		&   6.07     &      -22.09     &      2.83e+07     & 	 Matsuoka et al. 2018a     \\
SHELLQ J1416+0147		&   6.07     &      -23.27     &      8.39e+07     & 	 Matsuoka et al. 2018b     \\
SDSS J0842+1218			&   6.069     &      -26.91     &      2.40e+09     & 	 Banados et al. 2016     \\
SDSS J1630+4012			&   6.065     &      -26.19     &      1.23e+09     & 	 Banados et al. 2016     \\
SHELLQ J010603.68-003015.2	&   6.06     &      -23.53     &      1.06e+08     & 	 Matsuoka et al. 2019b     \\
SHELLQ J1201+0133 		&   6.06     &      -23.85     &      1.43e+08     & 	 Matsuoka et al. 2018a     \\
PSO J089.9394-15.5833		&   6.05     &      -26.93     &      2.44e+09     & 	 Banados et al. 2016     \\
ULAS J0828+2633			&   6.05     &      -26.37     &      1.45e+09     & 	 Banados et al. 2016     \\
SHELLQ J2223+0326		&   6.05     &      -25.20     &      4.96e+08     & 	 Matsuoka et al. 2018b     \\
CFHQS J2318-0246		&   6.05     &      -25.10     &      4.53e+08     & 	 Banados et al. 2016     \\
SHELLQ J0957+0053		&   6.05     &      -22.98     &      6.43e+07     & 	 Matsuoka et al. 2018b     \\
CFHQS J1641+3755		&   6.047     &      -25.67     &      7.66e+08     & 	 Banados et al. 2016     \\
HSC J1603+5510			&   6.041     &      -22.58     &      4.45e+07     & 	 Kashikawa et al. 2015        \\
SHELLQ J1429-0002 		&   6.04     &      -23.42     &      9.64e+07     & 	 Matsuoka et al. 2018a     \\
ULAS J1207+0630			&   6.04     &      -26.63     &      1.85e+09     & 	 Banados et al. 2016     \\
PSO J210.7277+40.4008		&   6.04     &      -25.86     &      9.12e+08     & 	 Banados et al. 2016     \\
SHELLQ J1400-0125		&   6.04     &      -23.70     &      1.24e+08     & 	 Matsuoka et al. 2018b     \\
SHELLQ J1400-0011		&   6.04     &      -22.95     &      6.25e+07     & 	 Matsuoka et al. 2018b     \\
ELAIS1091000446 		&   6.04     &      -22.64     &      4.70e+07     & 	 Banados et al. 2016     \\
SDSS J2054-0005			&   6.0391     &      -26.21     &      1.25e+09     & 	 Banados et al. 2016     \\
VDES J0408-5632 		&   6.03     &      -26.51     &      1.66e+09     & 	 Reed et al. 2017     \\
SHELLQ J0206-0255 		&   6.03     &      -24.91     &      3.80e+08     & 	 Matsuoka et al. 2018a     \\
SHELLQ J0202-0251 		&   6.03     &      -23.39     &      9.38e+07     & 	 Matsuoka et al. 2018a     \\
SHELLQ J1416+0015 		&   6.03     &      -22.39     &      3.73e+07     & 	 Matsuoka et al. 2018a     \\
SDSS J1137+3549			&   6.03     &      -27.36     &      3.63e+09     & 	 Banados et al. 2016     \\
PSO J333.9859+26.1081		&   6.03     &      -26.44     &      1.55e+09     & 	 Banados et al. 2016     \\
SHELLQ J1417+0117 		&   6.02     &      -22.83     &      5.60e+07     & 	 Matsuoka et al. 2018a     \\
SDSS J0818+1722			&   6.02     &      -27.52     &      4.20e+09     & 	 Banados et al. 2016     \\
VST-ATLAS J029.9915-36.5658	&   6.02     &      -27.00     &      2.60e+09     & 	 Banados et al. 2016     \\
SDSS J1257+6349			&   6.02     &      -26.27     &      1.33e+09     & 	 Banados et al. 2016     \\
VST-ATLAS J0142-3327 		&   6.02     &      -27.00     &      2.61e+09     & 	 Carnall et al. 2015         \\
SDSS J1306+0356			&   6.016     &      -26.81     &      2.18e+09     & 	 Banados et al. 2016     \\
SHELLQ J0902+0155 		&   6.01     &      -22.51     &      4.17e+07     & 	 Matsuoka et al. 2018a     \\
SHELLQ J0853+0139 		&   6.01     &      -22.51     &      4.17e+07     & 	 Matsuoka et al. 2018a     \\
\hline
\end{tabular}
\end{center}

\newpage
\begin{center}
\begin{tabular}{@{}l|c|c|c|c@{}}
\hline
NAME                         & Redshift  &   ${\rm M}_{1450}$    &   Mass$^{(a)}$	 &     Reference\\
                             &           &                      &   ${\rm M_\odot}$ &          \\
\hline
PSO J340.2041-18.6621		&   6.01     &      -26.42     &      1.52e+09     & 	 Banados et al. 2016     \\
SHELLQ J1219+0050		&   6.01     &      -23.85     &      1.43e+08     & 	 Matsuoka et al. 2018b     \\
HSC J1207-0005 			&   6.01     &      -22.62     &      4.61e+07     & 	 Banados et al. 2016     \\
CFHQS J0216-0455 		&   6.01     &      -22.49     &      4.09e+07     & 	 Banados et al. 2016     \\
HSC J2228+0128 			&   6.01     &      -22.41     &      3.80e+07     & 	 Banados et al. 2016     \\
SHELLQ J142611.33-012822.8	&   6.01     &      -23.75     &      1.30e+08     & 	 Matsuoka et al. 2019b     \\
CFHQS J0055+0146		&   6.006     &      -24.81     &      3.46e+08     & 	 Banados et al. 2016     \\
SDSS J2310+1855			&   6.0031     &      -27.80     &      5.44e+09     & 	 Banados et al. 2016     \\
VDES J2250-5015 		&   6.00     &      -26.80     &      2.16e+09     & 	 Reed et al. 2017     \\
PSO J007.0273+04.9571		&   6.00     &      -26.64     &      1.87e+09     & 	 Banados et al. 2016     \\
PSO J037.9706-28.8389		&   6.00     &      -26.23     &      1.28e+09     & 	 Banados et al. 2016     \\
SDSS J2356+0023			&   6.00     &      -25.50     &      6.55e+08     &        Banados et al. 2016     \\
\hline
\end{tabular}
\end{center}

  \vspace{\baselineskip}
  
\noindent $^{\rm (a)}$ Masses are obtained from the rest-frame UV
magnitude ${\rm M}_{1450}$ as: \\ Mass = 10$^{[(-{\rm M}_{1450} -
    3.459)/2.5]}$, which yields, on average, the published virial mass
estimates for those quasars for which virial masses are available.  We
caution that several of the least luminous quasars, discovered
predominantly in the SHELLQs survey, have Eddington ratios below unity
and lower than those of the brighter sources; the masses for these
least luminous sources are underestimated (see main text).

  \vspace{\baselineskip}

  \noindent $^{\rm (b)}$The quoted mass is a virial estimate from a broad
  emission line.

  \vspace{\baselineskip}

  \noindent $^{\rm (c)}$Quasars from prior to March 2016 were adopted from
  the compilation by Banados et al. (2016). Please see Table 7 of that
  paper for references to the original discoveries.

  \vspace{\baselineskip}

  \noindent $^{\rm (d)}$ ${\rm M}_{1450} $has not been published for this
  source.  This quasar is strongly lensed (by factor of 51.3); the
  quoted mass is the virial estimate, after a lensing correction.

\end{document}